\documentclass[aip,amsmath,amssymb,preprint]{revtex4-1}
\usepackage{graphicx}% Include figure files
\usepackage{dcolumn}% Align table columns on decimal point
\usepackage{bm}% bold math
\usepackage{amsmath}
\usepackage{mathrsfs}
\usepackage{subfigure}
\usepackage{threeparttable}
\usepackage{hyperref}
\usepackage{color}
\usepackage{CJK}
\usepackage[utf8]{inputenc}
\usepackage{multirow}
\usepackage{algorithm}
\usepackage{algpseudocode}

\draft % marks overfull lines with a black rule on the right

\begin{document}
\begin{CJK*}{UTF8}{gbsn}

\title{Phenomenological energy exchange of diatomic gases: Comparison of Pullin and Borgnakke-Larsen models in direct simulation Monte Carlo method} %Title of paper

\author{Hao Jin (金浩)}
\email[]{jinhao@mail.nwpu.edu.cn}
\affiliation{School of Aeronautics, Northwestern Polytechnical University, Xi'an, Shaanxi 710072, China}
\author{Sha Liu (刘沙)}
\email[Corresponding author: ]{shaliu@nwpu.edu.cn}
\affiliation{School of Aeronautics, Northwestern Polytechnical University, Xi'an, Shaanxi 710072, China}
\affiliation{Institute of Extreme Mechanics, Northwestern Polytechnical University, Xi'an, Shaanxi 710072, China}
\affiliation{National Key Laboratory of Aircraft Configuration Design, Northwestern Polytechnical University, Xi'an, Shaanxi 710072, China}
\author{Ningchao Ding (丁宁超)}
\affiliation{School of Aeronautics, Northwestern Polytechnical University, Xi'an, Shaanxi 710072, China}
\author{Sirui Yang (杨思睿)}
\affiliation{School of Aeronautics, Northwestern Polytechnical University, Xi'an, Shaanxi 710072, China}
\author{Huahua Cui (崔华华)}
\affiliation{Dawning Information Industry Co., Ltd, Binhai High-tech Industrial Development Zone, Tianjin 300384, China}
\author{Congshan Zhuo (卓丛山)}
\affiliation{School of Aeronautics, Northwestern Polytechnical University, Xi'an, Shaanxi 710072, China}
\affiliation{Institute of Extreme Mechanics, Northwestern Polytechnical University, Xi'an, Shaanxi 710072, China}
\affiliation{National Key Laboratory of Aircraft Configuration Design, Northwestern Polytechnical University, Xi'an, Shaanxi 710072, China}
\author{Chengwen Zhong (钟诚文)}
\affiliation{School of Aeronautics, Northwestern Polytechnical University, Xi'an, Shaanxi 710072, China}
\affiliation{Institute of Extreme Mechanics, Northwestern Polytechnical University, Xi'an, Shaanxi 710072, China}
\affiliation{National Key Laboratory of Aircraft Configuration Design, Northwestern Polytechnical University, Xi'an, Shaanxi 710072, China}

\date{\today}

\begin{abstract}
	In hypersonic rarefied flows, insufficient intermolecular collisions cause significant deviations between translational and rotational temperatures, leading to strong thermal nonequilibrium. This behavior differs markedly from continuum flows, where these temperatures are nearly identical and can be represented by a reduced translational-rotational temperature. The redistribution of energy among different modes substantially affects the flow-field structure and surface heating of hypersonic vehicles, making accurate modeling of energy exchange essential for rarefied flow simulations. For diatomic gases such as nitrogen and oxygen, the direct simulation Monte Carlo (DSMC) method commonly employs the Borgnakke-Larsen (BL) model to simulate translational-rotational energy exchange (relaxation) processes. Although widely used, the BL model lacks a rigorous theoretical foundation and assumes that only a fraction of collisions lead to rotational relaxation, limiting its physical realism. To address these shortcomings, Pullin introduced a kinetically consistent relaxation model into the gas kinetic theory. By employing the Beta function for energy partitioning, a concrete collision cross section that satisfies the detailed balance condition is constructed. In this study, a comparative investigation of the BL and Pullin models is performed within the DSMC framework, where both original and simplified equations are considered and parameterized by physical accommodated coefficient in the Beta function. A series of test cases--including zero-dimensional rotational relaxation of nitrogen, one-dimensional planar Couette flow and normal shock wave, two-dimensional hypersonic flow past a cylinder, and three-dimensional hypersonic flow around an X38-like vehicle--are performed to assess the accuracy and efficiency of these models. The results confirm the consistency between the Pullin and BL models. Owing to its rigorous theoretical foundation and accurate physical representation, the Pullin model is expected to provide substantial support for the extension of subsequent theoretical studies and numerical simulations. Moreover, in the highly rarefied flow regime (Knudsen number greater than 1, or altitudes above 100 km), the simplified Pullin model exhibits performance comparable to that of the BL model.
\end{abstract}

%\pacs{}% insert suggested PACS numbers in braces on next line

\maketitle %\maketitle must follow title, authors, abstract and 
% \clearpage
\end{CJK*}

\section{Introduction}\label{sec1_Introduction}
In hypersonic flows, diatomic gases (e.g., N$_2$ and O$_2$) undergo complex energy exchange processes associated with their internal degrees of freedom (DOFs), including rotational and vibrational modes. Under continuum conditions at moderate altitudes, collisions occur frequently enough that energy transfer among translational, rotational, and vibrational modes tends toward equipartition--particularly between translational and rotational modes--thereby maintaining an approximate local thermodynamic equilibrium \cite{zhang2022review}. At higher altitudes, however, the mean free path increases and collisional energy transfer becomes significantly slower. As a result, the relaxation times of internal energy modes may become comparable to characteristic flow timescales, giving rise to thermodynamic nonequilibrium\cite{anderson1989hypersonic,boyd2017nonequilibrium,schouler2020survey}. In this rarefied regime, the redistribution of energy among different modes strongly influences the flowfield structure, surface heating, and overall aerothermodynamic performance of hypersonic vehicles. Accurate modeling of such energy exchange processes is therefore essential in rarefied hypersonic gas dynamics.
    
To capture thermodynamic nonequilibrium phenomena in rarefied flows, a variety of numerical methods have been developed for solving the Boltzmann equation, which are generally categorized into deterministic approaches\cite{liu2014unified,hu2021gas,zhang2023multiscale} and stochastic particle approaches\cite{bird1994molecular,pfeiffer2018extending,xu2021unified,fei2022unified}. Among these methods, the direct simulation Monte Carlo (DSMC)\cite{bird1994molecular,bird2013DSMC} method has become the most widely used stochastic approach for rarefied nonequilibrium gas flows. The DSMC method decouples molecular motion and collisions within each time step: molecules are first advanced in free transport at constant velocity, after which representative collision pairs are selected through probabilistic sampling. By explicitly modeling molecular collision pairs, the DSMC method can efficiently incorporate complex physico-chemical processes--including translational-internal energy exchange\cite{kosyanchuk2021detailed,eckert2022enforcing}, chemical reactions\cite{trivedi2025simulations,gokul2024insights}, and ionization\cite{fang2020dsmc}--without significantly increasing the computational cost, making it particularly well suited for hypersonic nonequilibrium flow simulations.

Several phenomenological models\cite{borgnakke1975statistical,pullin1978kinetic,erofeev1995numerical,prasanth2012variable,macrossan2021rotation,lu2022universal} for translational-internal energy exchange have been proposed and implemented in DSMC simulations. Among them, the Borgnakke-Larsen (BL) model\cite{borgnakke1975statistical}, proposed in 1975, is the most widely adopted approach for modeling energy transfer between translational and internal modes in diatomic molecules. In the BL model, a fraction of colliding particles is randomly selected to undergo inelastic collisions. Elastic collisions involve only changes in particle velocities, while inelastic collisions redistribute post-collision energies between translational and internal modes, with the new energies sampled from an equilibrium distribution corresponding to the collision energy. The probability of inelastic collisions is determined by a relaxation collision number, which enables this model to reproduce experimentally observed relaxation rates\cite{haas1994rates}. When the temperature-dependent relaxation models were introduced--such as Parker's model for rotational energy\cite{boyd1993temperature} and the Millikan-White correlation for vibrational energy\cite{vijayakumar1999vibrational}--it was shown that the detailed balance principle is satisfied if the probability of energy exchange depends exclusively on collision invariants (total collisional energy)\cite{eckert2022enforcing}. The BL model is simple and computationally efficient; however, it is physically unrealistic, as only a fraction of collisions are inelastic. To improve its physical fidelity, a restricted energy exchange\cite{bird1994molecular} variant of the model has been proposed, in which all colliding particles undergo inelastic collisions. However, this modification compromises computational efficiency and has been shown not to satisfy the detailed balance principle\cite{pullin1978kinetic,bird1994molecular}. During energy exchange, the principle of detailed balance requires that, at equilibrium, the probability of a forward collision must equal that of the corresponding reverse collision. Violation of this condition may produce unphysical equilibrium states, leading to incorrect distributions of translational and internal energies or inaccurate relaxation rates.

On the other hand, in studies focused on constructing a collision kernel, Pullin proposed an alternative kinetic model in which every degree of freedom participates in the energy exchange during each collision\cite{pullin1978kinetic}, and the detailed balance principle is satisfied. This model provides a more physically realistic description of rotation-translation energy exchange. In Pullin's original model, the energy exchange between rotational and translational modes is represented by five Beta-distributed variates, which can be readily implemented in DSMC simulations. However, this model is computationally expensive. To reduce the computational cost, a simplified variant was introduced employing only three Beta-distributed variates, in which the sum of the rotational energies of the colliding pair is treated as a whole, and the post-collision rotational energy is uniformly distributed\cite{erofeev1995numerical}. Another simplified model also employs three Beta-distributed variates, in which only a portion of the translational energy participates in the exchange, and has been applied to vibration-translation energy exchange in DSMC simulations\cite{macrossan2021rotation}. Compared with the BL model, these models offer a more physically realistic description of energy exchange; however, their higher computational cost limit their practical use in DSMC simulations. Moreover, the free parameter in the Pullin model for VHS molecules has not been well established. Consequently, comparative studies on the accuracy and efficiency of the Pullin and BL models remain limited.

Recently, a new rotational-translational distribution function based on Pullin model has been proposed, in which the relationship between the model parameters and the relaxation process is established through the relaxation rates of macroscopic quantities\cite{Ding2602.05775}. In this work, we employ this relation to determine the model parameters for VHS molecules within Pullin's framework and implement the model in the DSMC method. The performance of Pullin's model and its simplified variant is then assessed against that of the BL model through a series of numerical simulations. The remainder of this paper is organized as follows. Section \ref{sec2_Pullin} provides a brief review of Pullin model, its simplified variant, and an energy partition parameter suitable for the VHS model. Section \ref{sec3_DSMC} introduces the collision procedure in the DSMC method and outlines the implementation of the Pullin model within DSMC framework, compared with the BL model. Section \ref{sec4_Numerical} presents numerical simulations of several typical rarefied gas flows to offer a direct comparison of accuracy and efficiency. Finally, concluding remarks are given in Section \ref{sec5_Conclusion}.

\section{Restricted energy exchange scheme}\label{sec2_Pullin}

\subsection{Review of the Pullin model}\label{subSec2_Pullin}
The microscopic state of the gas is described by the density distribution function $f\left( \boldsymbol{c},\boldsymbol{x},t,\boldsymbol{e} \right)$, where $t$, $\boldsymbol{x}$, $\boldsymbol{c}$, and $\boldsymbol{e}$ denote the time, position vector, particle velocity, and particle internal energy, respectively. In the absence of external forces, the evolution of $f$ is governed by the reduced generalized Boltzmann equation\cite{pullin1978kinetic}, given by
\begin{equation}
    \begin{aligned}
        \frac{\partial \left( nf \right)}{\partial t}+\boldsymbol{c}\frac{\partial \left( nf \right)}{\partial \boldsymbol{x}}=Q\left( f,f \right),
    \end{aligned} \label{eq_boltzmann}
\end{equation}
where $n$ is the number density, and the bilinear collision operator $Q\left(f,f\right)$ describes binary particle collisions. The operator takes the form as
\begin{equation}
    \begin{aligned}
        Q\left( f,f \right)={{n}^{2}}\int{\int{\cdots \int{\left[ f_{1}^{'}f_{2}^{'}J-{{f}_{1}}{{f}_{2}} \right]c_rI\left( \boldsymbol{e}|{\boldsymbol{e}^{'}} \right)d{\boldsymbol{e}^{'}}d{{\Omega }^{'}}d{\boldsymbol{c}_{2}}d{{e}_{2}}}}},
    \end{aligned} \label{eq_collision_operator}
\end{equation}
where $\boldsymbol{e}=\left( {{e}_{1}},{{e}_{2}},{{e}_{t}} \right)$ represents the components of the internal energy, the factor $J={{\left( {{e}_{1}}{{e}_{2}}/e_{1}^{'}e_{2}^{'} \right)}^{\zeta -1}}$ with $\zeta =\upsilon /2$, and $\upsilon$ denoting the internal degrees of freedom. Here, $c_r$ is the relative speed, $d{\Omega }^{'}$ is the unit solid angle, and $I\left( \boldsymbol{e}|{\boldsymbol{e}^{'}} \right)$ is the scattering kernel. The scattering kernel $I\left( \boldsymbol{e}|{\boldsymbol{e}^{'}} \right)$ satisfies the detailed balancing relation if inverse collisions exist, namely,
\begin{equation}
    \begin{aligned}
        e_{t}e_{1}^{\zeta -1}e_{2}^{\zeta -1}I\left( \boldsymbol{e}|{\boldsymbol{e}^{'}} \right)=e_{t}^{'}e_{1}^{'\zeta -1}e_{2}^{'\zeta -1}I\left( {\boldsymbol{e}^{'}}|\boldsymbol{e} \right).
    \end{aligned} \label{eq_detail}
\end{equation}

To simplify the description of momentum and energy exchange during collisions, the scattering kernel $I$ is assumed to separate into independent velocity and energy scattering processes, which is
\begin{equation}
    \begin{aligned}
        I\left( \boldsymbol{e}|{\boldsymbol{e}^{'}} \right)=R\left( \boldsymbol{e}|{\boldsymbol{e}^{'}} \right)\sigma \left( \chi  \right),
    \end{aligned} \label{eq_separate}
\end{equation}
where $\sigma$ is the cross section, $\chi$ is the polar deflection angle, and $R$ is the energy scattering kernel. Notice that the decoupling of scattering and internal energy exchange is also consistent with the BL and other phenomenological models. Substituting Eq.(\ref{eq_separate}) into the collision operator (Eq.(\ref{eq_collision_operator})) yields a model $Q$, in which the monatomic geometrical scattering properties during collisions are retained and the polyatomic effects (internal-translational energy exchange) are represented phenomenologically through an appropriate choice of $R$.

For the cross section $\sigma$, consider the inverse-power-law molecules with intermolecular potential $V=a/{{r}^{\alpha }}$, where $r$ is the particle separation and $a$ is a constant. For the energy scattering kernel $R$, substituting the separated form (Eq.(\ref{eq_separate})) of $I$ into the detailed balance relation (Eq.(\ref{eq_detail})) leads to the following condition
\begin{equation}
    \begin{aligned}
        e_{t}^{\eta -1}e_{1}^{\zeta -1}e_{2}^{\zeta -1}R\left( \boldsymbol{e}|{\boldsymbol{e}^{'}} \right)=e_{t}^{'\eta -1}e_{1}^{'\zeta -1}e_{2}^{'\zeta -1}R\left( {\boldsymbol{e}^{'}}|\boldsymbol{e} \right),
    \end{aligned}
\end{equation}
where $\eta =2-2/\alpha $. In addition, the energy scattering kernel $R$ must satisfy the following requirements:
(1) Energy conservation: ${{e}_{0}}={{e}_{1}}+{{e}_{2}}+{{e}_{t}}=e_{1}^{'}+e_{2}^{'}+e_{t}^{'}=e_{0}^{'}$.
(2) Non-negativity: $R\ge 0$.
(3) Normalization: for all $\boldsymbol{e}$, $\iint{R\left( \boldsymbol{e}|{\boldsymbol{e}^{'}} \right)d{\boldsymbol{e}^{'}}}=1$.

It is generally difficult to construct an explicit formulation for $R$. Pullin\cite{pullin1978kinetic} introduced an $l$-dimensional random vector $\boldsymbol{s}=\left( {{s}_{1}},{{s}_{2}},\cdots ,{{s}_{l}} \right)$ with probability density $h\left( \boldsymbol{s} \right)=\prod\nolimits_{i=1}^{l}{{{h}_{i}}\left( {{s}_{i}} \right)}$ such that $\iint{h\left( \boldsymbol{s} \right)d\boldsymbol{s}=1}$, where $d\boldsymbol{s}=d{{s}_{1}}d{{s}_{2}}\ldots d{{s}_{l}}$. Using this representation, an alternative expression for $R$ can be written as
\begin{equation}
    \begin{aligned}
        R\left( \boldsymbol{e}|{\boldsymbol{e}^{'}} \right)d{\boldsymbol{e}^{'}}=\iint_{v}{h\left( \boldsymbol{s} \right)d\boldsymbol{s}},
    \end{aligned}
\end{equation}
where $h(\boldsymbol{s})$ is defined by five Beta distribution functions as
\begin{equation}
    \begin{aligned}
        h(\boldsymbol{s})=\text{ }\left[ \prod\limits_{j=1}^{2}{\beta }\left\langle {{s}_{j}}|\phi \zeta ,(1-\phi )\zeta  \right\rangle  \right]\beta \left\langle {{s}_{3}}|\psi \eta ,(1-\psi )\eta  \right\rangle \beta \left\langle {{s}_{4}}|\phi \zeta ,\phi \zeta  \right\rangle \beta \left\langle {{s}_{5}}|2\phi \zeta ,\psi \eta  \right\rangle. \label{eq_hs}
    \end{aligned}
\end{equation}
Here, $0<\phi <1$ and $0<\psi <1$ are arbitrary functions (model parameters) of total energy $ {{e}_{0}}/k{{T}^{'}}$ and viscosity temperature index $\omega$, where ${T}^{'}$ is any reference temperature. The Beta distribution with parameters $\mu_{1}$ and $\mu_2$ is defined as
\begin{equation}
    \begin{aligned}
        \beta \left\langle z\left| {{\mu }_{1}},{{\mu }_{2}} \right. \right\rangle =\frac{1}{B\left( {{\mu }_{1}},{{\mu }_{2}} \right)}{{z}^{{{\mu }_{1}}-1}}{{\left( 1-z \right)}^{{{\mu }_{2}}-1}},
    \end{aligned}
\end{equation}
where ${B\left( {{\mu }_{1}},{{\mu }_{2}} \right)}$ is the complete Beta function.

With different values of $\mu_1$ and $\mu_2$, the Beta distribution can be used to collect and redistribute the energies among various degrees of freedom and between molecules. According to the collision operator Eqs.(\ref{eq_boltzmann}) and (\ref{eq_hs}), for inverse-power-law molecules, the post-collision molecular energies in the center-of-mass system can be expressed as
\begin{equation}
	\begin{aligned}
		& e _{1}^{'}=\left( 1-{{s}_{1}} \right){{e }_{1}}+{{s}_{4}}{{s}_{5}}{{e }_{a}}, \\ 
		& e _{2}^{'}=\left( 1-{{s}_{2}} \right){{e }_{2}}+\left( 1-{{s}_{4}} \right){{s}_{5}}{{e }_{a}}, \\ 
		& e _{t}^{'}=\left( 1-{{s}_{3}} \right){{e }_{t}}+\left( 1-{{s}_{5}} \right){{e }_{a}}, \\
		& {{e }_{a}}={{s}_{1}}{{e }_{1}}+{{s}_{2}}{{e }_{2}}+{{s}_{3}}{{e }_{t}},
	\end{aligned}
\end{equation}
where $e _{t}$ and $e _{i}$ denote the pre-collision translational and rotational energies of the molecules, while $e _{t}^{'}$ and $e _{i}^{'}$ represent the corresponding post-collision translational and rotational energies. The random variates ${s}_{i}$ are sampled from the Beta distributions given as
\begin{equation}
	\begin{aligned}
		& \beta \left\langle {{s}_{j}}\left| \phi \zeta ,\left( 1-\phi  \right)\zeta  \right. \right\rangle ,j=1,2, \\ 
		& \beta \left\langle {{s}_{3}}\left| \psi \eta ,\left( 1-\psi  \right)\eta  \right. \right\rangle,  \\ 
		& \beta \left\langle {{s}_{4}}\left| \phi \zeta ,\phi \zeta  \right. \right\rangle,  \\ 
		& \beta \left\langle {{s}_{5}}\left| 2\phi \zeta ,\psi \eta  \right. \right\rangle,  \\ 			
	\end{aligned}
\end{equation}
with $\phi$ and $\psi$ being model parameters.

Furthermore, the post-collision rotational energies tend to be equally distributed between the two molecules, and it can be assumed that they are effectively partitioned equally\cite{erofeev1995numerical}. Under this assumption, the rotational energies of particles $e_1$ and $e_2$ are first treated as a whole, and the total $e_r$ is subsequently redistributed equally between them. Consequently, the post-collision energies can be expressed as
\begin{equation}
	\begin{aligned}
		& e_{t}^{\prime }=\left( 1-{{s}_{1}} \right){{e}_{t}}+\left( 1-{{s}_{3}} \right)\left( {{s}_{1}}{{e}_{t}}+{{s}_{2}}{{e}_{r}} \right), \\ 
		& e_{r}^{\prime }=\left( 1-{{s}_{2}} \right){{e}_{r}}+{{s}_{3}}\left( {{s}_{1}}{{e}_{t}}+{{s}_{2}}{{e}_{r}} \right),  \\ 
		& {{e}_{r}}={{e}_{1}}+{{e}_{2}},e_{1}^{\prime }=\mathscr{R} e_{r}^{\prime },e_{2}^{\prime }=(1-\mathscr{R} )e_{r}^{\prime }, \\ 
	\end{aligned}
\end{equation}
where $\mathscr{R}$ is a uniformly distributed random number in $(0, 1)$, the random variates ${s}_{i}$ are sampled from Beta distributions as
\begin{equation}
    \begin{aligned}
        & \beta \left\langle {{s}_{1}}\left| \psi \eta ,\left( 1-\psi  \right)\eta  \right. \right\rangle , \\ 
        & \beta \left\langle {{s}_{2}}\left| 2\phi , \right.2\left( 1-\phi  \right) \right\rangle , \\ 
        & \beta \left\langle {{s}_{3}}\left| 2\phi , \right.\psi \eta  \right\rangle . \\
    \end{aligned}
\end{equation}
For this simplified variant of the Pullin model, the detailed balance principle is also satisfied\cite{erofeev1995numerical}.

Once the values of $\phi$ and $\psi$ are specified, both models above are fully determined, and the relation between these parameters and the relaxation process is discussed in the following subsection.

\subsection{Selection of model parameters}\label{subSec2_Parameters}
In the Pullin model, two model parameters, $\phi$ and $\psi$, must be specified within the calculation framework. Based on the Chapman-Enskog expansion of the macroscopic transport coefficients, a specific formulation of these parameters for hard-sphere (HS) molecules was proposed\cite{pullin1978kinetic}, i.e.,
\begin{equation}
    \left\{ \begin{aligned}
        & \psi =\phi ={{\phi }_{0}}, \, \omega <1, \\ 
        & \psi =\phi =0, \, \omega >1. \\ 
    \end{aligned} \right.
\end{equation}
When $\omega <1$, the relationship between the model parameter ${\phi }_{0}$ and rotational collision number $Z$ is given as
\begin{equation}
    {{\phi }_{0}}=\frac{8\left( 2+\upsilon  \right)}{5\pi }\frac{1}{Z}.
\end{equation}

For the VHS molecules, Pullin did not establish a direct relationship between the parameters $\phi$ and $\psi$ and the relaxation properties, which may limit the broader applicability of this model. Recently, an explicit formulation of these parameters is derived based on the equipartition theorem, in which the translational and rotational energies of two colliding particles are redistributed in proportion to their respective degrees of freedom\cite{Ding2602.05775}. Under this framework, the relaxation of translational and rotational temperatures is governed by
\begin{equation}
    \begin{aligned}
        \frac{\partial {{T}_{tr}}}{\partial t}=\frac{5\cdot {{4}^{\eta }}\sqrt{\pi }n{{\left( \frac{m}{k{{T}_{tr}}} \right)}^{\frac{3}{2}-\eta }}\left( T-{{T}_{tr}} \right)c_{r,ref}^{4-2\eta }\cdot {{\sigma }_{ref}}\Gamma \left( 1+\eta  \right)\phi \psi }{3\left( 2\phi +\eta \psi  \right)},
    \end{aligned}
\end{equation}
\begin{equation}
    \begin{aligned}
        \frac{\partial {{T}_{rot}}}{\partial t}=\frac{5\cdot {{4}^{\eta }}\sqrt{\pi }n{{\left( \frac{m}{k{{T}_{tr}}} \right)}^{\frac{3}{2}-\eta }}\left( T-{{T}_{rot}} \right)c_{r,ref}^{4-2\eta }\cdot {{\sigma }_{ref}}\Gamma \left( 1+\eta  \right)\phi \psi }{3\left( 2\phi +\eta \psi  \right)}, \label{eq_rot_relax}
    \end{aligned}
\end{equation}
where $m$ is the molecular mass, ${\sigma }_{ref}$ is the collision cross section, and $\Gamma$ is the gamma function. Meanwhile, the relaxation of the stress tensor is given by
\begin{equation}
    \begin{aligned}
        \frac{\partial {{p}_{\langle ij\rangle }}}{\partial t}=-\frac{{{2}^{1+2\eta }}\sqrt{\pi }n{{\left( \frac{m}{k{{T}_{tr}}} \right)}^{\frac{3}{2}-\eta }}c_{r,ref}^{4-2\eta }{{\sigma }_{ref}}\Gamma \left( 1+\eta  \right)\left( 1+\eta  \right)}{15}{{p}_{\langle ij\rangle }}. \label{eq_stress_relax}
    \end{aligned}
\end{equation}

According to the definition of relaxation rate for macroscopic variables, the Eqs.(\ref{eq_rot_relax}) and (\ref{eq_stress_relax}) can be rewritten in the standard form, i.e.,
\begin{equation}
    \begin{aligned}
        & \frac{\partial {{T}_{rot}}}{\partial t}=\frac{T-{{T}_{rot}}}{{{\tau }_{rot}}}, \\
        & \frac{\partial {{p}_{\langle ij\rangle }}}{\partial t}=-\frac{{{p}_{\langle ij\rangle }}}{{{\tau }_{tr}}},
    \end{aligned}
\end{equation}
where ${\tau }_{rot}$ and ${\tau }_{tr}$ denote the rotational and translational relaxation times, respectively. The rotational collsion number $Z$ is defined as $Z = {\tau }_{rot} /{{\tau }_{tr}}$. Therefore, based on the relaxation rate of temperature and stress tensor, the rotational collision number $Z$ (relax to the total temperature $T$) can be expressed as 
\begin{equation}
	{Z}=\frac{2\left( 1+\eta  \right)\left( 2\phi +\eta \psi  \right)}{25\phi \psi }.
\end{equation}
In Pullin's partition function Eq.(\ref{eq_hs}), the terms $\phi \zeta$ and $\psi \eta$ denote the fractions of internal and translational energies, respectively. According to the equipartition theorem, the energy associated with each mode is distributed in proportion to its degrees of freedom. Accordingly, for a diatomic gas, the relation between $\psi$ and $\phi$ can be determined as follows
\begin{equation}
	\psi =\frac{3}{2\eta }\phi.
\end{equation}
Consequently, the model parameters $\psi$ and $\phi$ can be directly determined from the rotational collision number $Z$ as
\begin{equation}
	\begin{aligned}
		& \phi =\frac{14\eta \left( 1+\eta  \right)}{75 Z}, \\ 
		& \psi =\frac{7\left( 1+\eta  \right)}{25 Z}. \\ 
	\end{aligned}
\end{equation}
It should be noted that the values of $\phi$ and $\psi$ must remain within the range $\left( 0,1 \right)$, which imposes a lower bound on the rotational collision number $Z$. For nitrogen, this constraint requires that $Z$ must be greater than 0.906752. From a physical perspective, the rotational collision number is typically greater than unity, indicating that the present formulation is physically reasonable. Although these parameters are derived from the Pullin model, they are likewise employed in the simplified variant of the Pullin model used in this study.

\section{Implementation of the energy exchange model in the DSMC method}\label{sec3_DSMC}

\subsection{The direct simulation Monte Carlo method}\label{subSec3_DSMC}
The DSMC method is a stochastic, particle-based numerical approach for solving the Boltzmann equation\cite{bird1970direct}, in which the velocity distribution function is approximated by a large number of simulated particles. Using an operator splitting scheme, the generalized Boltzmann equation Eq.(\ref{eq_boltzmann}) is decomposed into free-transport and collision steps, i.e.,
\begin{equation}
	\left\{ \begin{aligned}
	 	& \frac{\partial \left( nf \right)}{\partial t}+\boldsymbol{c}\frac{\partial \left( nf \right)}{\partial \boldsymbol{x}}=0, \\ 
	 	& \frac{\partial \left( nf \right)}{\partial t}=Q\left( f,f \right). \\ 
	\end{aligned} \right.
\end{equation}
In the DSMC method, the free-transport step corresponds to free molecular motion, while the collision step simulates intermolecular interactions through stochastic sampling. During the free-transport step, the motion of each particle is computed using a Lagrangian tracking approach, in which each particle is advanced according to its instantaneous velocity ${\mathbf{c}}_{i}$ over a discrete time interval $\Delta t$, i.e.,
\begin{equation}
    \begin{aligned}
        &{{\mathbf{r}}^{n+1}_{i}}={{\mathbf{r}}^{n}_{i}}+{{\mathbf{c}}_{i}}\Delta t, \\
        &d\left(m_i \mathbf{c}_i\right) / d t=\mathbf{C}\left(\mathbf{c}_i,e_i\right),
    \label{free transport}
    \end{aligned}
\end{equation}
where subscript ``$i$'' denotes the particle index, $\mathbf{r}_i$ and $\mathbf{c}_i$ represent the position and velocity of the particle $i$, respectively, $m$ is the mass of the particle, and $\mathbf{C}$ represents the binary collision process each
particle undergoes during the time step. As a core component of the DSMC method, the collision algorithm determines the post-collision velocities and energies of selected particle pairs based on their pre-collision states. Among various approaches, the No Time Counter (NTC) scheme\cite{bird1994molecular, 2013The} is the most widely adopted. In this scheme, the number of candidate collision pairs, $N_c$, within each time step $\Delta t$ is estimated using the maximum collision probability. The value of $N_c$ is given by
\begin{equation}
	{{N}_{c}}=\frac{1}{2}N\left( N-1 \right)F_N{{\left( {{\sigma }_{T}}{{c}_{r}} \right)}_{\max }}\Delta t/{{V}_{c}}
\end{equation}
where $N$ is the number of simulated particles in the computational cell, $F_N$ denotes the ratio of real molecules to simulated particles, $\sigma_T$ represents the total collision cross section, $c_r$ is the relative velocity between two colliding particles, and $V_c$ is the volume of the computational cell. Since the number of collisions must be an integer, the actual number of collision pairs tested, $N_{c,\mathrm{test}}$, is taken as $\lfloor N_{c} \rfloor$ with an additional collision included with probability $N_{c} - \lfloor N_{c} \rfloor$. Each of the ${N}_{c,\mathrm{test}}$ pairs of particles is selected at random regardless of position in the cell, and then the candidate pairs are chosen to collide with a probability
\begin{equation}
	p_c=\frac{{{\sigma }_{T}}{{c}_{r}}}{{{\left( {{\sigma }_{T}}{{c}_{r}} \right)}_{\max }}},
\end{equation}
where ${{\sigma }_{T}}{{c}_{r}}$ represents the product of the total collision cross-section and the relative velocity of the selected particle pair. The quantity ${\left({{\sigma }_{T}}{{c}_{r}} \right)}_{\max }$ is initially assigned a suitable reference value within the computational cell and is subsequently updated if ${{\sigma }_{T}}{{c}_{r}}$ for any selected particle pair exceeds this threshold. By comparing the collision probability $p_c$ with a uniformly distributed random number $\mathscr{R} \in (0, 1)$, an acceptance-rejection (AR) procedure is used to determine whether the candidate particle pair actually collides. If accepted, this particle pair undergoes either an elastic collision (the change of particle speed) or an inelastic collision (the exchange of translational energy and internal energy), in which translational energy is exchanged with internal energy and redistributed between the different degrees of freedom.

\subsection{Implementation of the Borgnakke-Larsen model}\label{subSec3_BL}
In the DSMC method, the BL model\cite{borgnakke1975statistical,bird1994molecular} is commonly employed to describe energy exchange between translational and internal modes during collisions. Within the BL framework, the total energy $e_c$ of a collision pair, consisting of both translational and internal components, is strictly conserved. After the collision, the internal energy is redistributed according to an equilibrium distribution based on the combined translational and internal energies. The relaxation rate is governed by the ratio of inelastic to elastic collisions, ensuring consistency with the prescribed macroscopic relaxation behavior. In practice, the inelastic collision probability $p_{i}$ is typically defined as $p_{i} = 1/Z_{\mathrm{BL}}$, where $Z_{\mathrm{BL}}$ denotes the model-specific relaxation number. It should be emphasized, however, that the definition of $Z_{\mathrm{BL}}$ may differ from the collision number $Z$ employed in other models, such as Pullin's model. In the present work, the analysis is restricted to rotational energy, and the serial BL procedure\cite{lu2022universal} is adopted. Specifically, each particle in a collision pair is independently tested for inelastic participation. When particle A is selected, the collision type is determined based on the inelastic collision probability $p_i$. If an inelastic collision occurs, the collision energy is defined as ${e_c} = e_t + e_1$. The post-collision rotational energy $e_i$ lies within the interval $[0, e_c]$ and is generated by sampling from a uniform random variate $\mathscr{R}_1$. The probability of a given rotational energy, normalized by its maximum value, is expressed as
\begin{equation}
    \frac{p}{p_{\max }}=\left(\frac{\zeta_{\mathrm{rot}} / 2 +1 / 2-\omega}{\zeta_{\mathrm{rot}} / 2 - 1} \frac{e_i}{e_c}\right)^{\zeta_{\mathrm{rot}} / 2 - 1}\left[\frac{\zeta_{\mathrm{rot}}/2+1 / 2-\omega}{3 / 2-\omega}\left(1-\frac{e_i}{e_c}\right)\right]^{3 / 2-\omega},
\end{equation}
where $\zeta_{\mathrm{rot}}$ is the rotational degrees of freedom. The post-collision rotational energy is then given by $e_i = \mathscr{R}_1 e_c$. Another random number $\mathscr{R}_2$ is generated for comparison. If $\mathscr{R}_2 \leq p/p_{\max }$, the rotational energy $e_i$ is accepted as the post-collision rotational energy ${e_1}'$ of particle A; otherwise, a new value of $e_i$ is sampled until acceptance. The post-collision translational energy of particle A is then determined by $e_t' = e_c - e_1'$. Subsequently, particle B is subjected to the same probabilistic criterion $p_i$ to decide whether its collision is elastic or inelastic. The redistributed translational energy from the first step is combined with the rotational energy of particle B to define a new collision energy ${e_c}' = e_t' + e_2$. The post-collision rotational energy of particle B is sampled from a uniform random variate $\mathscr{R}_3$, until acceptance. The final post-collision rotational energy of particle B is then given by $e_2' = \mathscr{R}_3 e_c'$, and the post-collision translational energy is given by $e_t'' = {e_c}' - e_2'$. Therefore, the magnitude of the post-collision relative velocity $c_{r}^{'}$ between the molecules can be expressed as $c_{r}^{'}=\sqrt{2e_{t}^{''}/{{m}_{r}}}$, where $m_r$ is the reduced mass of the collision pair. The post-collision velocities of particle pair are updated as follows
\begin{equation}
    \begin{aligned}
        & \mathbf{c}_{1}^{\prime}=\frac{1}{2}[(\mathbf{c}_1+\mathbf{c}_2)-c_{r}^{'}\mathbf{\omega}], \\
        & \mathbf{c}_{2}^{\prime}=\frac{1}{2}[(\mathbf{c}_1+\mathbf{c}_2)+c_{r}^{'}\mathbf{\omega}],
    \end{aligned}
\end{equation}
where $\mathbf{c}_1$ and $\mathbf{c}_2$ denote the pre-collision velocities of the two colliding particles A and B, $\mathbf{c}_{1}^{\prime}$ and $\mathbf{c}_{2}^{\prime}$ represent their post-collision velocities, and $\mathbf{\omega}$ is a unit vector that defines the direction of the post-collision relative velocity. The vector $\mathbf{\omega}$ performs a random walk on the unit sphere, which depends on the selected collision model. For the VHS model, $\mathbf{\omega}$ is given as
\begin{equation}
    \mathbf{\omega}={{\left( \cos \chi ,\sin \chi \cos \theta ,\sin \chi \sin \theta  \right)}^{T}},
\end{equation}
where, in polar coordinates, the cosine of deflection angle $\chi$ and azimuth angle $\theta$ are uniformly distributed over the intervals $[-1, 1]$ and $[0, 2\pi]$, respectively. That is,
\begin{equation}
    \cos \chi =2{{\mathscr{R}}_{4}}-1,\text{ }\theta =2\pi {{\mathscr{R}}_{5}}.
\end{equation}

Although the BL model has been widely applied and shows good agreement in practical simulations, it is physically inconsistent because only a subset of collisions is treated as inelastic. To enhance the physical fidelity of the model, Bird proposed a modified variant of the BL model \cite{bird1994molecular}, in which all collisions are regarded as inelastic, but only a prescribed fraction $p_i$ of the calculated change in rotational energy is transferred during each collision. In practical implementation, this means that all colliding particles are treated as undergoing inelastic collisions, thereby eliminating the need for probabilistic selection. Specifically, the post-collision rotational energy of particle A is calculated as $e_1'' = e_1(1-p_i)+e_1' p_i$, and the corresponding post-collision translational energy for the collision pair is then given by $e_t' = e_c - e_1''$. Subsequently, the post-collision rotational energy of particle B is computed as $e_2'' = e_2 (1-p_i)+e_2' p_i$, and the final post-collision translational energy for the collision pair is given by $e_t'' = {e_c}' - e_2''$. While this modification preserves the relaxation rate, it does not satisfy the principle of detailed balance\cite{pullin1978kinetic}.

\subsection{Implementation of the Pullin model}\label{subSec3_Pullin}
\begin{figure}
    \centering
    \includegraphics[width=1.0\textwidth]{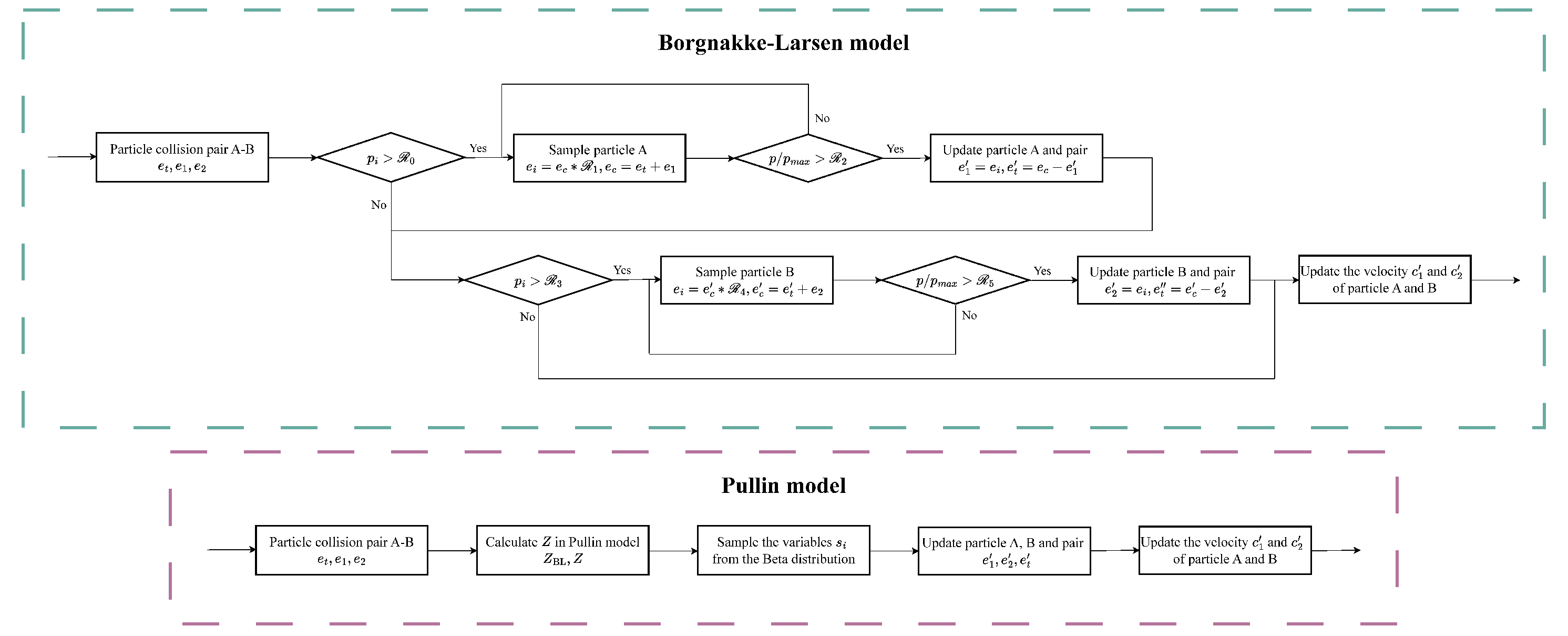}
    \caption{\label{theory-flowchart} Comparison of the implementation procedures of the BL and Pullin models within the DSMC framework. Top: BL model; bottom: Pullin model.}
\end{figure}

The Pullin model is straightforward to implement within the DSMC framework, requiring modifications only to the colliding particles. It allows for the computation of the translational and rotational energies of each particle after collision, while the remaining parts of the Pullin model remain consistent with the conventional DSMC method. The computational procedures for the Pullin and BL models within the DSMC framework are illustrated in the Fig.\ref{theory-flowchart}. It should be noted that the definition of rotational collision number in the Pullin model $Z$ differs from that in the BL model $Z_{\mathrm{BL}}$; to ensure consistent results, this difference must be analyzed first. Moreover, both the Pullin model and its simplified variant rely solely on sampling from Beta distributions (five or three), with only the input parameters differing. Therefore, the following section provides a detailed discussion of the rotational collision number in the Pullin model and the associated Beta distribution sampling procedure.

In the Pullin model, the rotational collision number is defined directly in a form consistent with experimental measurements\cite{parker1959rotational, carnevale1967ultrasonic}, i.e., 
\begin{equation}
    Z = \frac{\tau_{rot}}{\tau_{tr}},
\end{equation}
where $\tau_{tr}$ denotes the translational relaxation time, defined as $\mu /p$, with $\mu$ being the viscosity coefficient and $p$ the pressure. In the BL model, the definition of the rotational collision number $Z_{\mathrm{BL}}$ is consistent with that of the Pullin model; however, the definition of the translational relaxation time $\tau_{tr}$ differs and depends on the employed collision model. For the widely used VHS model, the translational relaxation time is given by
\begin{equation}
    {{\tau }_{tr,\mathrm{VHS}}}=\frac{(5-2\omega )(7-2\omega )}{30}\frac{\mu }{p}.
\end{equation}
To enable a consistent comparison between the Pullin and BL models, the collision number $Z$ is expressed in terms of the collision number $Z_{\mathrm{BL}}$ through the relationship
\begin{equation}
    Z=\frac{{{\tau }_{rot}}}{{{\tau }_{tr}}}=\frac{{{\tau }_{rot}}}{{{\tau }_{tr,\text{VHS}}}}\cdot \frac{{{\tau }_{tr,\text{VHS}}}}{{{\tau }_{tr}}}=\frac{(5-2\omega)(7-2\omega )}{30}{{Z}_{\text{BL}}}. \label{eq_Z_and_ZBL}
\end{equation}
If the Pullin model is employed in the DSMC method, the rotational collision number $Z_{\mathrm{BL}}$ must be converted to the corresponding rotational collision number $Z$ in the Pullin model.

Several algorithms have been proposed for the computer generation of random variates following the $\beta\left\langle z \mid \mu_1, \mu_2 \right\rangle$ distribution. Among the most widely used are J{\"o}hnk's method \cite{johnk1964erzeugung} and Cheng's algorithm \cite{cheng1978generating}, both of which are applicable for all shape parameters ${{\mu }_{1}}>0,{{\mu }_{2}} >0$ and capable of generating Beta-distributed variates over a wide range of parameter values. J{\"o}hnk's method, in particular, is simple and effective, relying on the transformation of uniform random numbers. The procedure can be summarized as follows: two independent $\text{Uniform}(0,1)$ random numbers ${\mathscr{R}_{01}}$ and ${\mathscr{R}_{02}}$ are first generated, and then transformed via $V_1 = {\mathscr{R}_{01}}^{1/{{\mu }_{1}}}$ and $V_2 = {\mathscr{R}_{02}}^{1/{{\mu }_{2}}}$. The sum $W = V_1 + V_2$ is then computed; If $W \leq 1$, the Beta-distributed random variate is obtained as $z = V_1 / W$; otherwise, the process is repeated. This algorithm ensures that the resulting random variate $z$ follows the $\beta\left\langle z \mid \mu_1, \mu_2 \right\rangle$ distribution. Its generality and ease of implementation make it well suited for use in the Pullin model within the DSMC framework.

\section{Numerical simulations}\label{sec4_Numerical}
To evaluate the predictive capability of the Pullin model across a wide range of nonequilibrium phenomena, five typical test cases are simulated and analysed: zero-dimensional (0-D) rotational relaxation of nitrogen, one-dimensional (1-D) planar Couette flow and normal shock structure, two-dimensional (2-D) hypersonic flow around a circular cylinder, and three-dimensional (3-D) hypersonic flow around an X38-like vehicle. The Pullin model is implemented within the DSMC framework, and the results are compared with those obtained using the BL model. Unless otherwise specified, all Pullin and BL results discussed in the following are obtained from the DSMC codes DSMC0R and DS1 developed by Bird \cite{bird1994molecular}, or from the open-source solver SPARTA \cite{plimpton2019direct}.

For nitrogen flow, the variable hard sphere (VHS) model is adopted, in which the viscosity $\mu$ depends on the translational temperature $T_{tr}$ according to the power-law relation
\begin{equation}
    \mu=\mu_{ref}\left(\frac{T_{tr}}{T_{ref}}\right)^\omega.
\end{equation}
In all subsequent simulations, the reference viscosity is ${{\mu }_{ref}}=1.656\times {{10}^{-5}}\mathrm{Pa}\cdot s$ with a power-law exponent $\omega =0.74$, and the corresponding reference diameter $d_{ref}=4.17\times{{10}^{-10}m}$ at $T_{ref}= 273 \mathrm{K}$. Unless otherwise stated, the freestream Knudsen number $\mathrm{Kn}$ is defined as $\mathrm{Kn}={\lambda}/{L_{ref}}$, where the mean free path $\lambda$ for the VHS model is expressed as
\begin{equation}
    \lambda =\frac{1}{\sqrt{2}\pi d_{ref}^{2}n{{\left( \frac{{{T}_{ref}}}{T_{tr}} \right)}^{\omega -0.5}}}.
\end{equation}

\subsection{Rotational relaxation of nitrogen}\label{subSec4_relaxation}
The rotational relaxation of nitrogen is a typical 0-D nonequilibrium flow problem, widely employed to examine the energy exchange between translational and rotational modes\cite{bird1994molecular,prasanth2012variable}. The system is initialized in a strongly nonequilibrium state: 100,000 simulated particles are confined within a single isolated computational cell, with a translational temperature of $T_{tr}=500 \mathrm{K}$ and a rotational temperature of $T_{rot}=0 \mathrm{K}$, corresponding to zero rotational energy. As relaxation proceeds, the system asymptotically approaches thermal equilibrium at $T_{eq}=300 \mathrm{K}$. In the present simulation, the rotational collision number in the BL model is maintained at a constant value of $Z_{\mathrm{BL}}=5$, while the corresponding parameter $Z$ in the Pullin model is evaluated from Eq.~(\ref{eq_Z_and_ZBL}).

In this simulation, the time evolution of the rotational and translational temperatures can be derived analytically\cite{bird1994molecular}. The predicted value of the rotational temperature is given by
\begin{equation}
    {{T}_{rot}}=300[ 1-\exp \left( -\upsilon t/5 \right)],
\end{equation}
while the translational temperature is obtained analogously as
\begin{equation}
    {{T}_{tr}}=300+200\exp \left( -\upsilon t/5 \right),
\end{equation}
where $\upsilon$ denotes the collision frequency. During relaxation, the translational temperature $T_{tr}$ decreases monotonically, whereas the rotational temperature $T_{rot}$ increases monotonically, with both asymptotically approaching $300 \mathrm{K}$. At equilibrium, the molecular speed distribution $f_{\beta c}$ takes the Maxwell form
\begin{equation}
    f_{\beta c}=\frac{4}{\sqrt{\pi}}\beta^{2}c^{2}\exp(-\beta^{2}c^{2}),
\end{equation}
while the rotational energy distribution $f_{e_{rot}/(kT)}$ of nitrogen is given by
\begin{equation}
    f_{e_{rot}/(kT)}=\exp\left(-{e_{rot}}/{kT}\right).
\end{equation}

\begin{figure}
    \centering
    \includegraphics[width=0.45\textwidth]{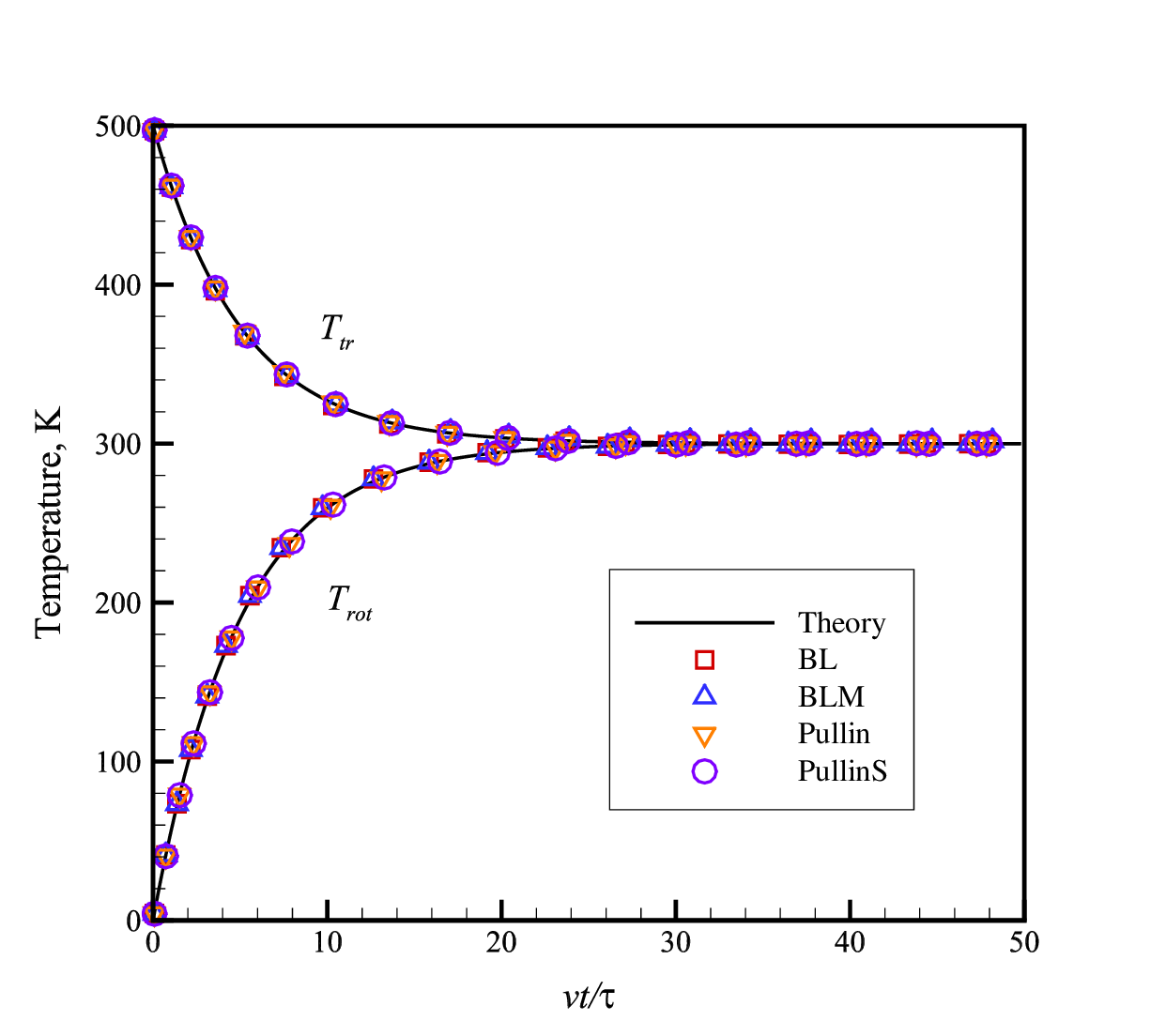}
    \caption{\label{0D-Distribution-Temp} Rotational relaxation in nitrogen gas. Squares: BL model; triangles: modified BL model; gradients: Pullin model; circles: simplified Pullin model; solid line: analytical solution.}
\end{figure}

\begin{figure}
    \centering
    \subfigure[]{\label{0D-Distribution-Vel}\includegraphics[width=0.45\textwidth]{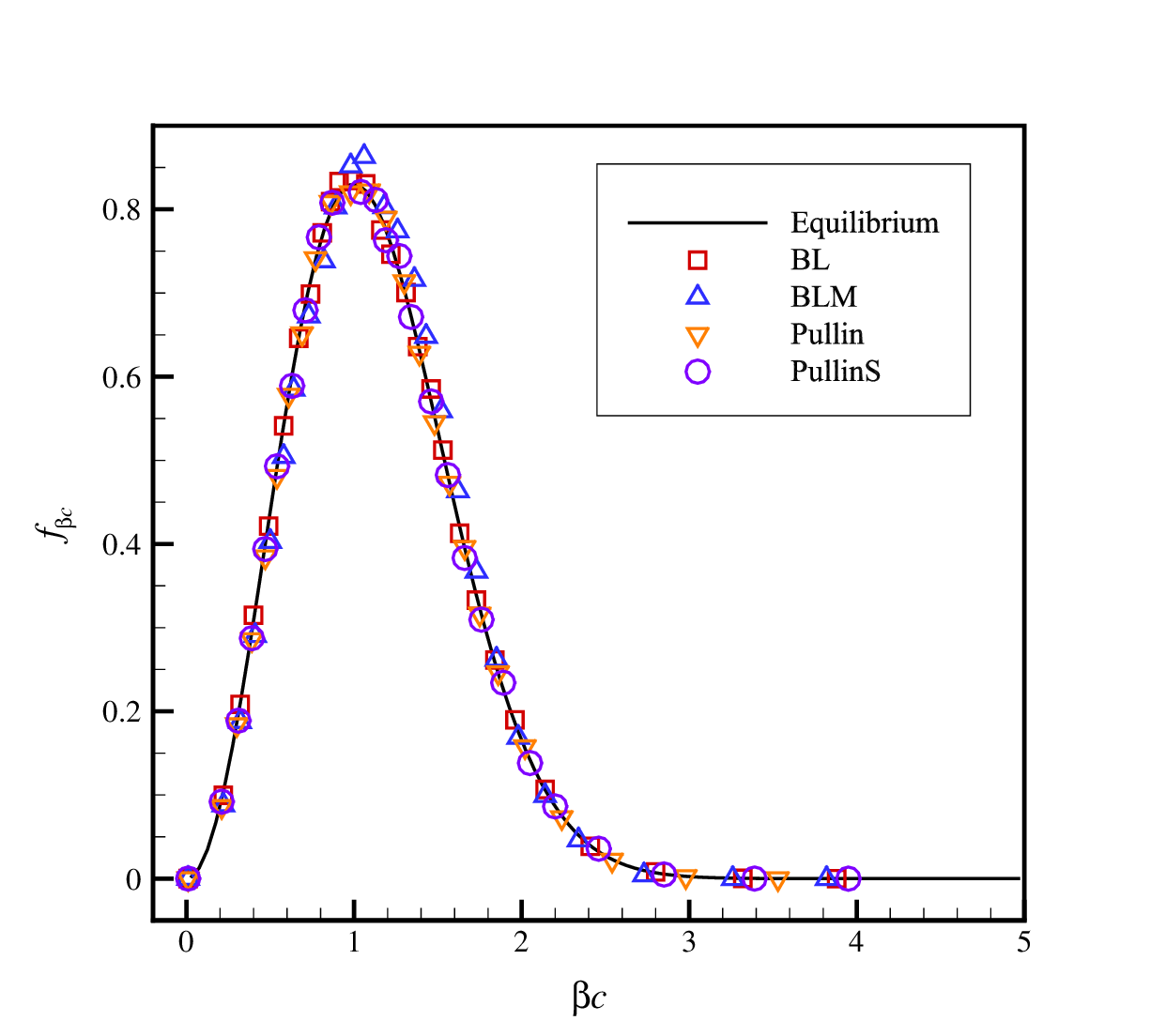}}
	\subfigure[]{\label{0D-Distribution-Erot}\includegraphics[width=0.45\textwidth]{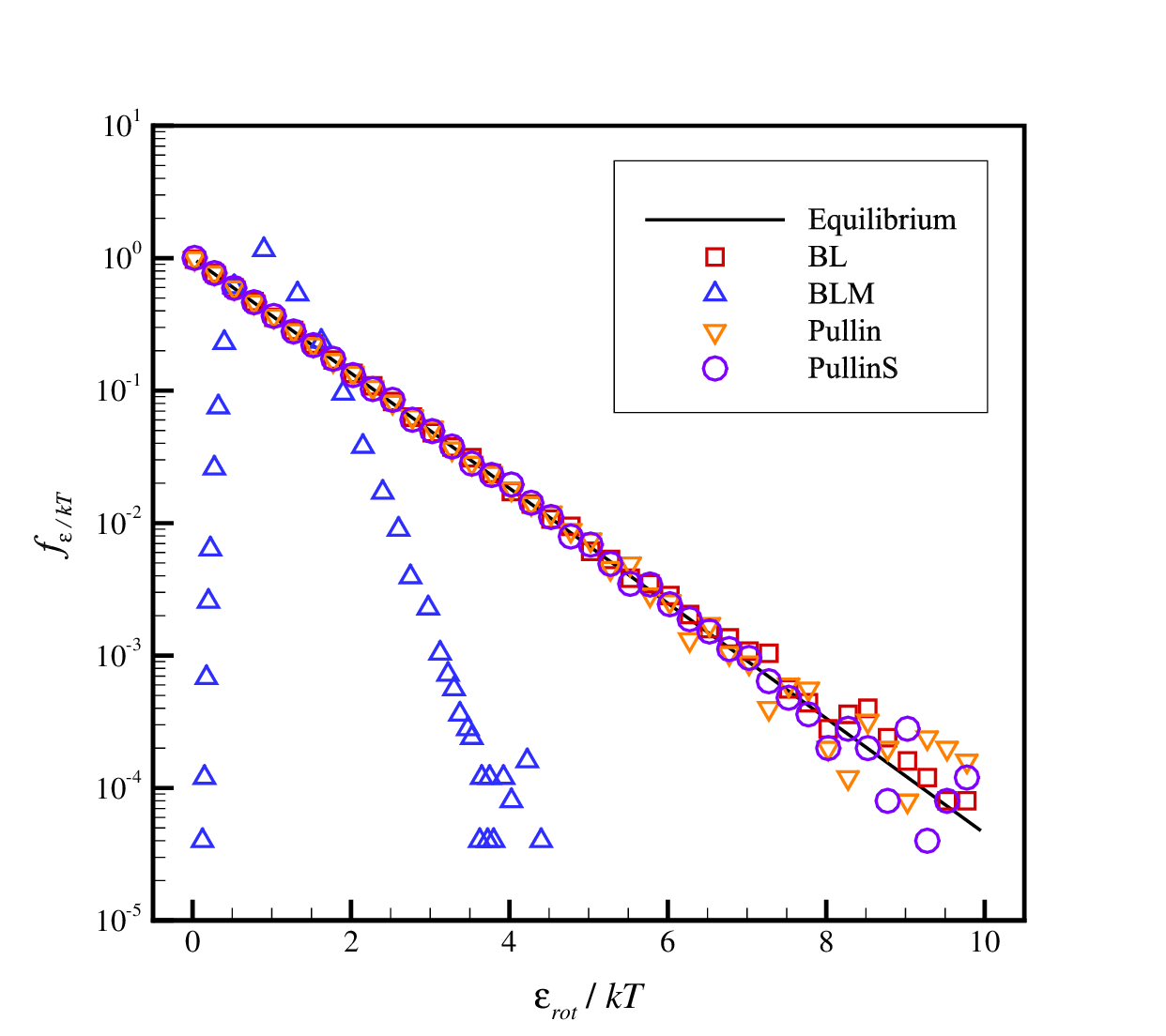}}
    \caption{\label{0D-distribution} Equilibrium distributions of nitrogen molecules: (a) molecular speeds and (b) rotational energies. Squares: BL model; triangles: modified BL model; gradients: Pullin model; circles: simplified Pullin model; solid line: theoretical solution.}
\end{figure}

Figure \ref{0D-Distribution-Temp} illustrates the temporal evolution of the rotational temperature during the relaxation process. The results from the Pullin model and its simplified variant agree well with the analytical solution. For comparison, results from the BL model and its modified variant are also included; both exhibit reasonable agreement with the Pullin model and the analytical solution. At equilibrium, the velocity distribution functions from the four different models are shown in Fig.~\ref{0D-Distribution-Vel} and compared with the Maxwell distribution $f_{\beta c}$ at the equilibrium temperature $T_{eq}$, with all four models demonstrating satisfactory agreement. Figure \ref{0D-Distribution-Erot} presents the rotational energy distribution functions on a logarithmic scale. The Pullin model, its simplified variant, and the BL model exhibit an exponential dependence on the rotational energy and agree closely with the theoretical solution $f_{e_{rot}/(kT)}$, whereas the modified BL model fails to reproduce the theoretical distribution at equilibrium, primarily because it does not satisfy the principle of detailed balance. These results demonstrate that the Pullin model, its simplified variant, and the BL model accurately capture the nonequilibrium energy exchange between translational and rotational modes.

The computational times for all test cases obtained using different methods are summarized in Table \ref{0D-time}. All computations are performed using a single core of an AMD Ryzen 9 5900X processor. The results show that the Pullin model requires approximately $40\%$ more computational time than the BL model, primarily due to the additional sampling of Beta distributions. The simplified Pullin model, which involves fewer Beta distribution samplings, reduces the computational time by about $7.7\%$ compared to the full Pullin model while maintaining comparable accuracy. Overall, both the Pullin model and its simplified variant are computationally more demanding than the BL model but offer improved physical fidelity in simulating energy exchange processes.

\begin{table}
    \centering
    \caption{\label{0D-time} Comparison of computational times for different rotational energy relaxation models.}
    \begin{tabular}{p{140pt}<{\centering} p{70pt}<{\centering} p{70pt}<{\centering} p{70pt}<{\centering} p{70pt}<{\centering}}
        \hline
        \hline
        Exchange model & BL & BLM & Pullin & PullinS \\
        \hline
        Total CPU time ($s$) & 424.549 & 470.266 & 594.484 & 548.813 \\
        \hline
        \hline
    \end{tabular}
\end{table}

\subsection{Planar Couette flow}\label{subSec4_couette}
The planar Couette flow\cite{wu2015kinetic} is considered to assess the performance of the Pullin model in shear-driven transport. The configuration consists of a lower plate at $x = 0$ and a parallel upper plate separated by a distance of 10 mean free paths ($h = 10 \lambda$). The plates move in opposite directions, each with a velocity of $U_p = 336.89 m/s$. The Knudsen number is set to $\mathrm{Kn} = 0.1$, with the reference length defined as the plate separation $h$. The working gas is nitrogen, modeled using the VHS model. The initial state is uniform, with a number density of $n_0 = 2.69 \times 10^{25} m^{-3}$ and translational and rotational temperatures of $T_0 = 273 \mathrm{K}$. Both plates are maintained at $T_{wall} = 273 \mathrm{K}$ and modeled as fully diffusive walls. The rotational collision number in the BL model is set to $Z_{\mathrm{BL}} = 5$, while the corresponding value $Z$ in the Pullin model is calculated from Eq.~(\ref{eq_Z_and_ZBL}). The computational domain is divided into 300 uniform cells along the vertical direction, with each computational cell initially containing approximately 2,000 simulated particles. The results are averaged over 10,000 time steps after the system reaches steady state.

\begin{figure}
	\centering
	\subfigure[]{\label{Couette-den}\includegraphics[width=0.45\textwidth]{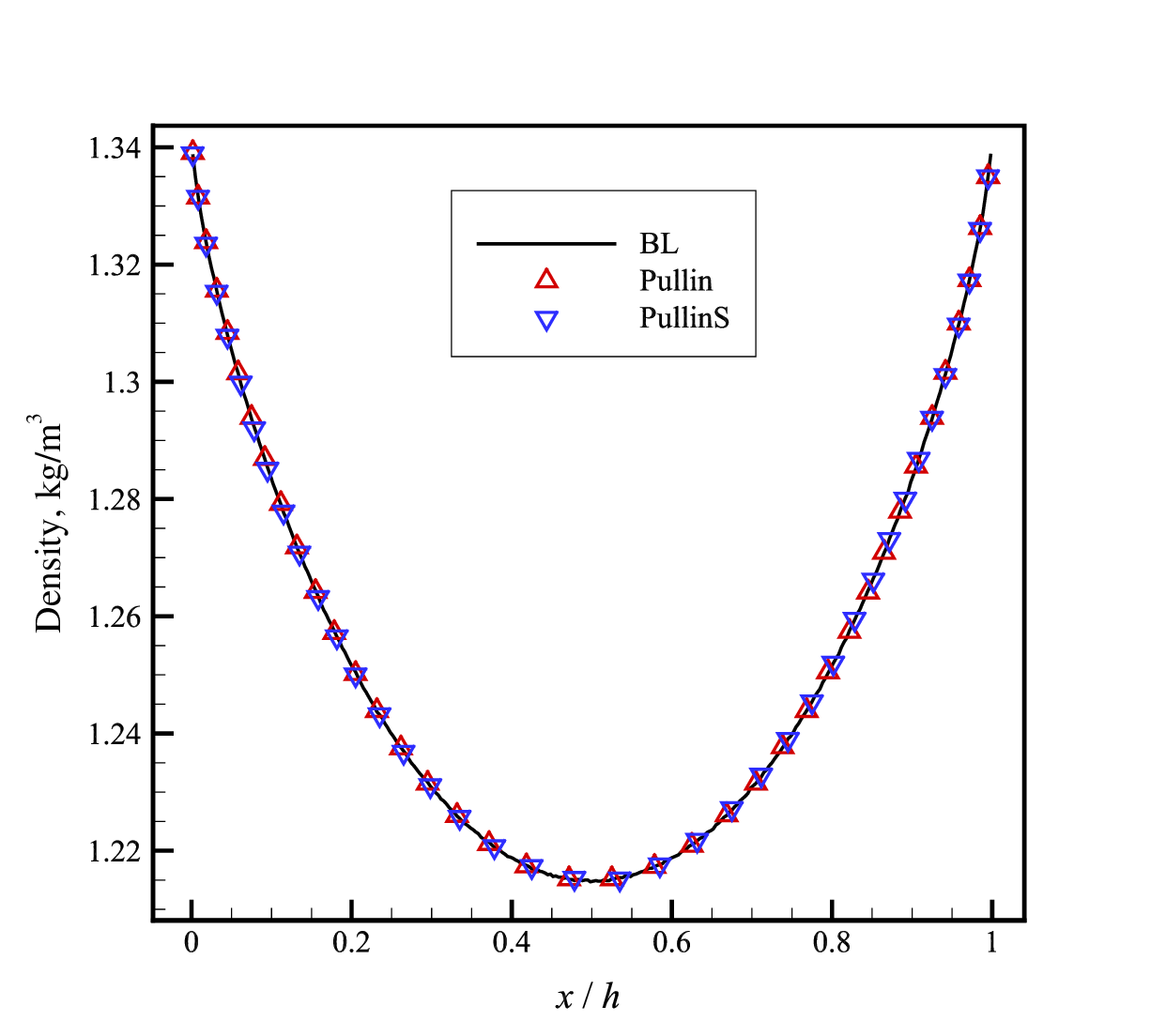}}
	\subfigure[]{\label{Couette-vel}\includegraphics[width=0.45\textwidth]{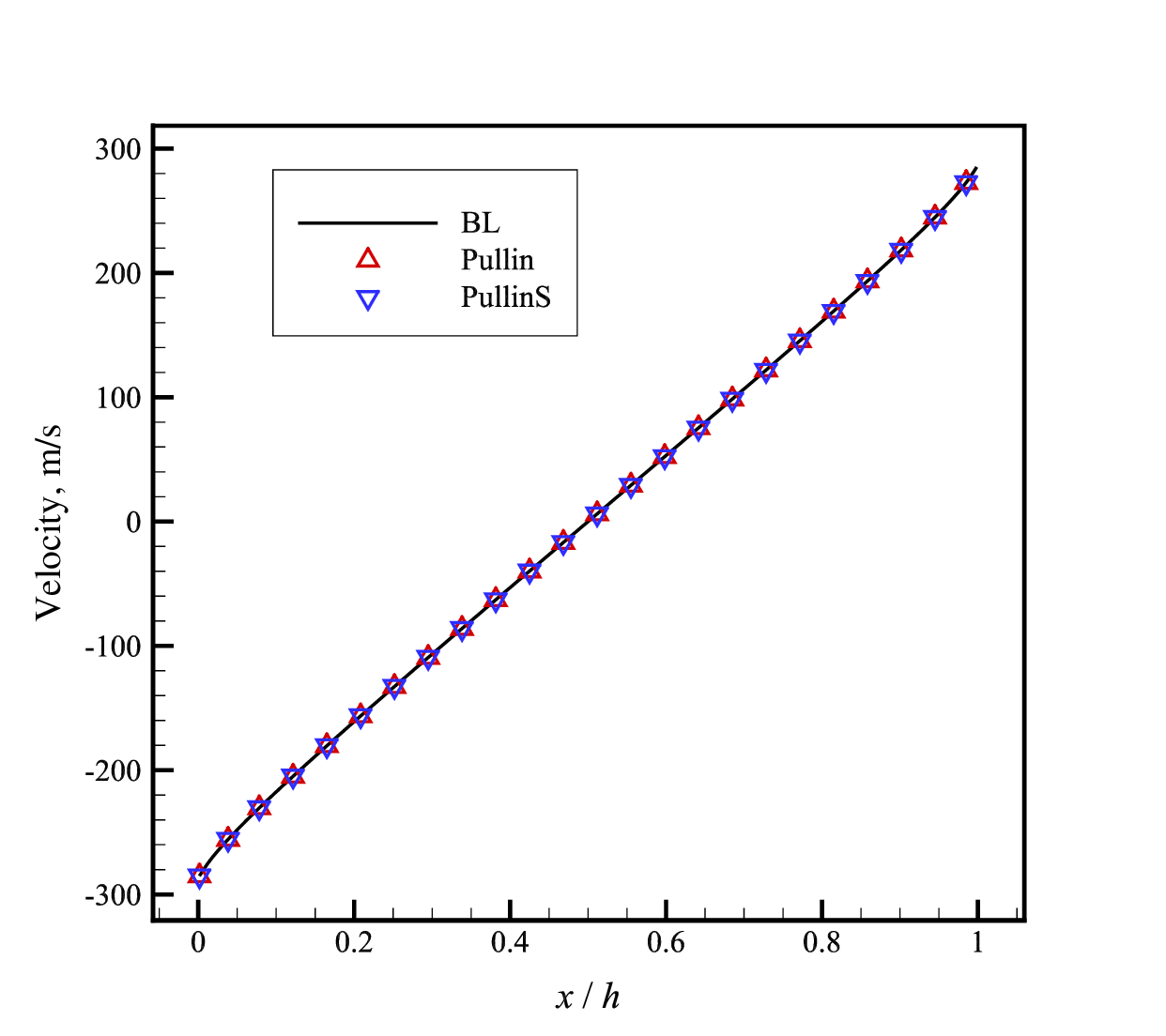}}
	\subfigure[]{\label{Couette-T}\includegraphics[width=0.45\textwidth]{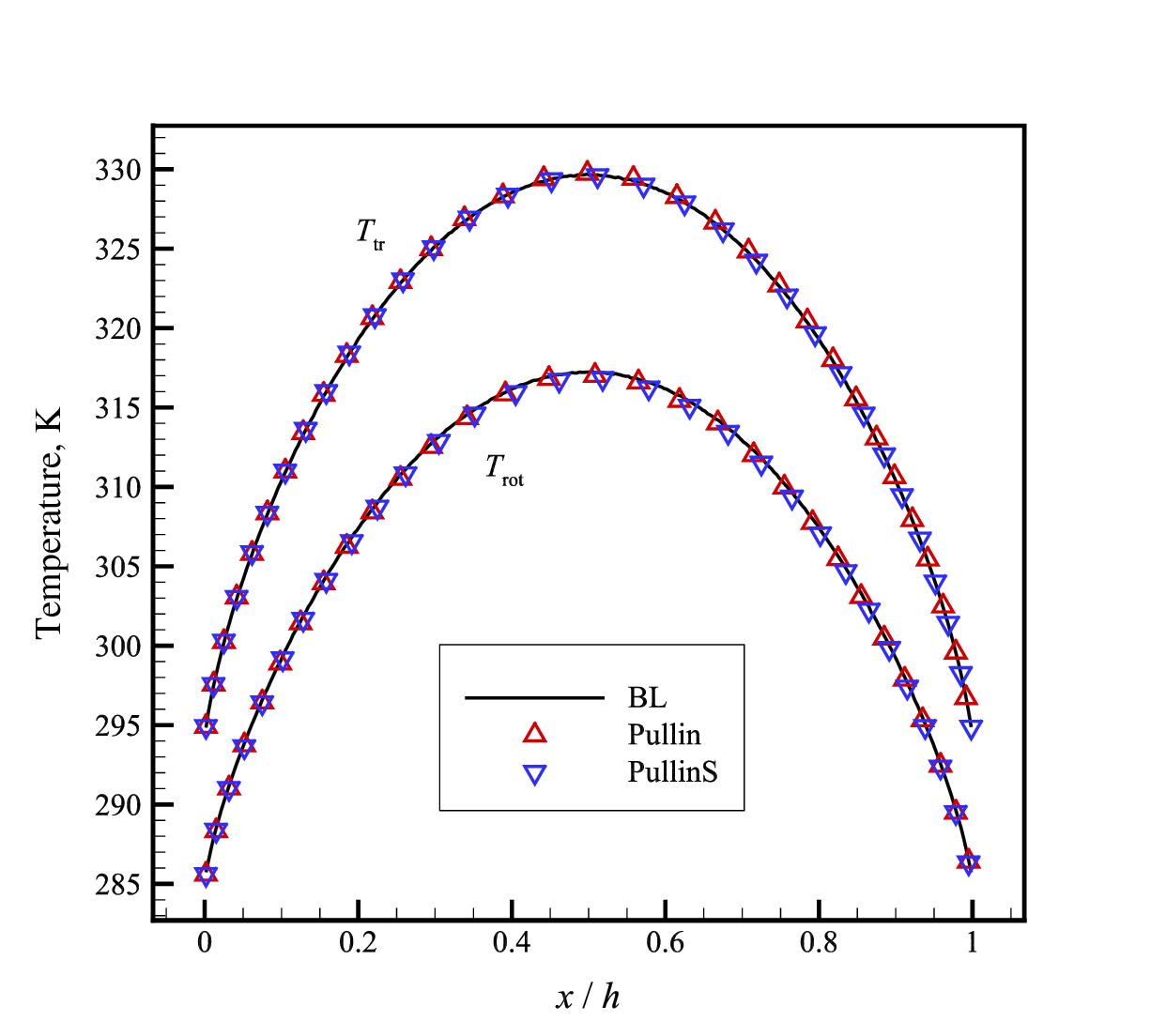}}
	\subfigure[]{\label{Couette-Q}\includegraphics[width=0.45\textwidth]{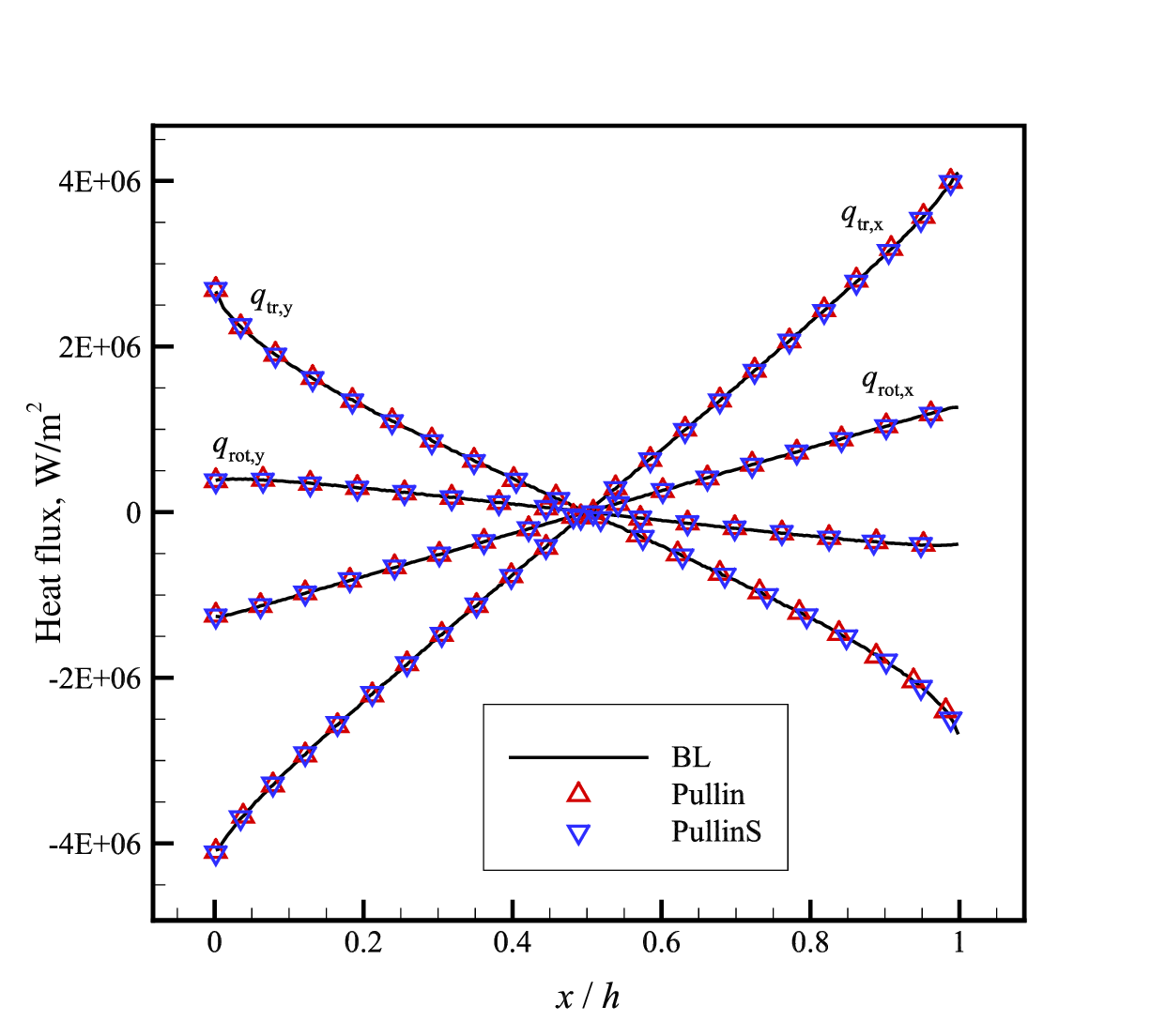}}
	\caption{\label{1D-Couette} Couette flow profiles at $\mathrm{Kn}=0.1$: (a) density; (b) velocity; (c) temperature; (d) heat flux. Triangles: Pullin model; gradients: simplified Pullin model; solid line: BL model.}
\end{figure}

Figures \ref{Couette-den} and \ref{Couette-vel} show the density and velocity profiles, respectively, while Fig.~\ref{Couette-T} presents the translational temperature profile. The heat flux profile is given in Fig.~\ref{Couette-Q}. Across all flow variables, the results from the Pullin model and its simplified variant agree well with those from the BL model. In particular, the Pullin model reproduces the heat flux profile with high accuracy, a quantity especially sensitive to nonequilibrium effects. These results demonstrate that the Pullin model provides a reliable description of shear-driven transport phenomena in rarefied gas flows.

\subsection{Normal shock wave}\label{subSec4_shock}
The normal shock wave serves as a fundamental test case for validating the Pullin model under strongly nonequilibrium conditions, as sharp gradients of velocity, density, and temperature develop over only a few mean free paths. Robben and Talbot \cite{robben1966experimental} reported experimental data for nitrogen shock waves, including measurements of both rotational temperature and density. Additional density-profile measurements were obtained by Alsmeyer \cite{alsmeyer1976density}, and these datasets have since been widely used for numerical validation, as demonstrated by Boyd \cite{boyd1990rotational}. 

Following Alsmeyer's experimental setup, the present study simulates the normal shock wave in nitrogen at upstream Mach numbers of $\mathrm{Ma} = 1.53$, $2.0$, $6.1$, and $10.0$. The upstream conditions are specified as follows: number density $n_1 = 1.0 \times 10^{20} m^{-3}$, translational and rotational temperatures $T_{tr,1} = T_{rot,1} = 300 \mathrm{K}$, and velocity $u_1$ determined from the Mach number. The corresponding downstream conditions are obtained from the Rankine-Hugoniot relations and summarized in Table \ref{1D-Shockwave}, which also lists the rotational collision numbers $Z_{BL}$ and viscosity indices $\omega$ employed in the simulations. The corresponding value $Z$ in the Pullin model is calculated from Eq.~(\ref{eq_Z_and_ZBL}). The computational domain extends from $x = -50 \lambda_1$ to $x = 50 \lambda_1$, where $\lambda_1$ is the upstream mean free path, and is discretized into 2,000 uniform cells, with each cell initially containing approximately 2,500 simulated particles. The upstream boundary is treated as an inflow condition, while the downstream boundary is modeled as a specularly reflecting wall moving at the downstream velocity $u_2$.

\begin{table}
    \centering
    \caption{\label{1D-Shockwave} The upstream and downstream parameters of the normal shock wave.}
    \begin{tabular}{{p{30pt}<{\centering} p{60pt}<{\centering} p{60pt}<{\centering} p{30pt}<{\centering} p{60pt}<{\centering} p{50pt}<{\centering} p{50pt}<{\centering} p{40pt}<{\centering} p{40pt}<{\centering}}}
        \hline
        \hline
        $\mathrm{Ma}$ & $n_1 (m^{-3})$ & $u_1 (m/s)$ & $T_1 (\mathrm{K})$ & $n_2 (m^{-3})$ & $u_2 (m/s)$ & $T_2 (\mathrm{K})$ & $Z_{BL}$ & $\omega$ \\
        \hline
        1.53 & 1.0E20 & 540.37 & 300 & 1.92E20 & 282.43 & 402.09 & 4.0 & 0.72\\
        2.0 & 1.0E20 & 706.37 & 300 & 2.67E20 & 264.89 & 506.25 & 4.0 & 0.72\\
        6.1 & 1.0E20 & 2154.43 & 300 & 5.29E20 & 407.32 & 2452.8 & 4.4 & 0.72\\
        10.0 & 1.0E20 & 3531.86 & 300 & 5.72E20 & 618.08 & 6116.25 & 5.0 & 0.72\\
        \hline
        \hline
    \end{tabular}
\end{table}

\begin{figure}
	\centering
	\subfigure[]{\label{Shockwave-Ma1.53-T}\includegraphics[width=0.45\textwidth]{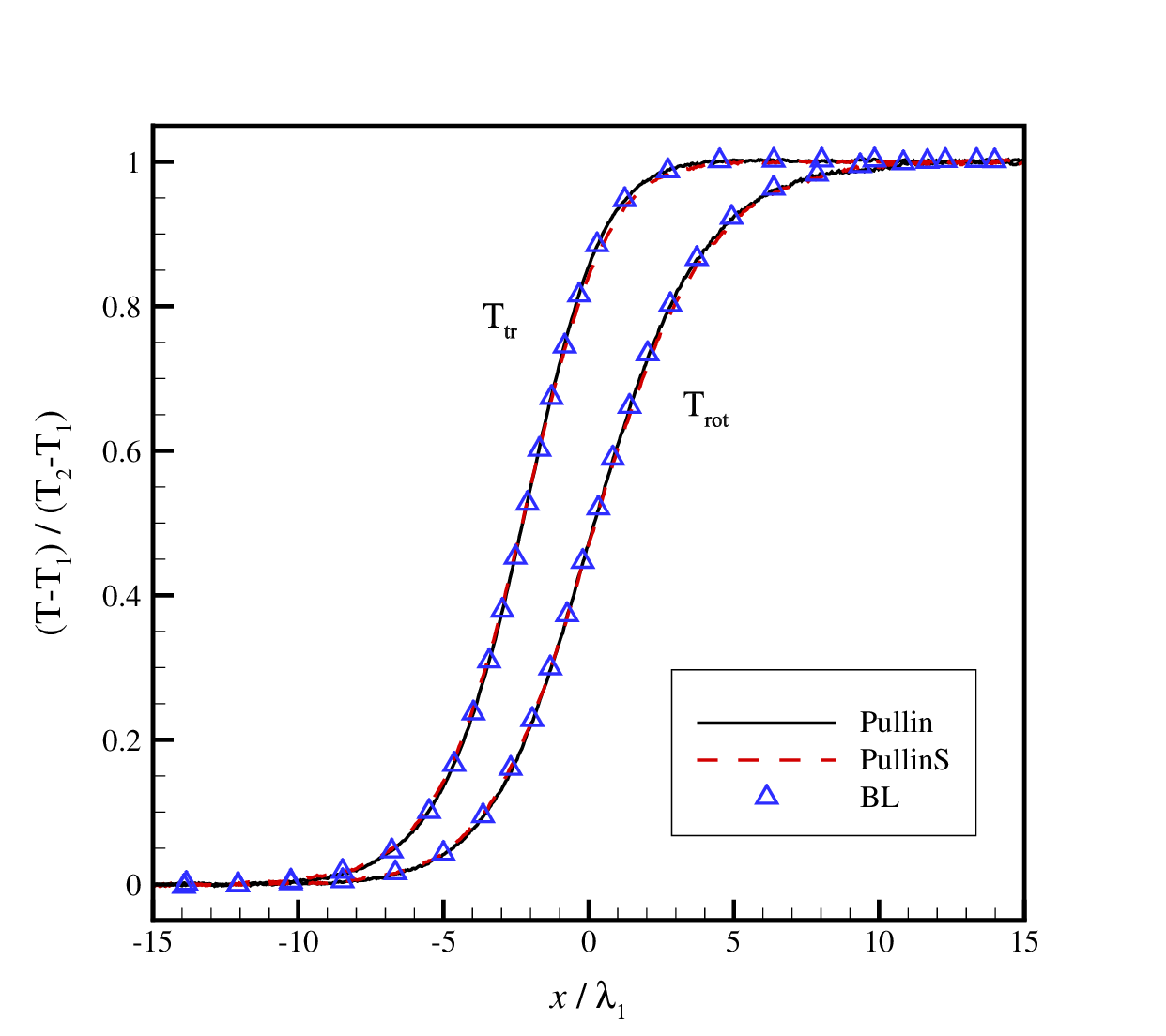}}
	\subfigure[]{\label{Shockwave-Ma2-T}\includegraphics[width=0.45\textwidth]{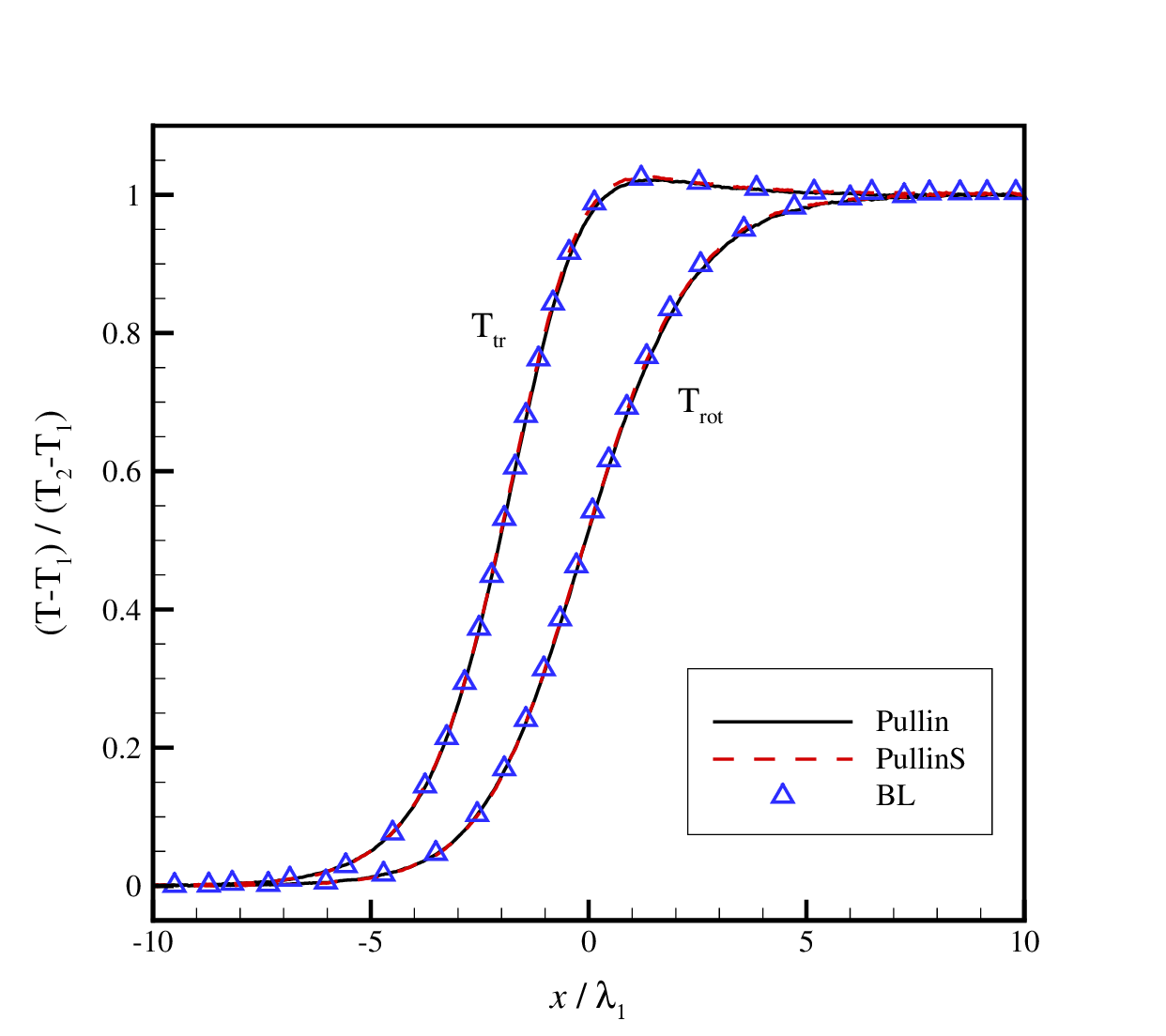}}
    \subfigure[]{\label{Shockwave-Ma6.1-T}\includegraphics[width=0.45\textwidth]{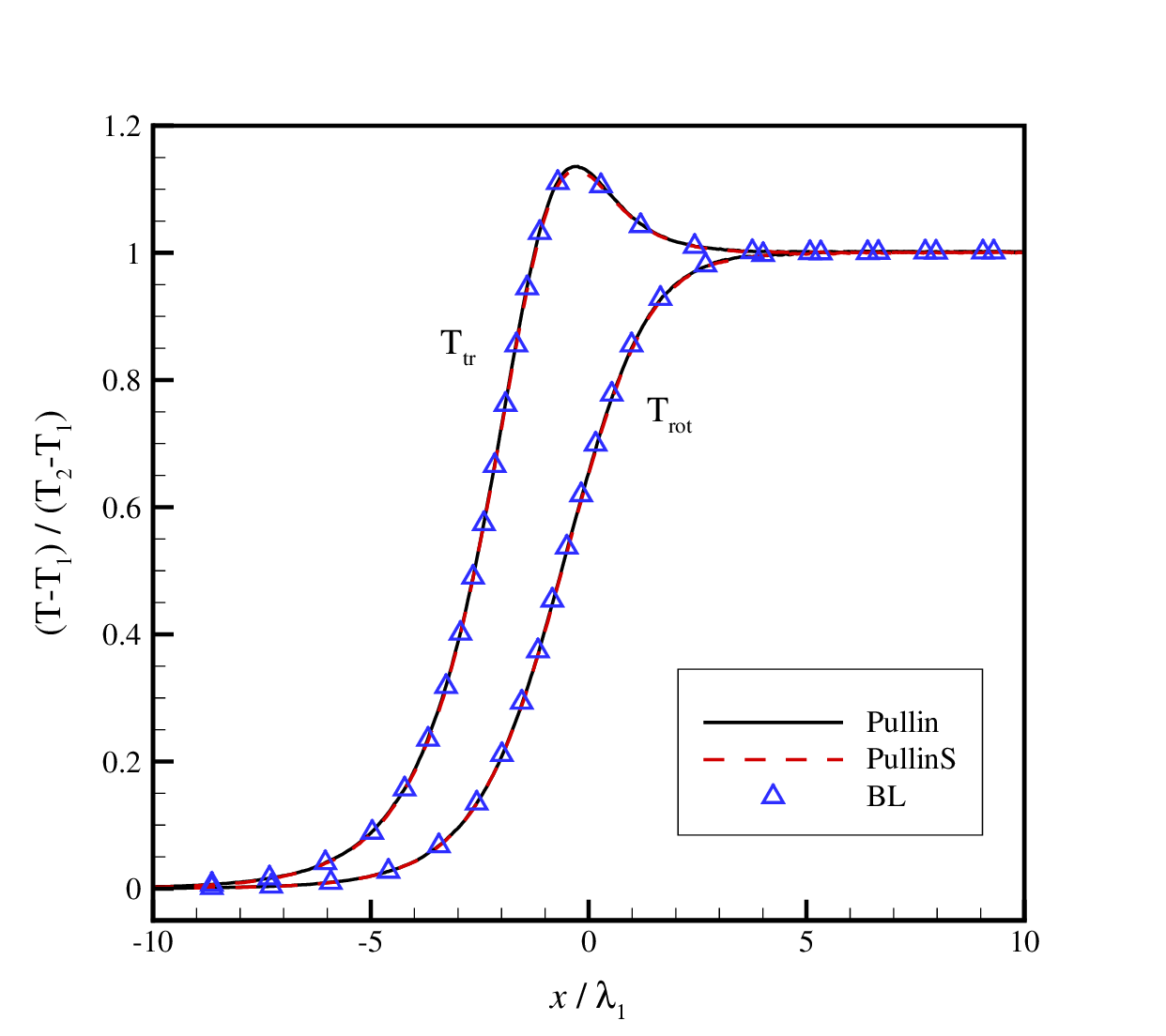}}
    \subfigure[]{\label{Shockwave-Ma10-T}\includegraphics[width=0.45\textwidth]{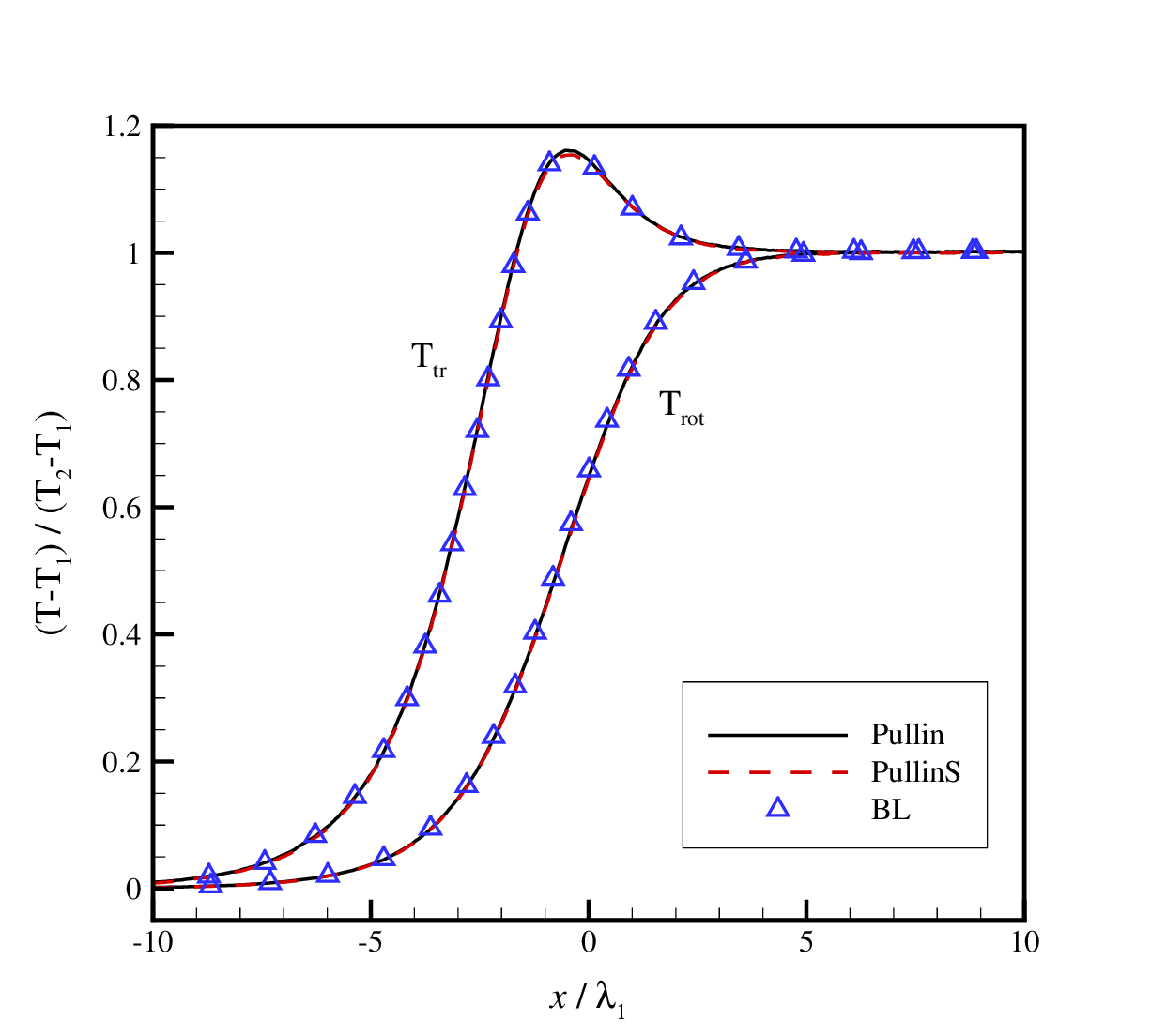}}
	\caption{\label{Shockwave-T} Temperature profiles of shock waves at different Mach numbers: (a) $\mathrm{Ma}=1.53$; (b) $\mathrm{Ma}=2.0$; (c) $\mathrm{Ma}=6.1$; (d) $\mathrm{Ma}=10.0$. Solid line: Pullin model; dashed line: simplified Pullin model; triangles: BL model.}
\end{figure}

\begin{figure}
	\centering
	\subfigure[]{\label{Shockwave-Ma1.53-rou}\includegraphics[width=0.45\textwidth]{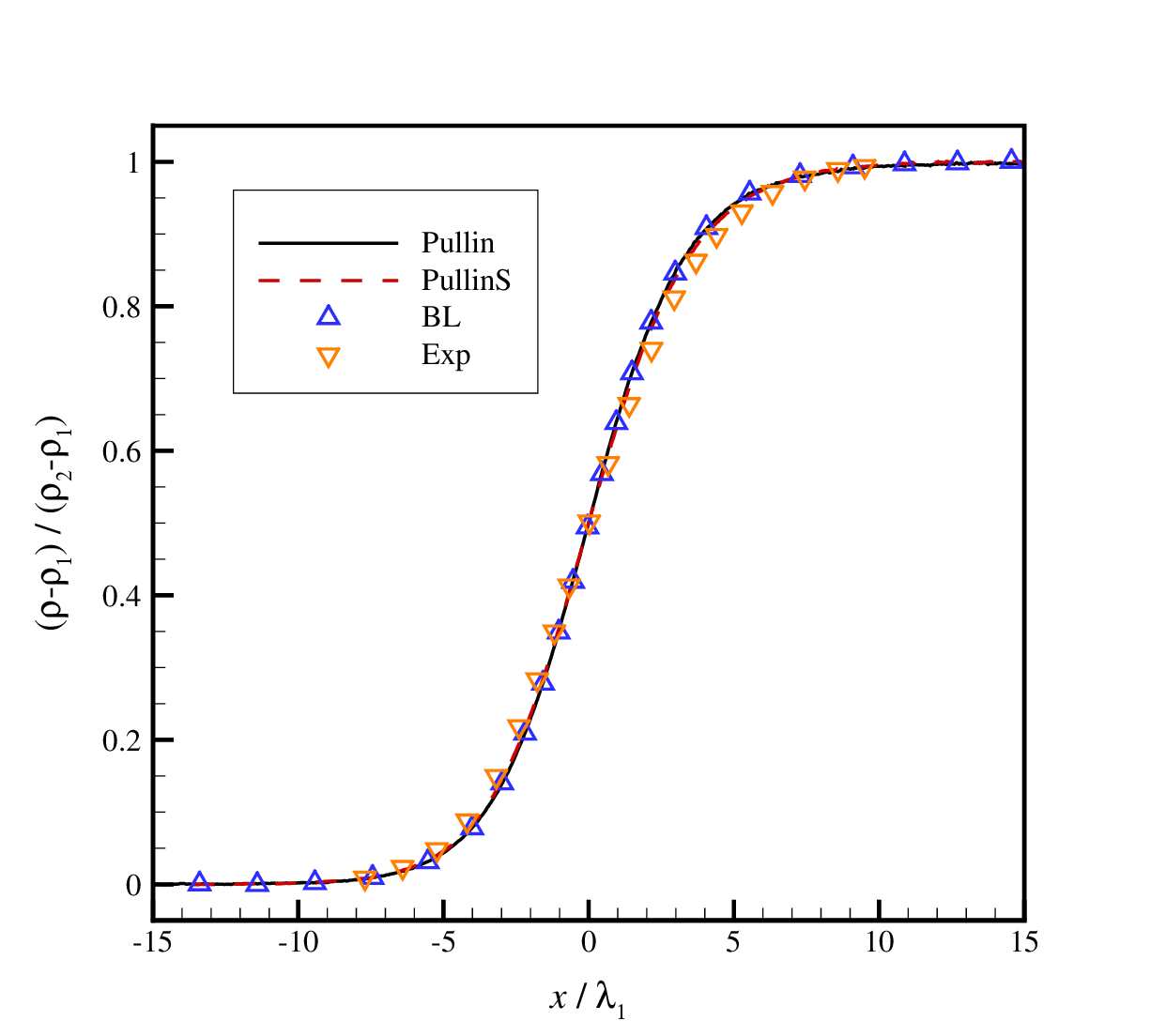}}
	\subfigure[]{\label{Shockwave-Ma2-rou}\includegraphics[width=0.45\textwidth]{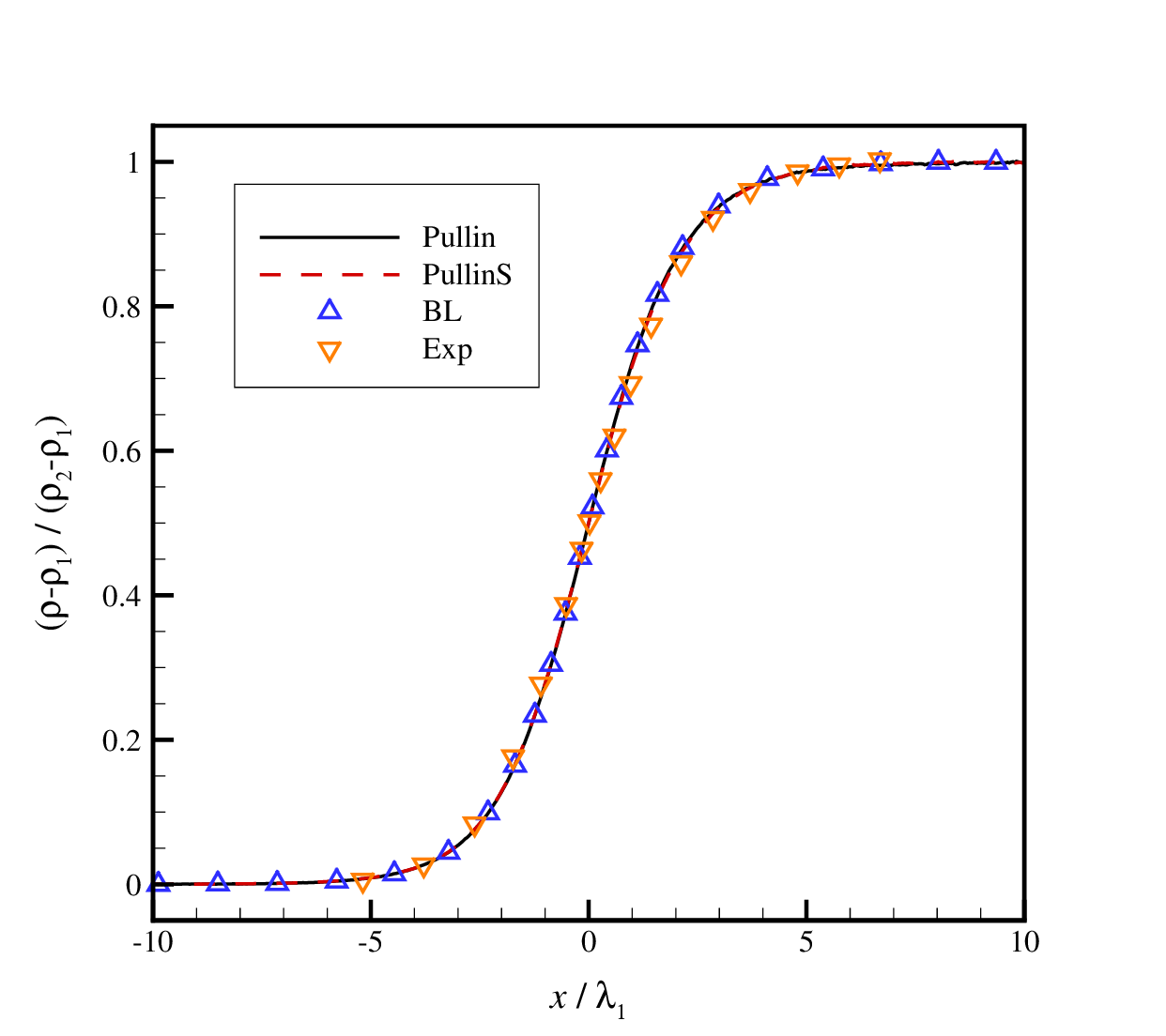}}
    \subfigure[]{\label{Shockwave-Ma6.1-rou}\includegraphics[width=0.45\textwidth]{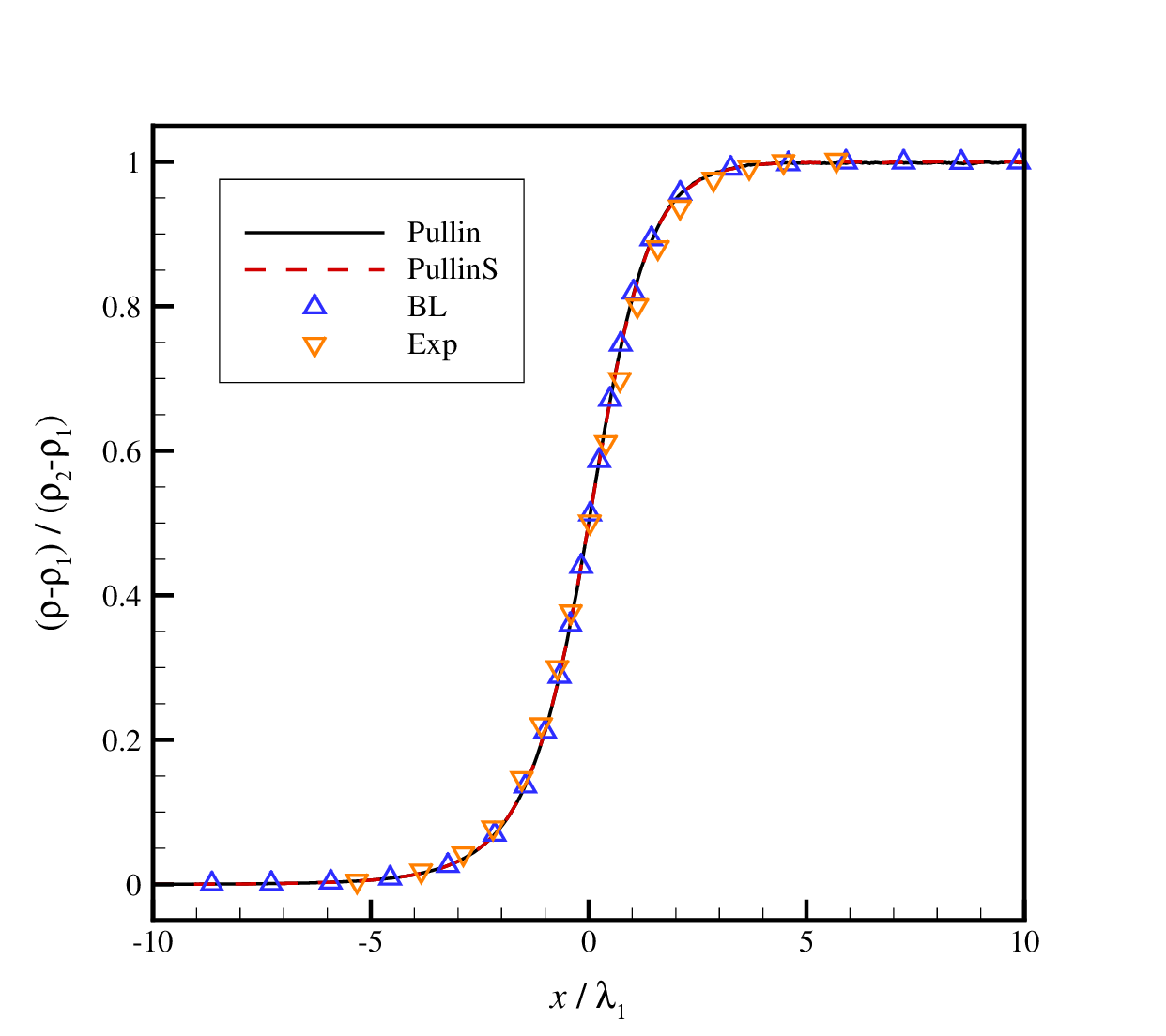}}
    \subfigure[]{\label{Shockwave-Ma10-rou}\includegraphics[width=0.45\textwidth]{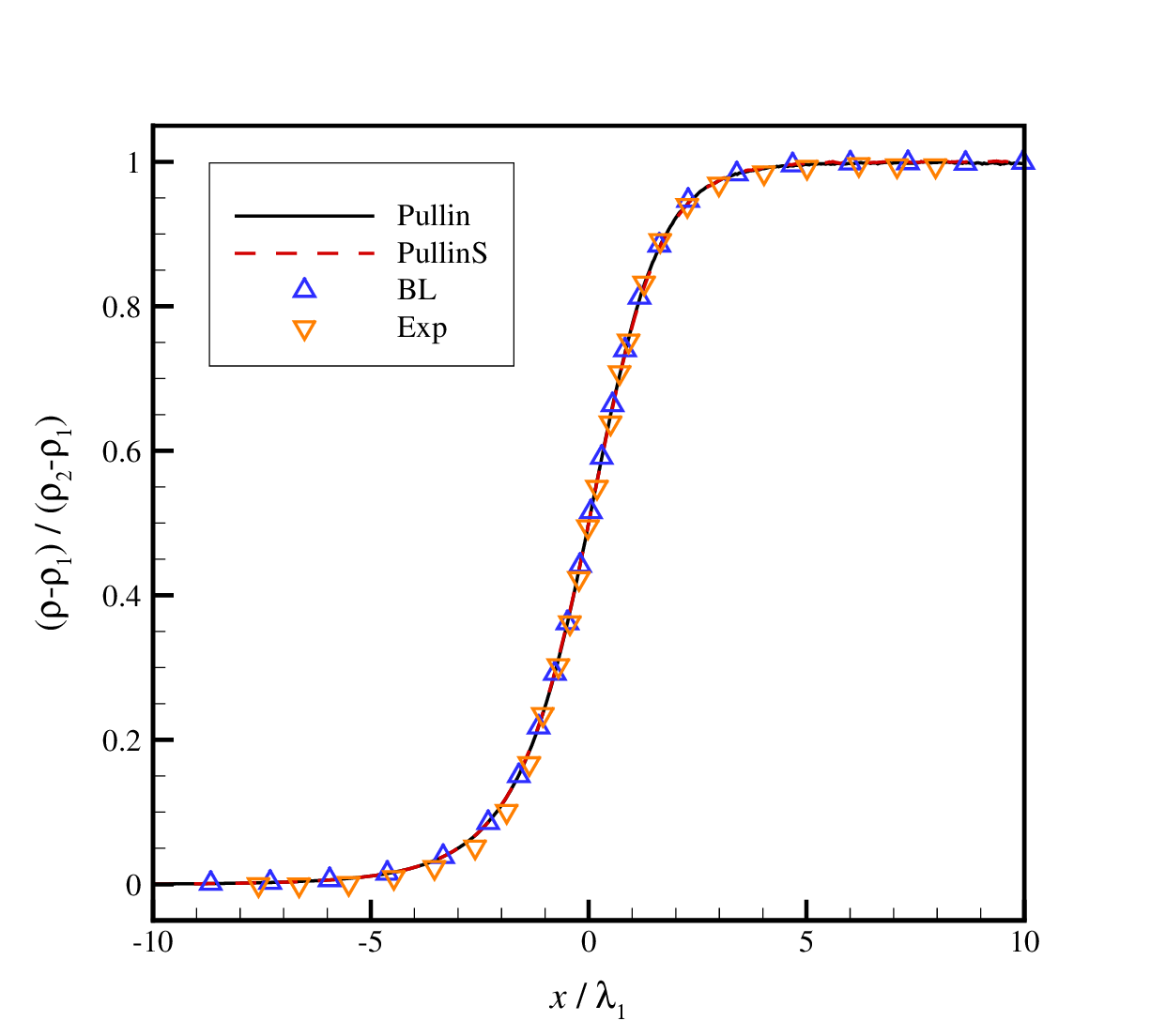}}
	\caption{\label{Shockwave-rou} Density profiles of shock waves at different Mach numbers: (a) $\mathrm{Ma}=1.53$; (b) $\mathrm{Ma}=2.0$; (c) $\mathrm{Ma}=6.1$; (d) $\mathrm{Ma}=10.0$. Solid line: Pullin model; dashed line: simplified Pullin model; triangles: BL model; gradients: experimental data from Alsmeyer \cite{alsmeyer1976density}.}
\end{figure}

Figure \ref{Shockwave-T} presents the translational and rotational temperature profiles across the shock wave for different Mach numbers. The $x$ coordinate is normalized by the upstream mean free path $\lambda_1$, defined according to the hard-sphere (HS) model as
\begin{equation}
    \lambda_1=\frac{16\mu_1}{5\rho_1\left(2\pi RT_1\right)^{1/2}},
\end{equation}
where $R$ is the gas constant, $\mu_1$ and $\rho_1$ are the upstream viscosity and density, respectively. The upstream viscosity $\mu_1$ is calculated using the reference viscosity $\mu_{\rm ref} = 1.656\times 10^{-5}\ \mathrm{Pa\cdot s}$ and reference temperature $T_{\rm ref} = 273\ \mathrm{K}$. The translational temperature predicted by the Pullin model differs slightly from that of its simplified variant; however, both are generally in good agreement with the results of the BL model. As the upstream Mach number increases, the shock wave becomes thicker and nonequilibrium effects become more pronounced. Figure \ref{Shockwave-rou} shows the corresponding density profiles across the shock wave. The results from the Pullin model and its simplified variant agree well with those from the BL model as well as with experimental data from Alsmeyer \cite{alsmeyer1976density}. Overall, these results demonstrate that the Pullin model and its simplified variant accurately capture the complex nonequilibrium phenomena in normal shock waves.

\subsection{Hypersonic flow past a cylinder}\label{subSec4_cylinder}
The hypersonic flow past a cylinder is a classical benchmark for assessing energy-exchange models in rarefied gas dynamics, characterized by a high-density region ahead of the cylinder and a rarefied wake downstream. In this study, hypersonic flow past a cylinder is simulated at a freestream Mach number of $\mathrm{Ma} = 5$ for four Knudsen numbers: $\mathrm{Kn} = 0.01$, $0.1$, $1.0$, and $10.0$. The cylinder radius, chosen as the characteristic length, is set to $L_{ref}=R=1m$. The working gas is nitrogen, with freestream number densities of $1.29438 \times 10^{20} m^{-3}$, $1.29438 \times 10^{19} m^{-3}$, $1.29438 \times 10^{18} m^{-3}$, and $1.29438 \times 10^{17} m^{-3}$ corresponding to the four Knudsen numbers. Both translational and rotational temperatures are initialized at $T_{tr} = T_{rot} = 273 \mathrm{K}$, and the freestream velocity is $V = 1684.59 m/s$. A fully diffusive wall boundary condition is applied, maintaining a constant wall temperature of $T_{wall} = 500 \mathrm{K}$. The rotational collision number in the BL model is set to $Z_{\mathrm{BL}} = 5$, while the corresponding value $Z$ in the Pullin model is obtained from Eq.~(\ref{eq_Z_and_ZBL}). Each computational cell is initialized with about 20 simulated particles, and the results are averaged over 40,000 time steps after the flow reaches steady state.

\begin{figure}
	\centering
	\subfigure[]{\label{CylKn001-Ttr}\includegraphics[width=0.45\textwidth]{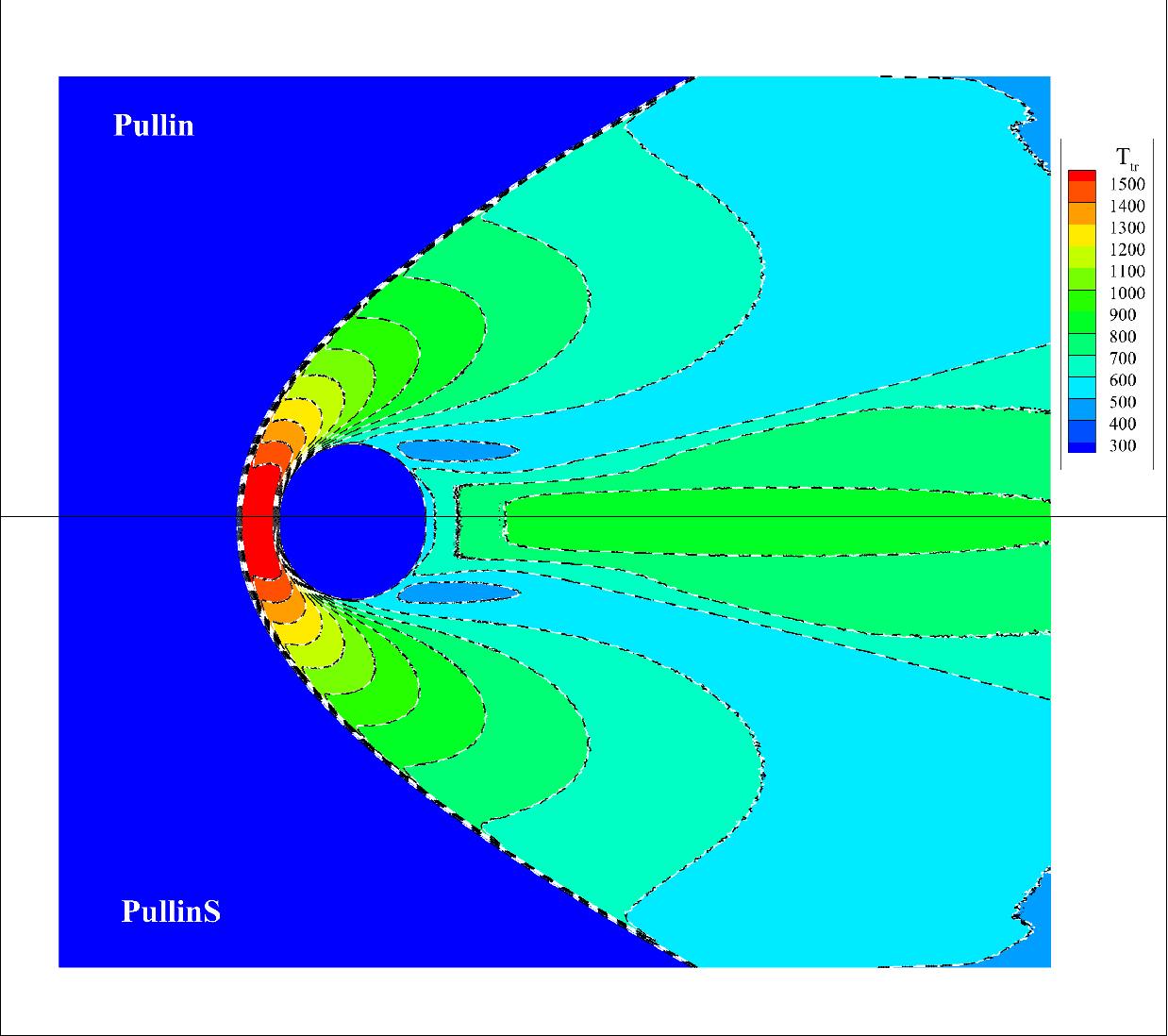}}
	\subfigure[]{\label{CylKn001-Trot}\includegraphics[width=0.45\textwidth]{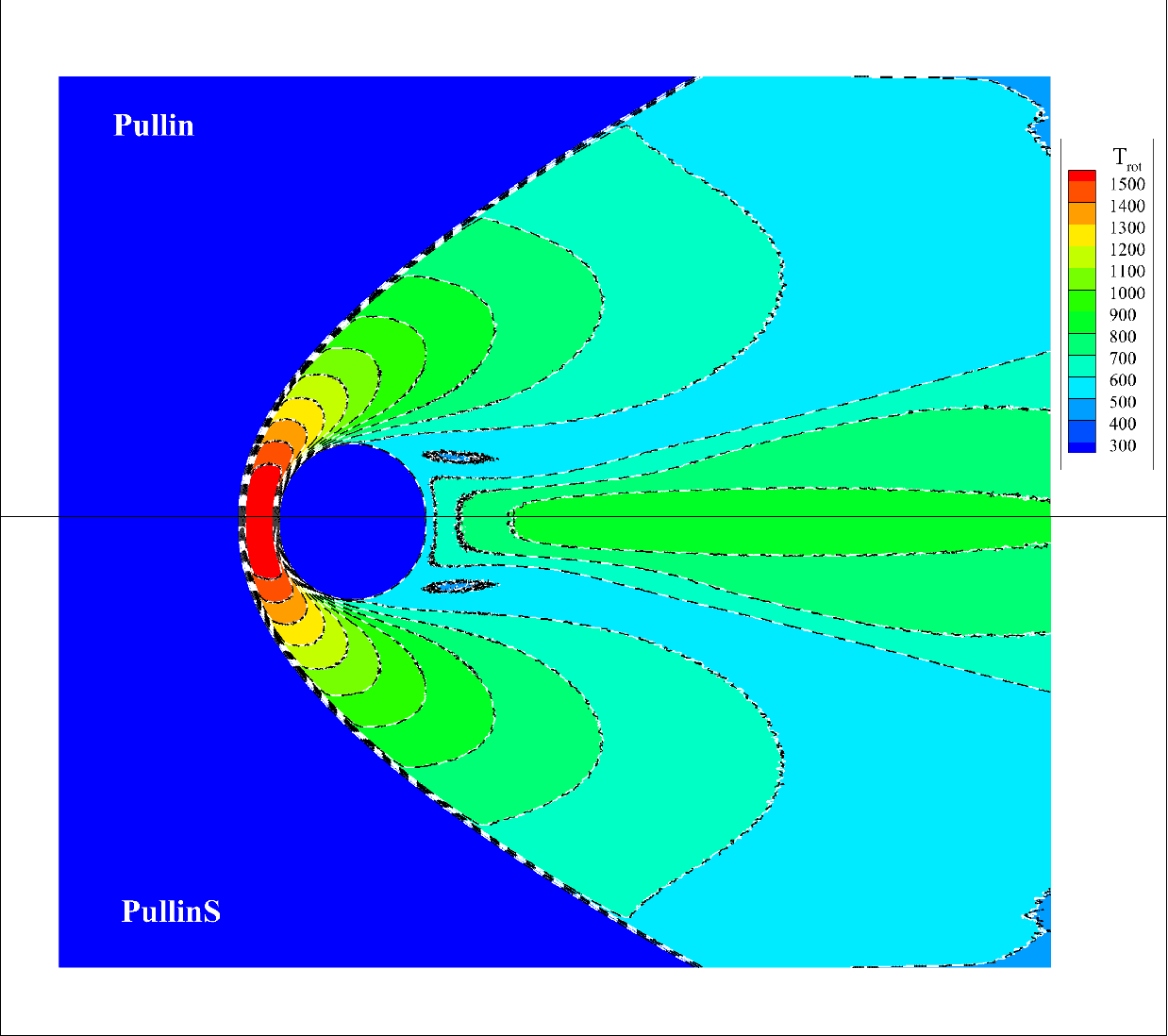}}
	\caption{\label{CylKn001-T} Temperature contours of hypersonic cylinder flow at $\mathrm{Ma}=5$ and $\mathrm{Kn}=0.01$: (a) translational temperature; (b) rotational temperature. Upper solid line: Pullin model; lower solid line: simplified Pullin model; dashed line: BL model.}
\end{figure}

\begin{figure}
	\centering
	\subfigure[]{\label{CylKn01-Ttr}\includegraphics[width=0.45\textwidth]{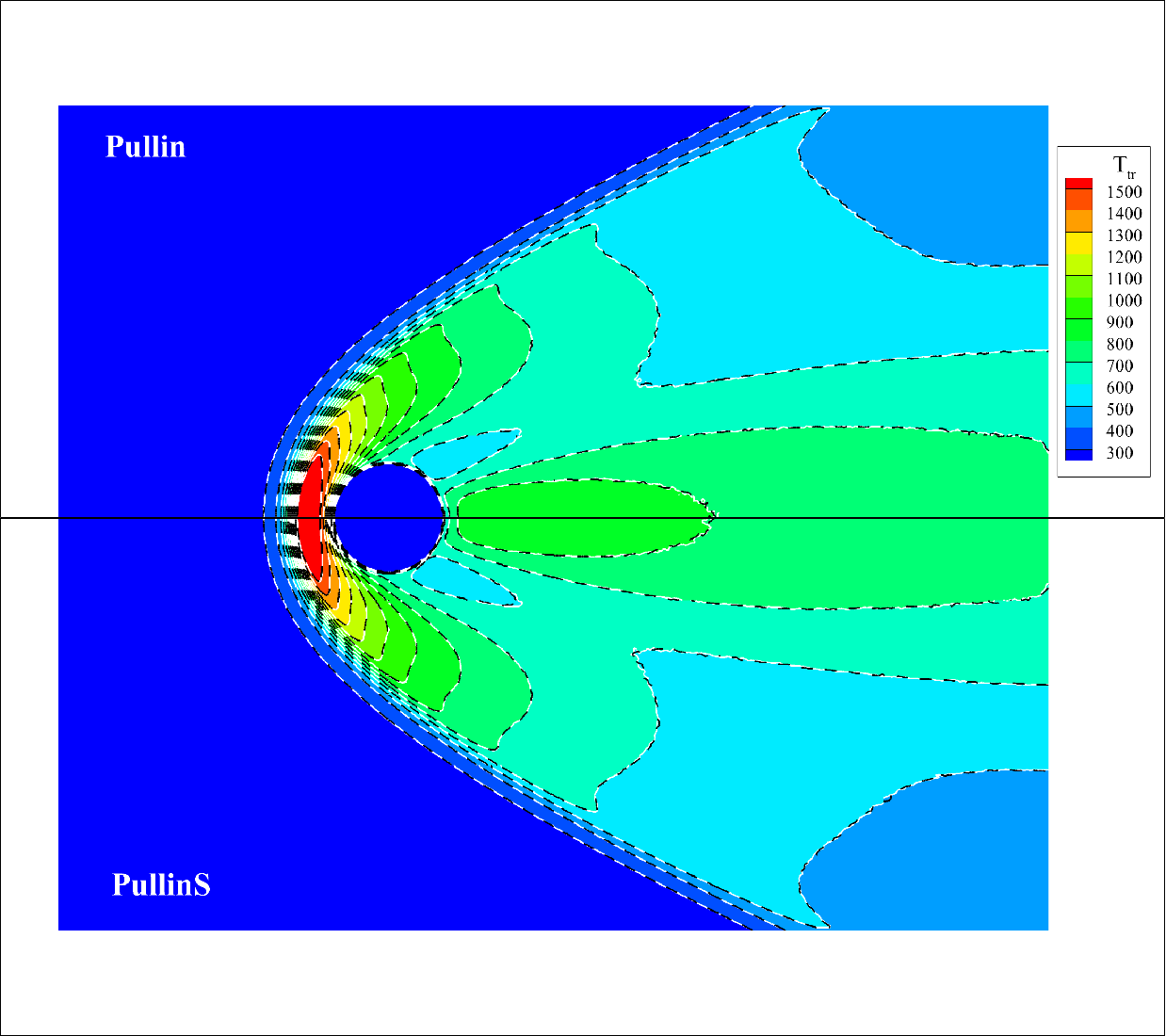}}
	\subfigure[]{\label{CylKn01-Trot}\includegraphics[width=0.45\textwidth]{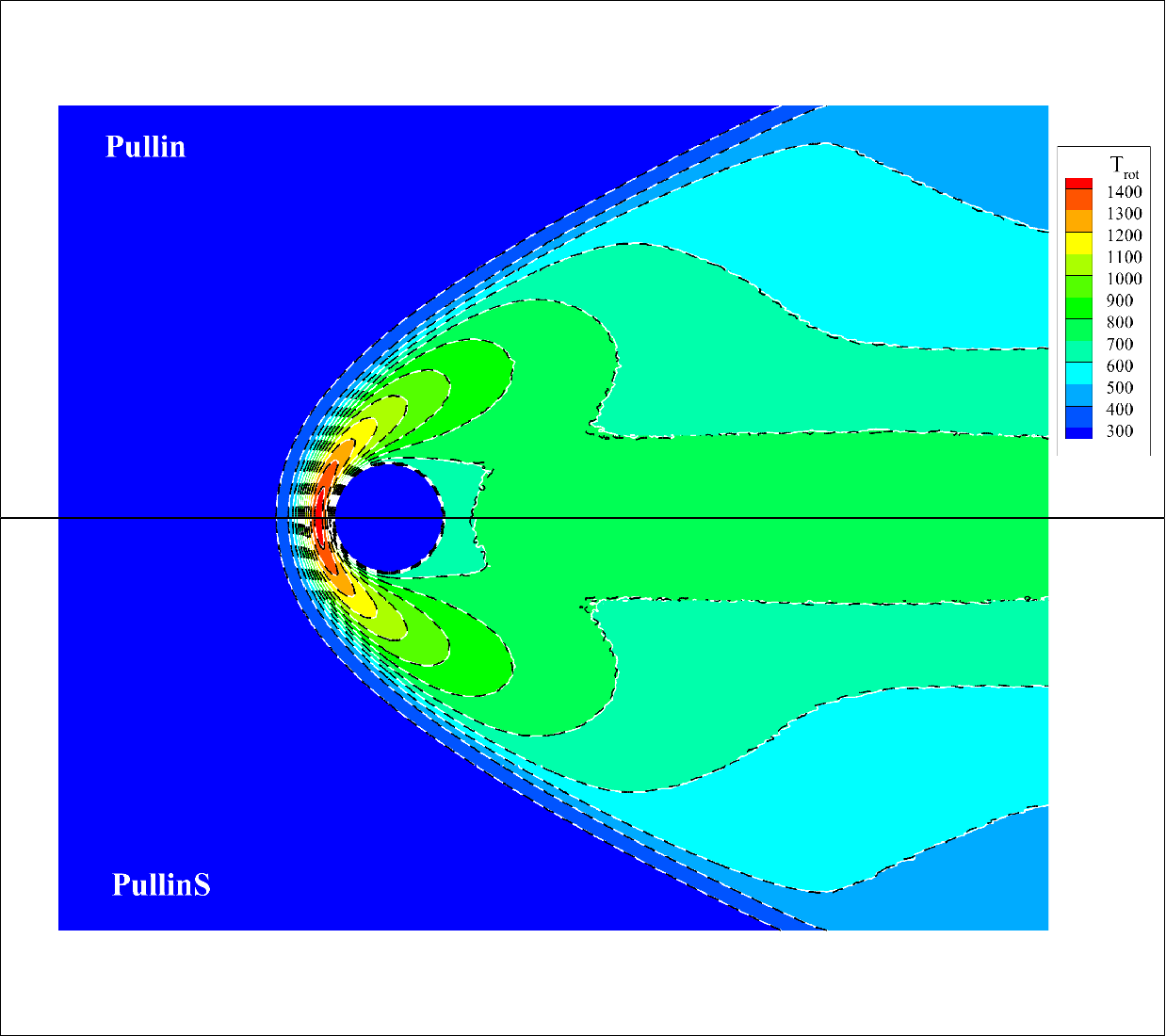}}
	\caption{\label{CylKn01-T} Temperature contours of hypersonic cylinder flow at $\mathrm{Ma}=5$ and $\mathrm{Kn}=0.1$: (a) translational temperature; (b) rotational temperature. Upper solid line: Pullin model; lower solid line: simplified Pullin model; dashed line: BL model.}
\end{figure}

\begin{figure}
	\centering
	\subfigure[]{\label{CylKn1-Ttr}\includegraphics[width=0.45\textwidth]{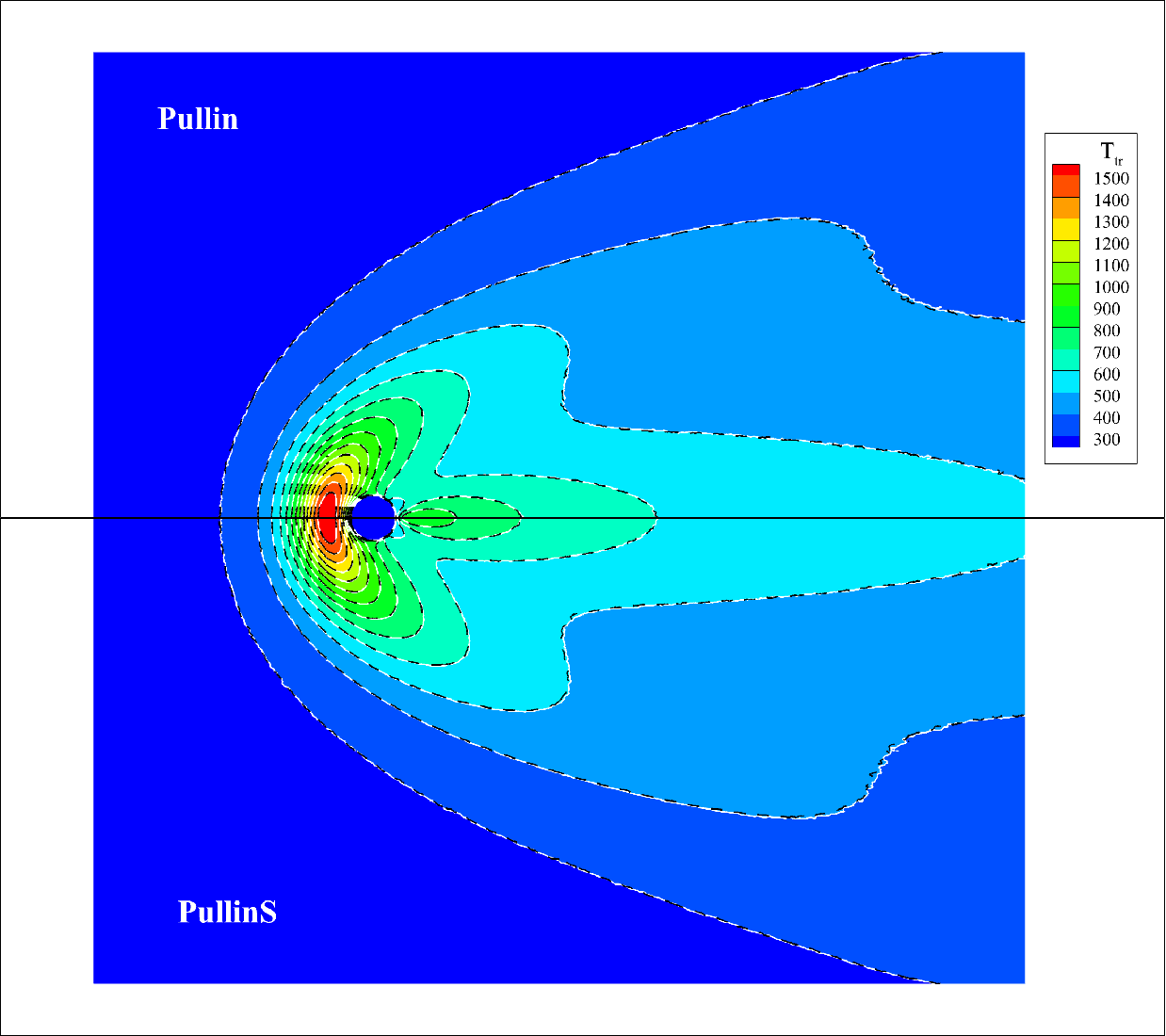}}
	\subfigure[]{\label{CylKn1-Trot}\includegraphics[width=0.45\textwidth]{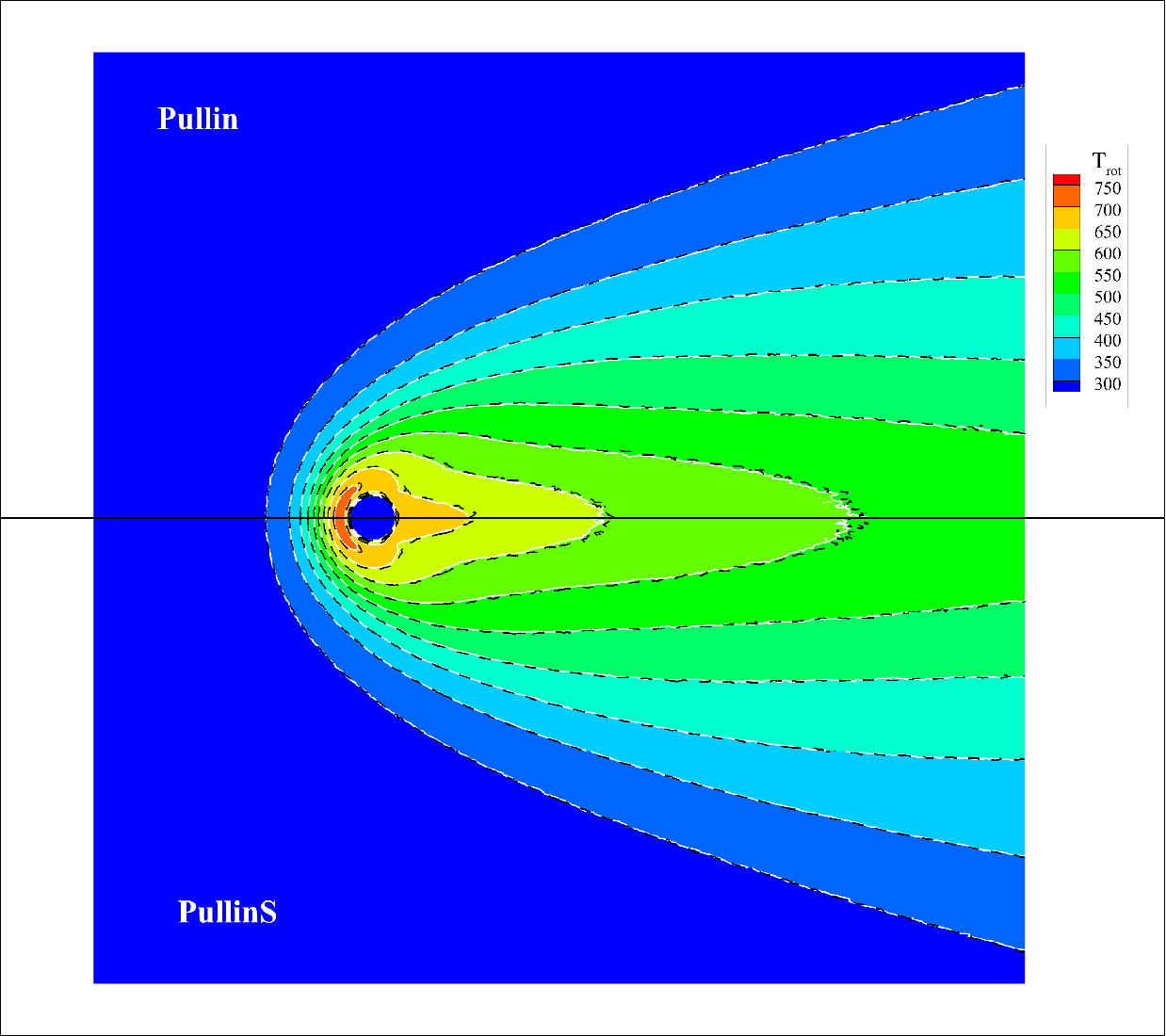}}
	\caption{\label{CylKn1-T} Temperature contours of hypersonic cylinder flow at $\mathrm{Ma}=5$ and $\mathrm{Kn}=1.0$: (a) translational temperature; (b) rotational temperature. Upper solid line: Pullin model; lower solid line: simplified Pullin model; dashed line: BL model.}
\end{figure}

\begin{figure}
	\centering
	\subfigure[]{\label{CylKn10-Ttr}\includegraphics[width=0.45\textwidth]{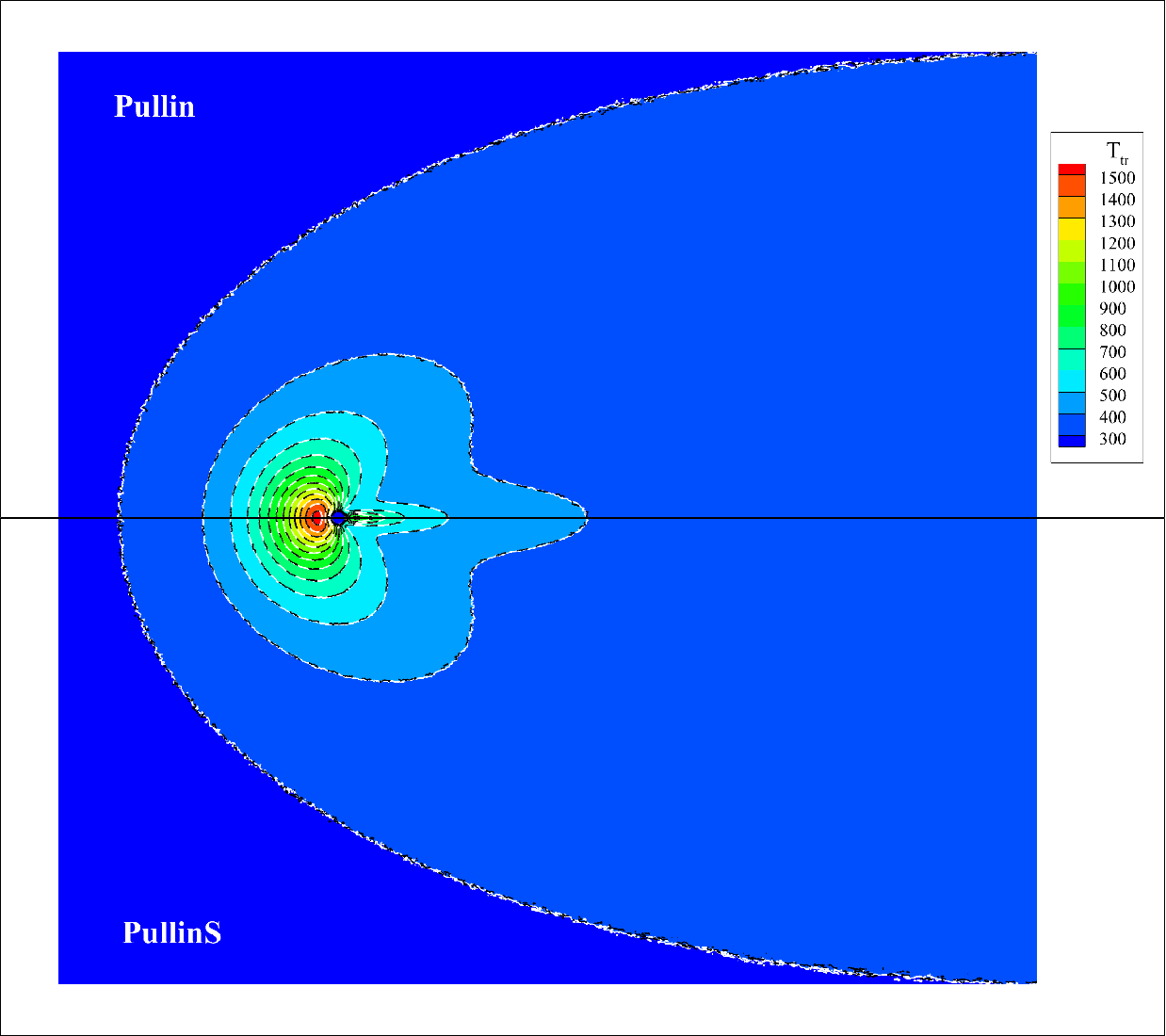}}
	\subfigure[]{\label{CylKn10-Trot}\includegraphics[width=0.45\textwidth]{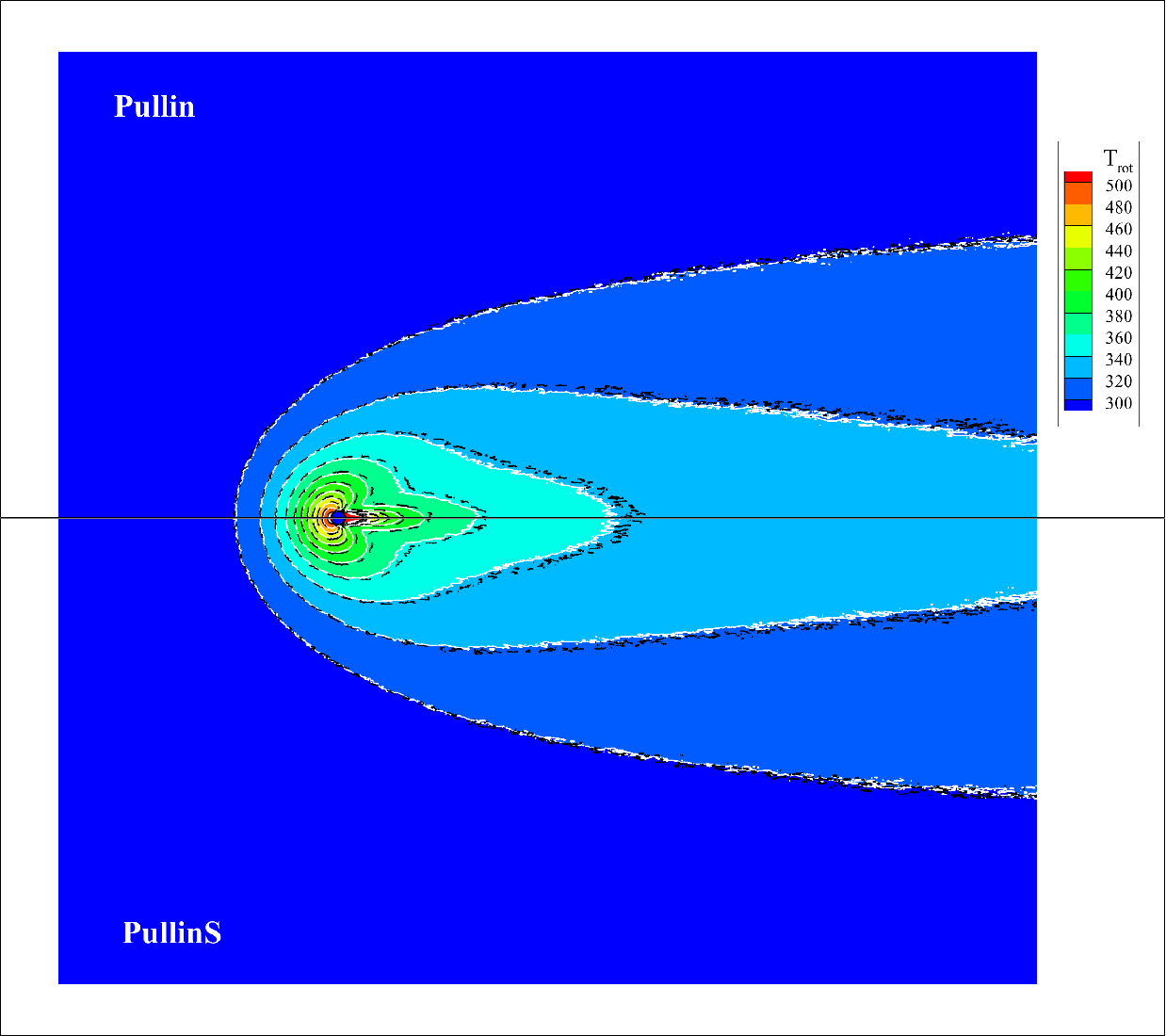}}
	\caption{\label{CylKn10-T} Temperature contours of hypersonic cylinder flow at $\mathrm{Ma}=5$ and $\mathrm{Kn}=10$: (a) translational temperature; (b) rotational temperature. Upper solid line: Pullin model; lower solid line: simplified Pullin model; dashed line: BL model.}
\end{figure}

Figures \ref{CylKn001-T}-\ref{CylKn10-T} compare the translational and rotational temperature contours predicted by the three internal energy relaxation models at four freestream Knudsen numbers. Across all four Knudsen numbers, the translational temperature results from the three models agree well. Similarly, for $\mathrm{Kn} = 0.01$ and $0.1$, the rotational temperature predictions are also consistent. However, differences emerge as the Knudsen number increases: at $\mathrm{Kn} = 1$ and $10$, the Pullin and BL models exhibit slight discrepancies in the rotational temperature in the wake region behind the cylinder, whereas the Pullin model and its simplified variant remain nearly identical. These flowfield results further indicate that at $\mathrm{Kn} = 0.01$, the translational and rotational temperatures are almost equal, suggesting negligible thermal nonequilibrium. As the Knudsen number increases, the difference between translational and rotational temperatures grows, reflecting stronger nonequilibrium effects. Notably, the rotational temperature in the wake progressively rises with Kn, and at $\mathrm{Kn} = 10$, it becomes comparable to--or even slightly higher than--that at the stagnation point, demonstrating that nonequilibrium effects are significant in highly rarefied hypersonic flows.

\begin{figure}
	\centering
	\subfigure[]{\label{CylKn001-pres}\includegraphics[width=0.45\textwidth]{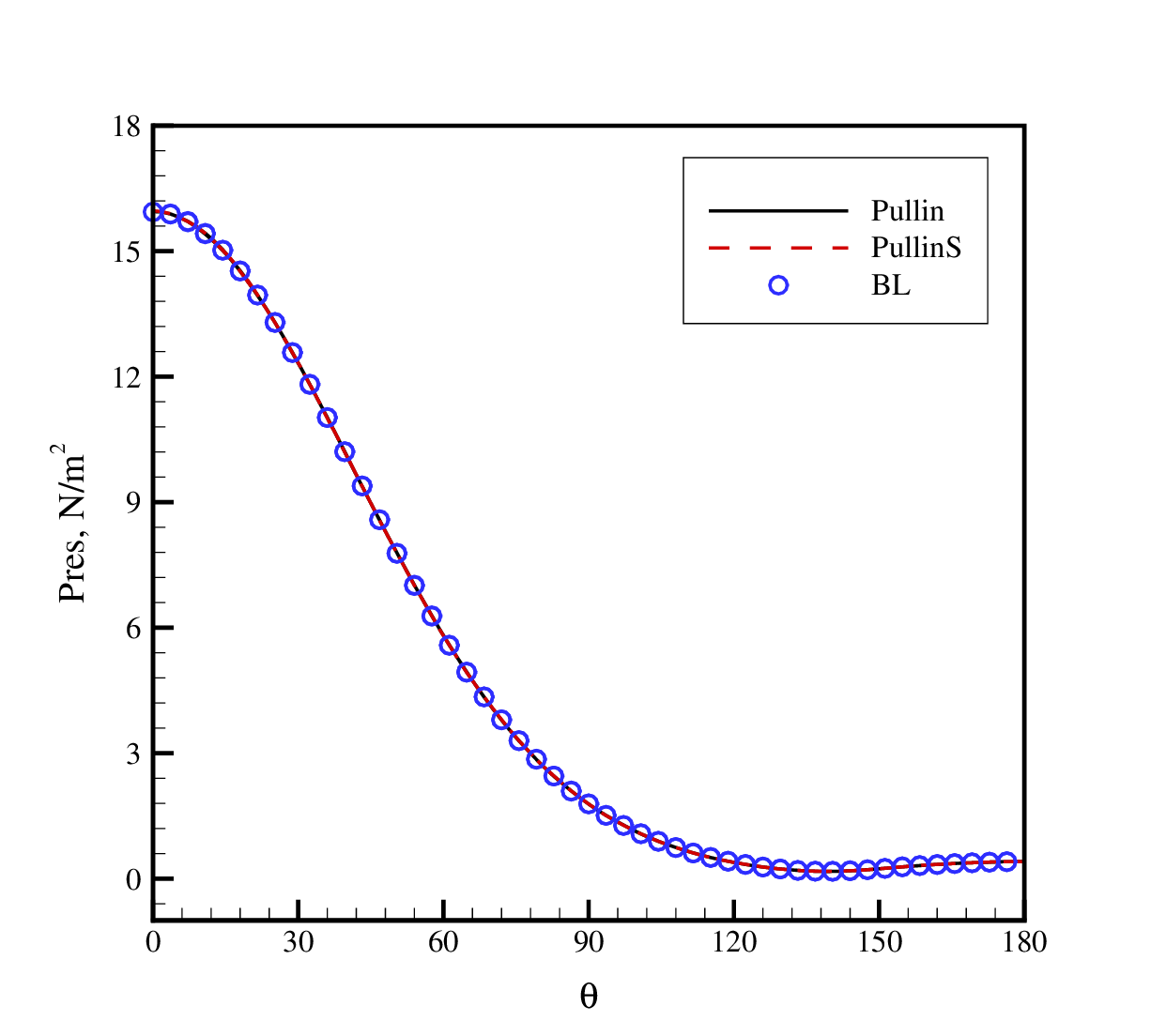}}
	\subfigure[]{\label{CylKn01-pres}\includegraphics[width=0.45\textwidth]{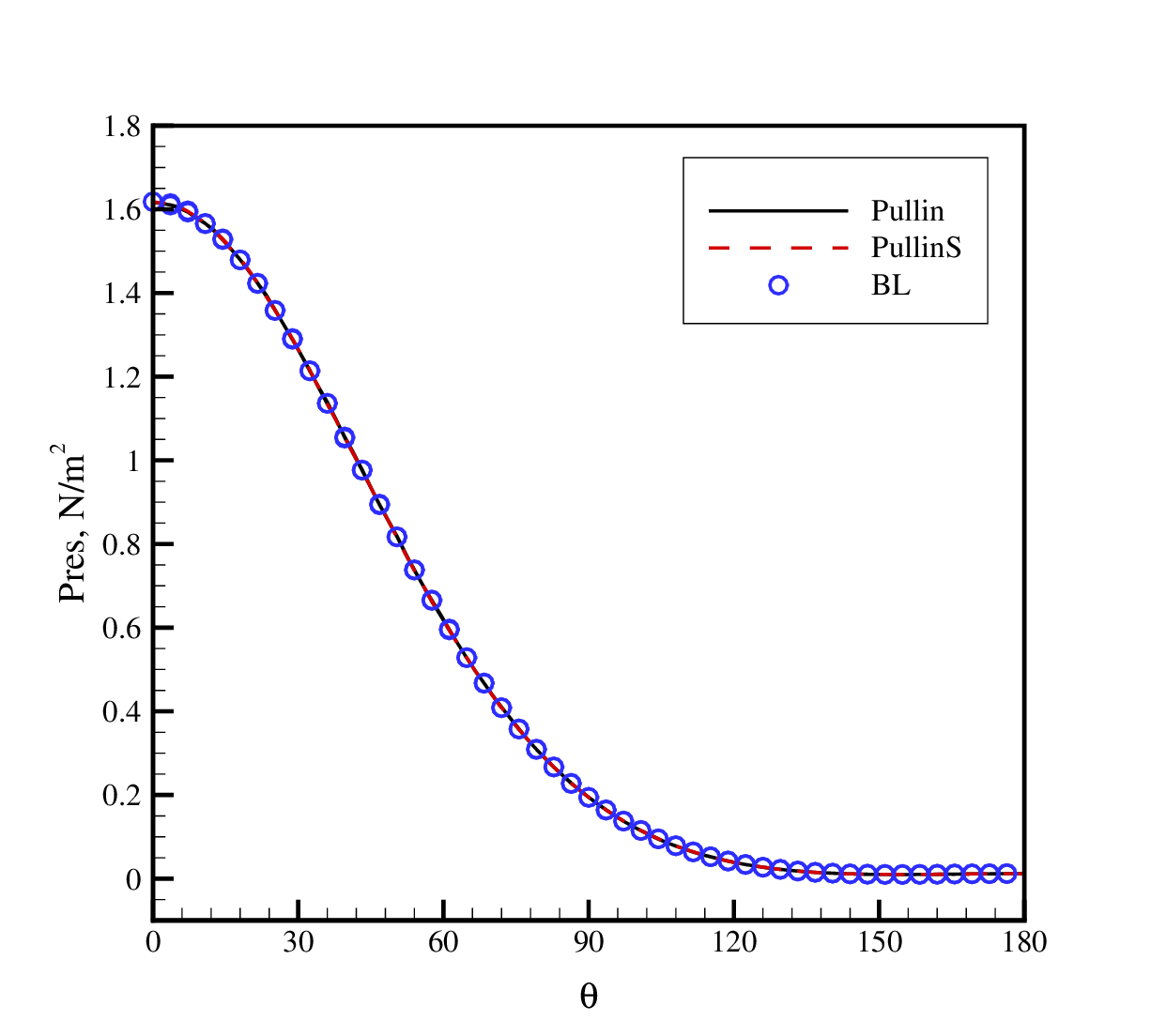}}
	\subfigure[]{\label{CylKn1-pres}\includegraphics[width=0.45\textwidth]{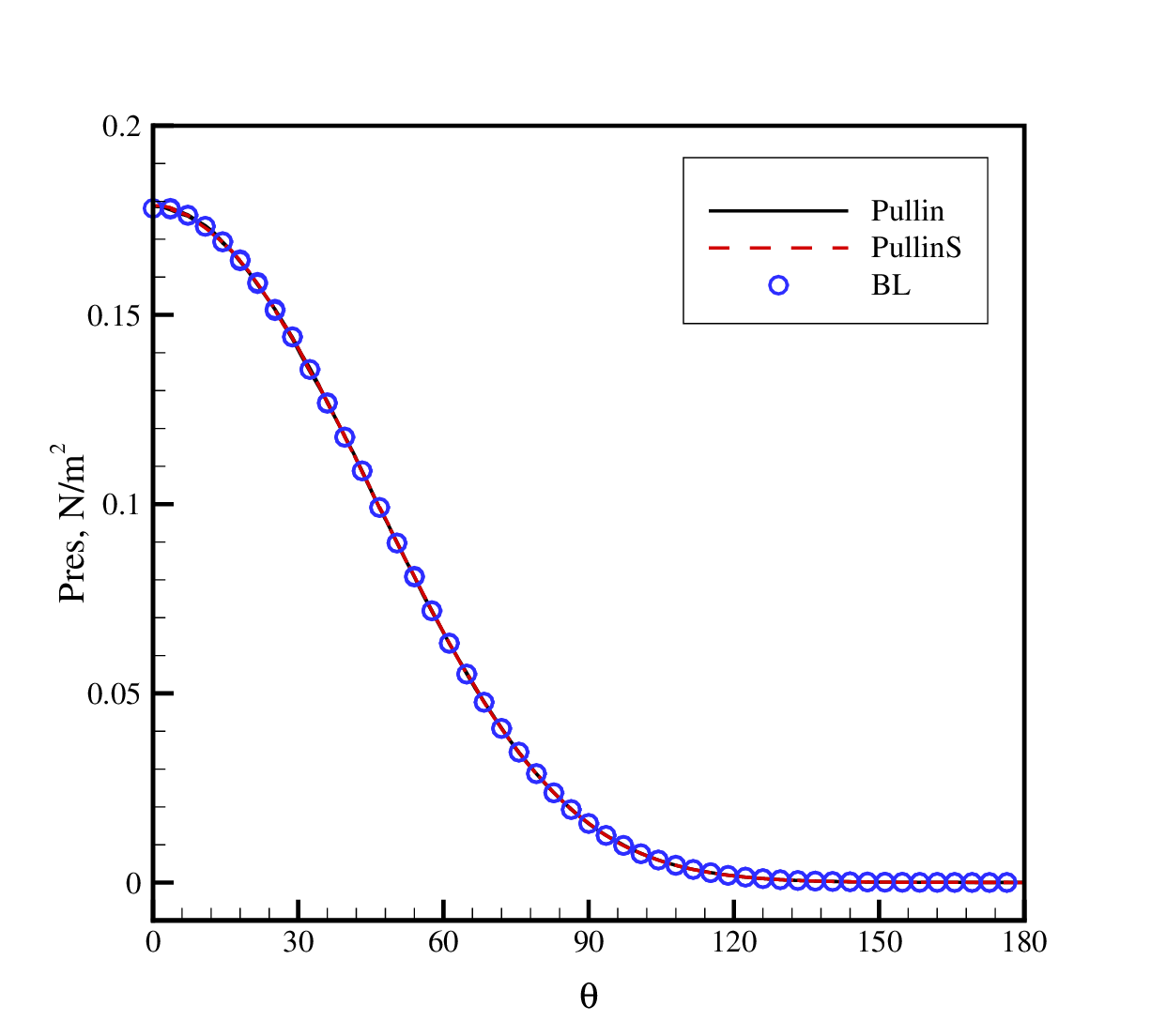}}
	\subfigure[]{\label{CylKn10-pres}\includegraphics[width=0.45\textwidth]{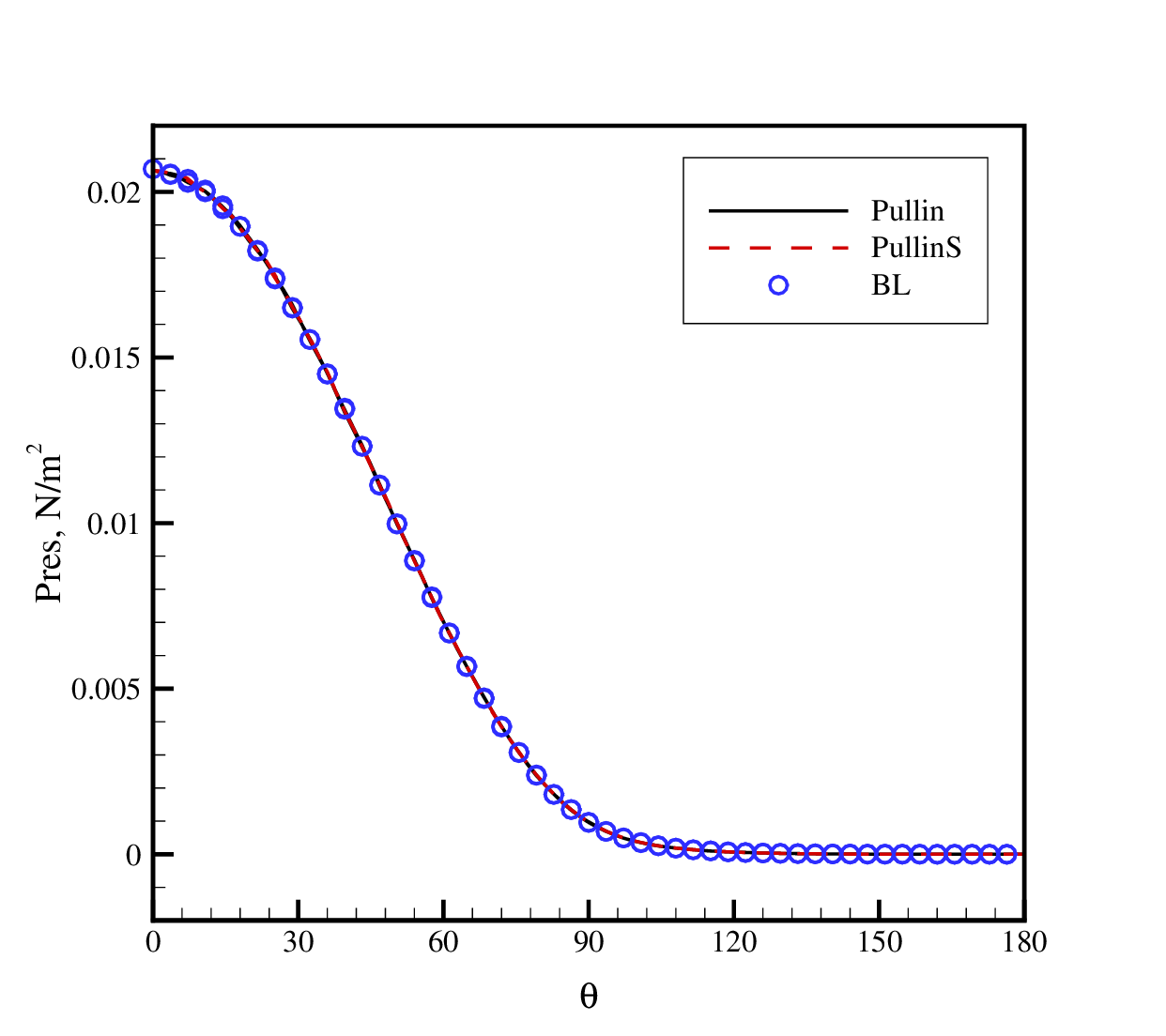}}
	\caption{\label{Cyl-pres} Surface pressure results of hypersonic cylinder flow at $\mathrm{Ma}=5$: (a) $\mathrm{Kn}=0.01$, (b) $\mathrm{Kn}=0.1$, (c) $\mathrm{Kn}=1.0$, (d) $\mathrm{Kn}=10.0$. Solid line: Pullin model; dashed line: simplified Pullin model; circles: BL model.}
\end{figure}

\begin{figure}
	\centering
	\subfigure[]{\label{CylKn001-tau}\includegraphics[width=0.45\textwidth]{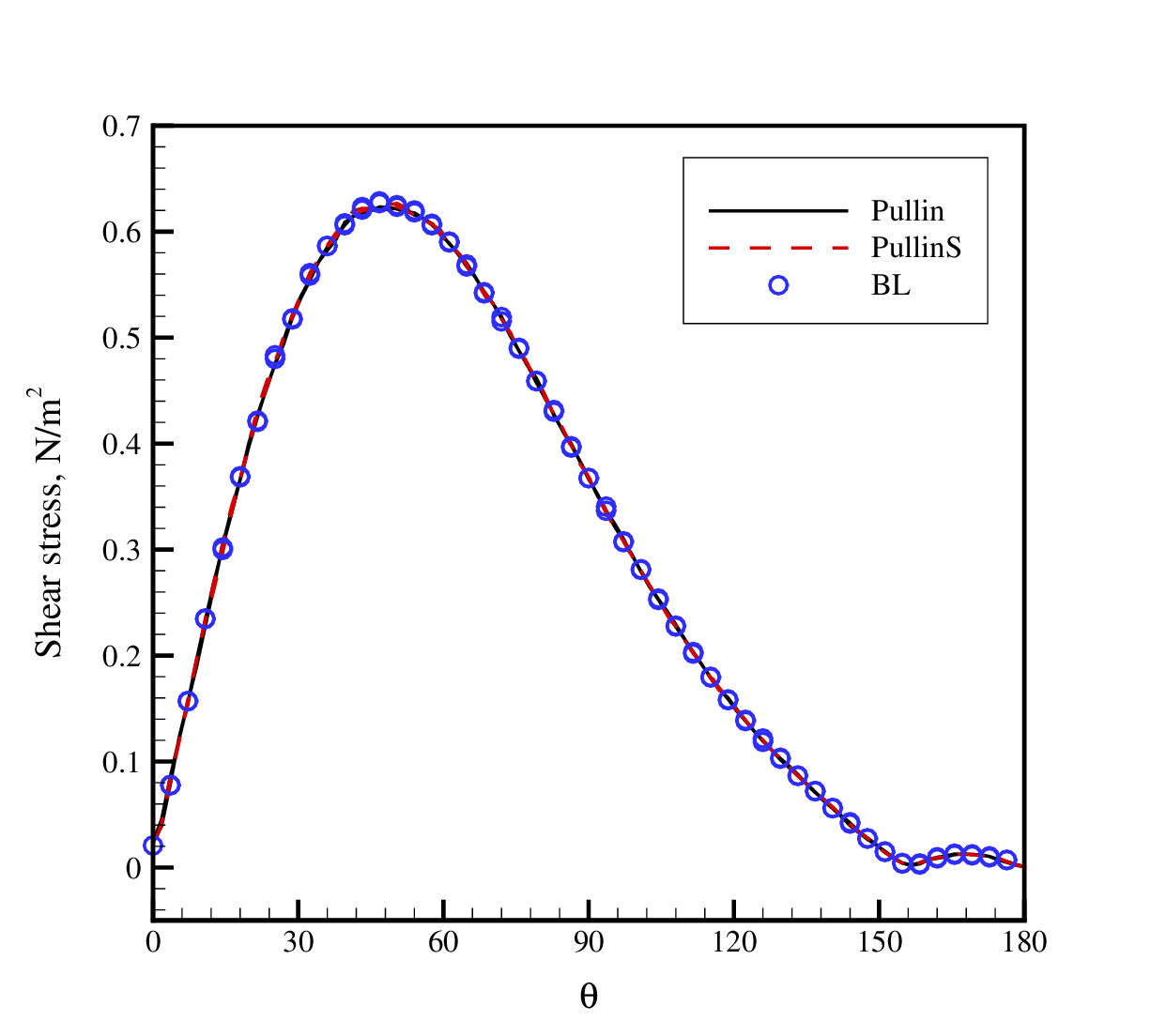}}
	\subfigure[]{\label{CylKn01-tau}\includegraphics[width=0.45\textwidth]{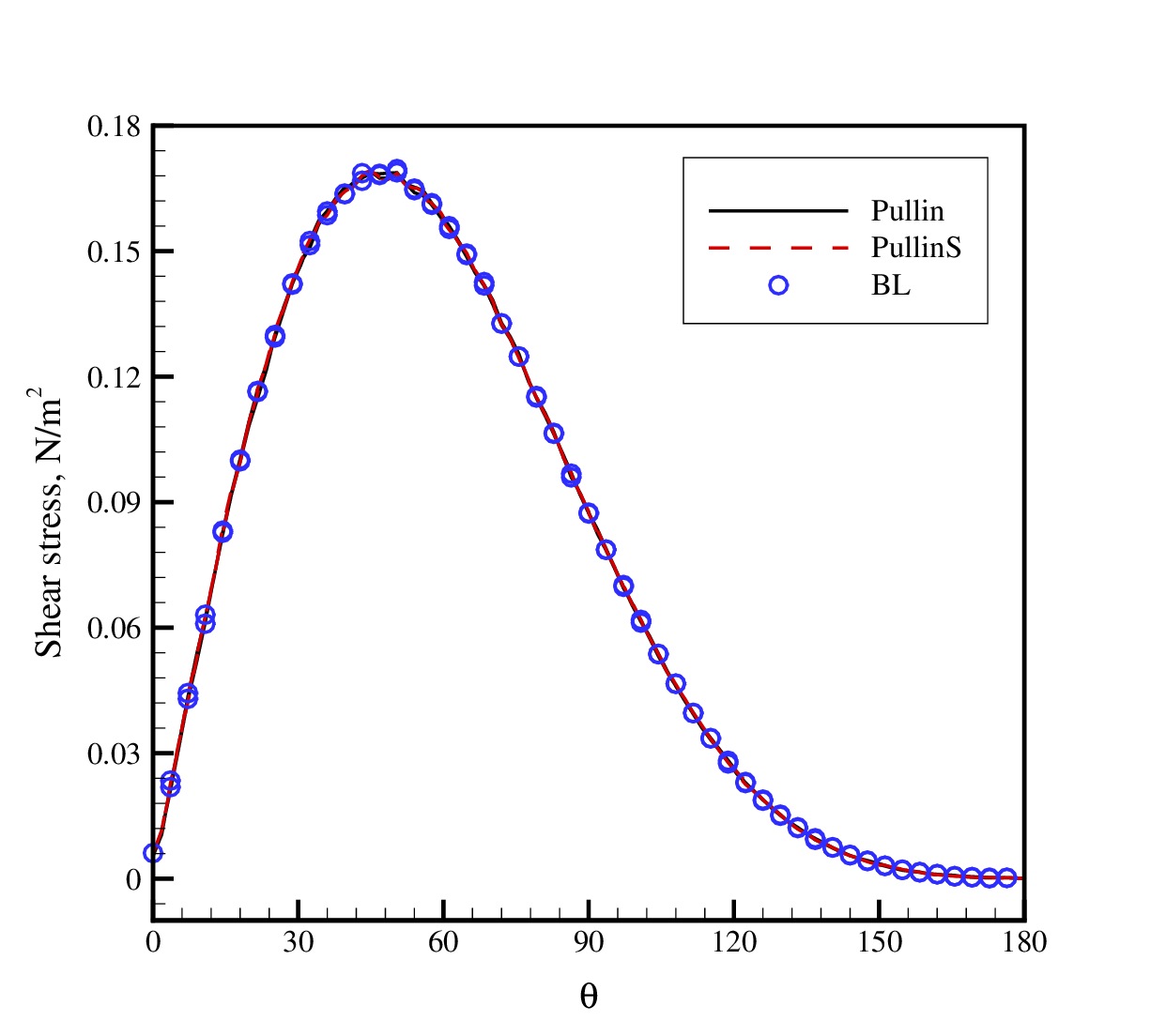}}
	\subfigure[]{\label{CylKn1-tau}\includegraphics[width=0.45\textwidth]{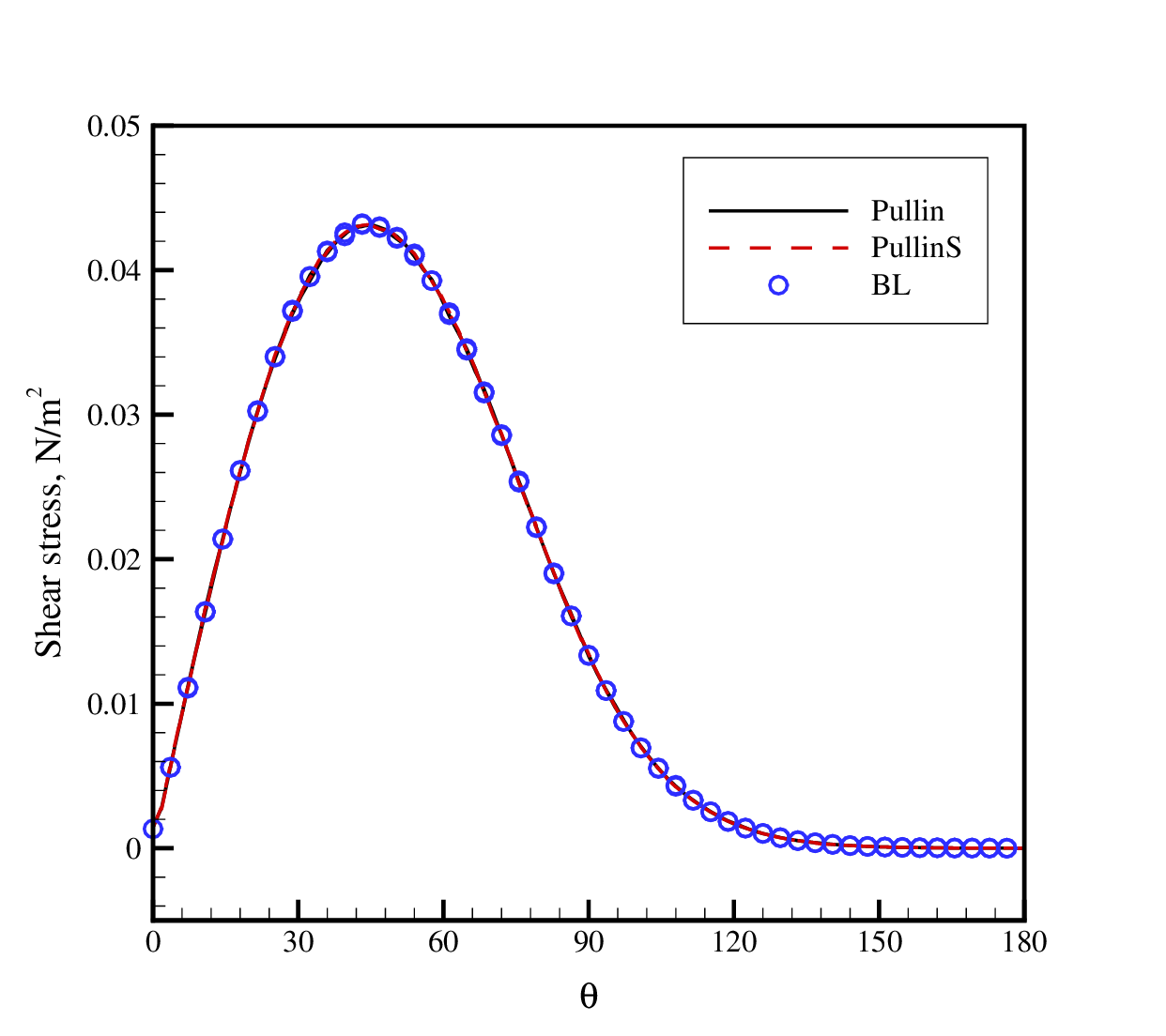}}
	\subfigure[]{\label{CylKn10-tau}\includegraphics[width=0.45\textwidth]{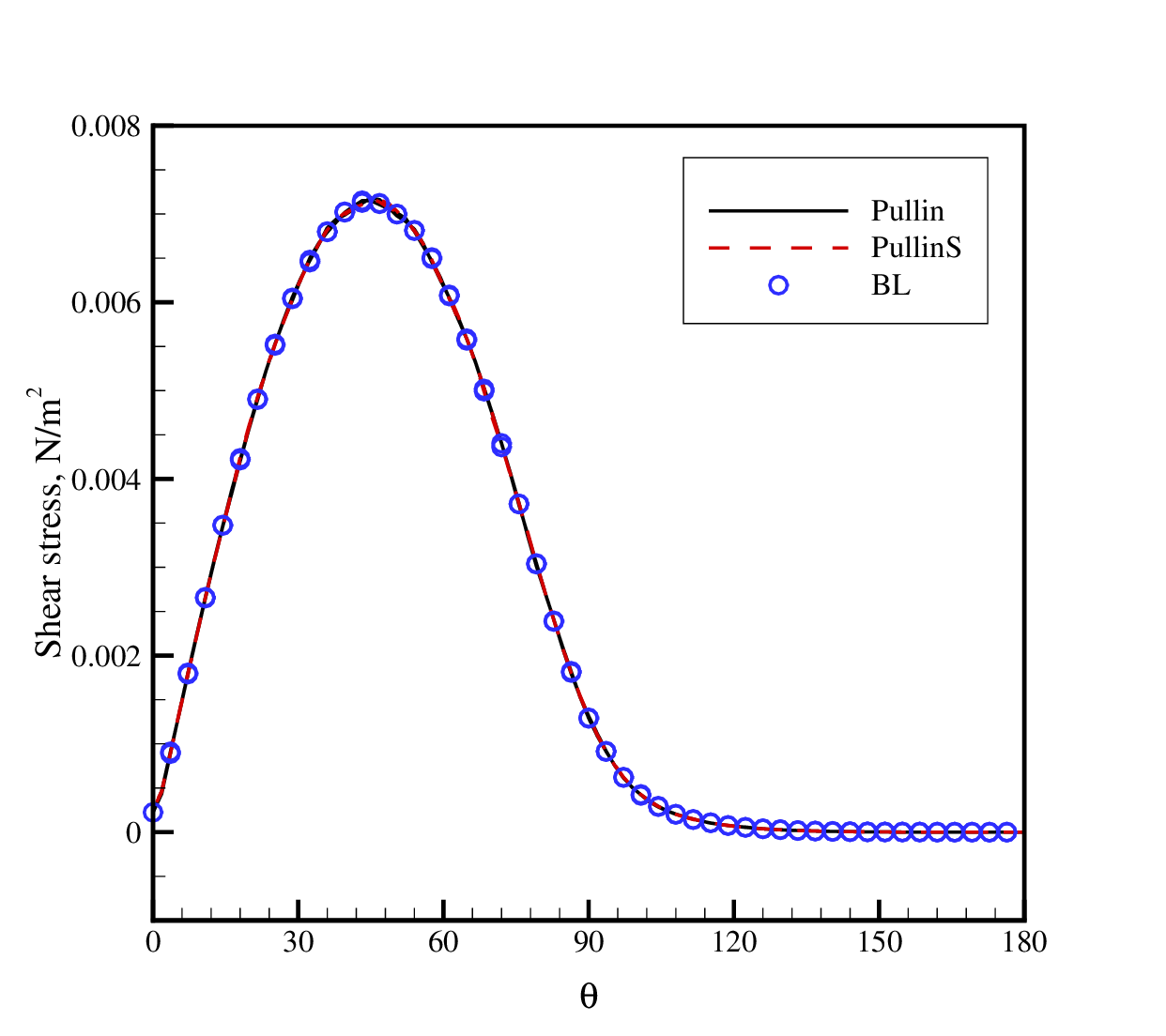}}
	\caption{\label{Cyl-tau} Surface shear stress results of hypersonic cylinder flow at $\mathrm{Ma}=5$: (a) $\mathrm{Kn}=0.01$, (b) $\mathrm{Kn}=0.1$, (c) $\mathrm{Kn}=1.0$, (d) $\mathrm{Kn}=10.0$. Solid line: Pullin model; dashed line: simplified Pullin model; circles: BL model.}
\end{figure}

\begin{figure}
	\centering
	\subfigure[]{\label{CylKn001-q-2}\includegraphics[width=0.45\textwidth]{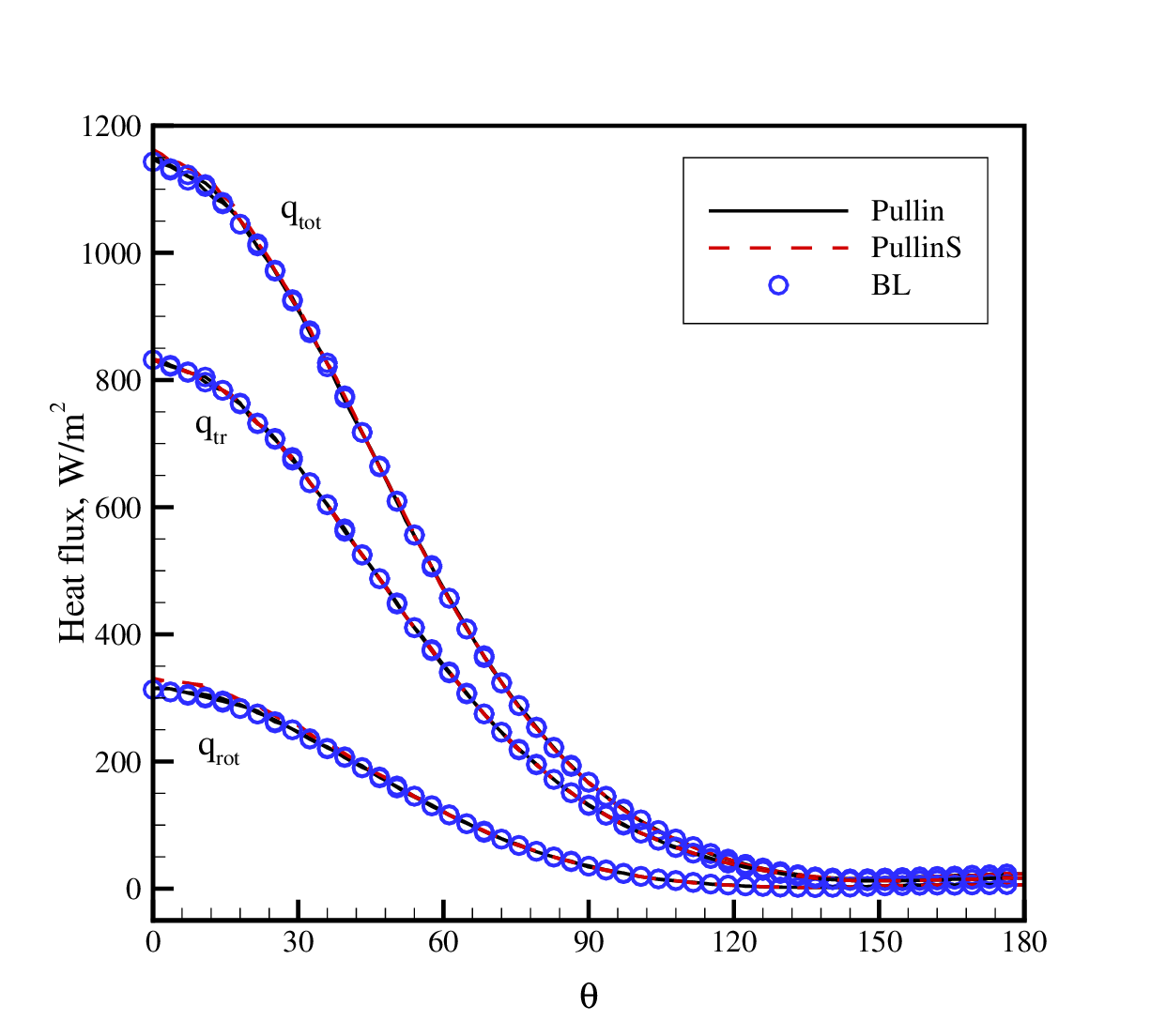}}
	\subfigure[]{\label{CylKn01-q-2}\includegraphics[width=0.45\textwidth]{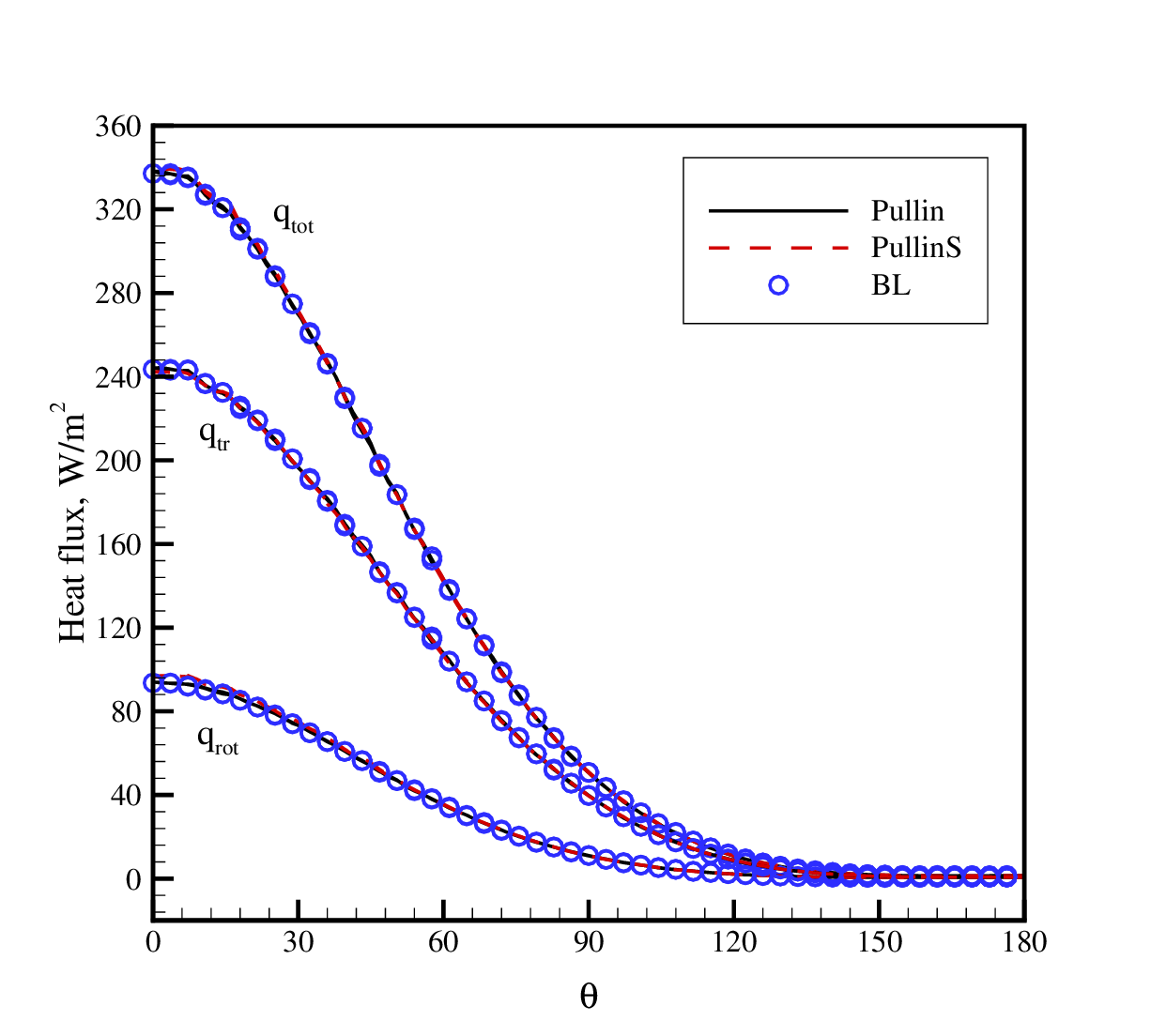}}
	\subfigure[]{\label{CylKn1-q-2}\includegraphics[width=0.45\textwidth]{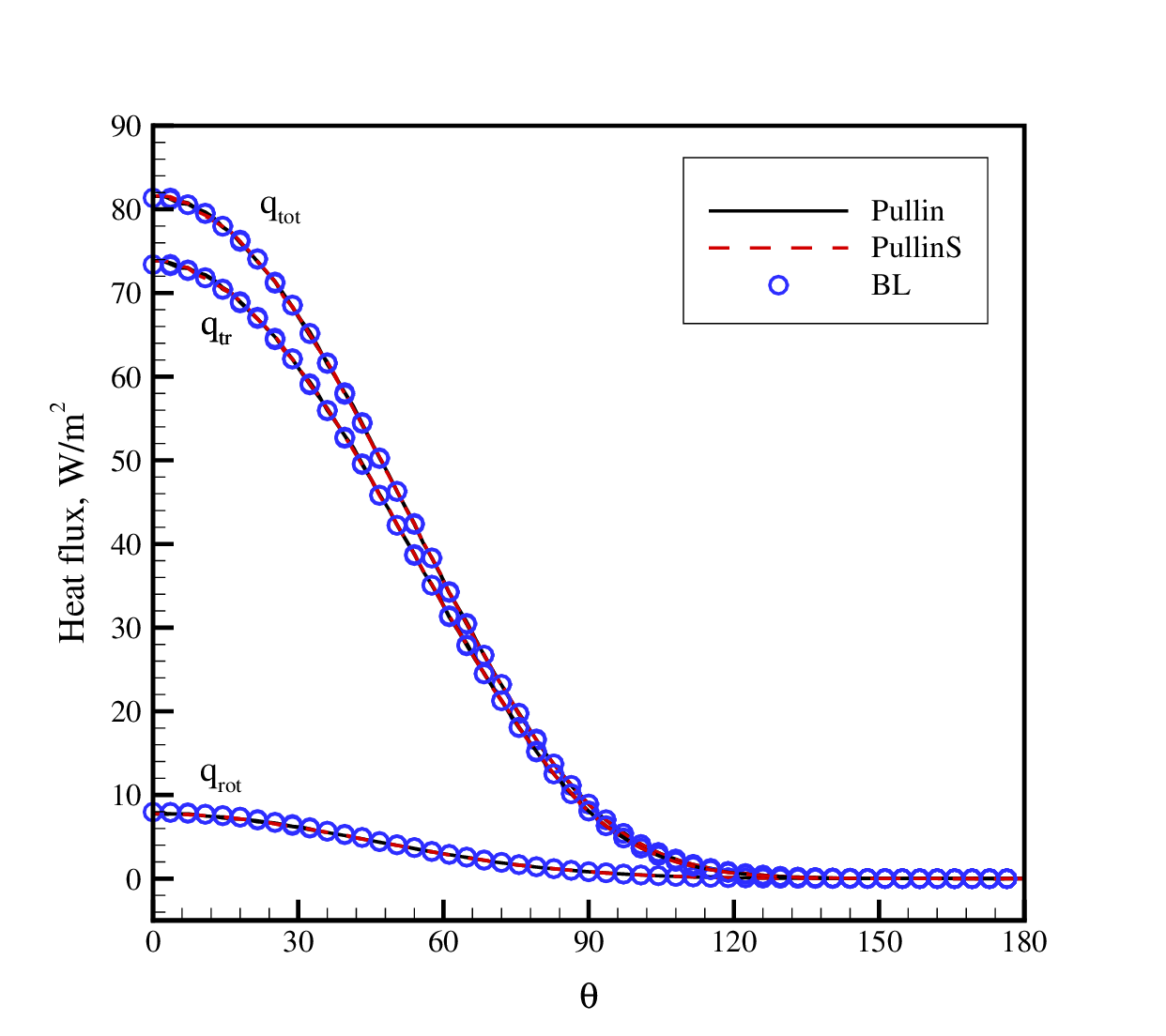}}
	\subfigure[]{\label{CylKn10-q-2}\includegraphics[width=0.45\textwidth]{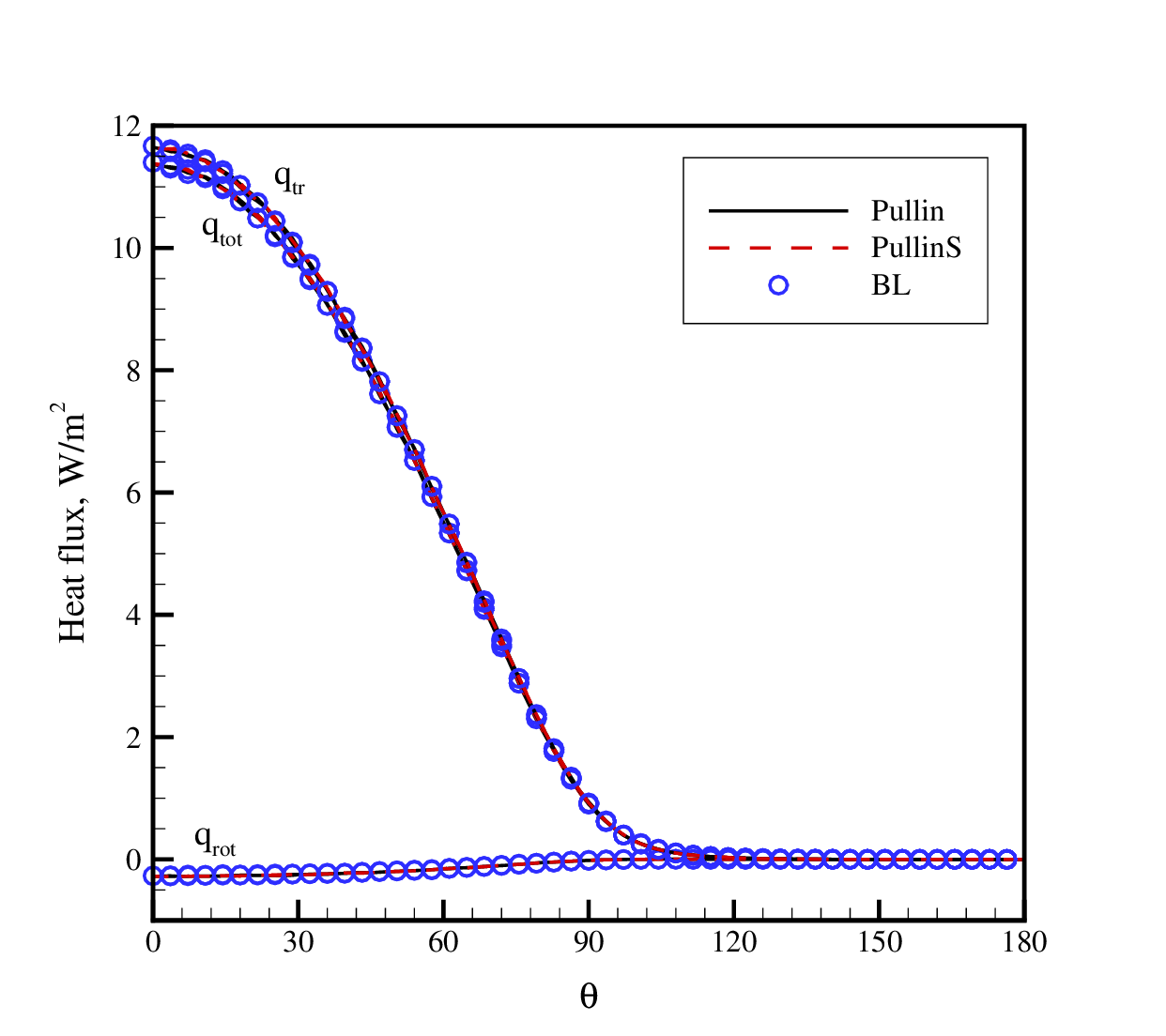}}
	\caption{\label{Cyl-q-2} Surface heat flux results of hypersonic cylinder flow at $\mathrm{Ma}=5$: (a) $\mathrm{Kn}=0.01$, (b) $\mathrm{Kn}=0.1$, (c) $\mathrm{Kn}=1.0$, (d) $\mathrm{Kn}=10.0$. Solid line: Pullin model; dashed line: simplified Pullin model; circles: BL model.}
\end{figure}

Figure \ref{Cyl-pres} presents a comparison of the surface pressure at different Knudsen numbers, while Figure \ref{Cyl-tau} shows the corresponding surface shear stress. It can be observed that the Pullin model and its simplified variant are in good agreement with the BL model reference solutions. Figure \ref{Cyl-q-2} compares the surface heat fluxes for different Knudsen numbers, including the translational and rotational components. Across all four Knudsen numbers, the results from the three models are in good agreement. As the freestream Knudsen number increases, the flow becomes increasingly nonequilibrium, resulting in a gradual decrease in the rotational heat flux and a corresponding increase in the contribution of translational heat flux to the total heat flux. At $\mathrm{Kn} = 10$, the rotational heat flux even becomes negative, reflecting the strong nonequilibrium effects that dominate heat transfer in highly rarefied flows.

The computational times for all test cases are summarized in Table \ref{0D-time}. All computations are performed on the Computing Center in Xi'an using high-performance computing nodes, each equipped with dual Hygon 7285H 32C CPUs (2.5 GHz) and 256 GB memory. The non-linear variation of computational time with Knudsen number is primarily attributed to differences in the computational domain size and the ratio of grid spacing to the mean free path. In the near-continuum regime ($\mathrm{Kn} = 0.01$), the Pullin model is $43.26\%$ slower than the BL model, while its simplified variant is $32.62\%$ slower, limiting the broad applicability of the Pullin model. With increasing rarefaction, the efficiency difference between the Pullin and BL models decreases, and under highly rarefied conditions ($\mathrm{Kn} = 1$ and $10$), their computational costs are comparable.

\begin{table}
    \centering
    \caption{\label{Time of 2D-Cyl} Comparison of the computational times required by different models for hypersonic flow past a cylinder.}
    \begin{threeparttable}
        \begin{tabular}{p{60pt}<{\centering} p{60pt}<{\centering}  p{85pt}<{\centering} p{55pt}<{\centering} p{60pt}<{\centering} p{55pt}<{\centering} p{55pt}<{\centering}}
            \hline
            \hline
            Case & Models & No. of particles & $N_{\mathrm{step}}$ & CPU cores & Time(h) & CPU hours \\
            \hline
            \multirow{3}*{Kn=$10^{-2}$} & Pullin & 36 932 194 & 80 000 & 60 & 10.1 & 606 \\
            ~ & PullinS & 36 936 927 & 80 000 & 60 & 9.35 & 561\\
            ~ & BL & 36 935 474 & 80 000 & 60 & 7.05 & 423\\
            \hline
            \multirow{3}*{Kn=$10^{-1}$} & Pullin & 6 767 054 & 80 000 & 30 & 1.39 & 41.7 \\
            ~ & PullinS & 6 768 071 & 80 000 & 30 & 1.28 & 38.4\\
            ~ & BL & 6 766 777 & 80 000 & 30 & 1.07 & 32.1\\
            \hline
            \multirow{3}*{Kn=$1$} & Pullin & 13 656 471 & 80 000 & 30 & 2.14 & 64.2 \\
            ~ & PullinS & 13 659 587 & 80 000 & 30 & 2.13 & 63.9 \\
            ~ & BL & 13 656 578 & 80 000 & 30 & 2.1 & 63 \\
            \hline
            \multirow{3}*{Kn=$10$} & Pullin & 45 656 769 & 80 000 & 60 & 4.76 & 285.6 \\
            ~ & PullinS & 45 656 113 & 80 000 & 60 & 4.76 & 285.6 \\
            ~ & BL & 45 654 716 & 80 000 & 60 & 4.73 & 283.8 \\
            \hline
            \hline
        \end{tabular}
    \end{threeparttable}
\end{table}

\subsection{Hypersonic rarefied flow around an X38-like vehicle}\label{subSec4_X38}
\begin{figure}
    \centering
    \includegraphics[width=0.8\textwidth, trim = 1mm 1mm 1mm 1mm, clip]{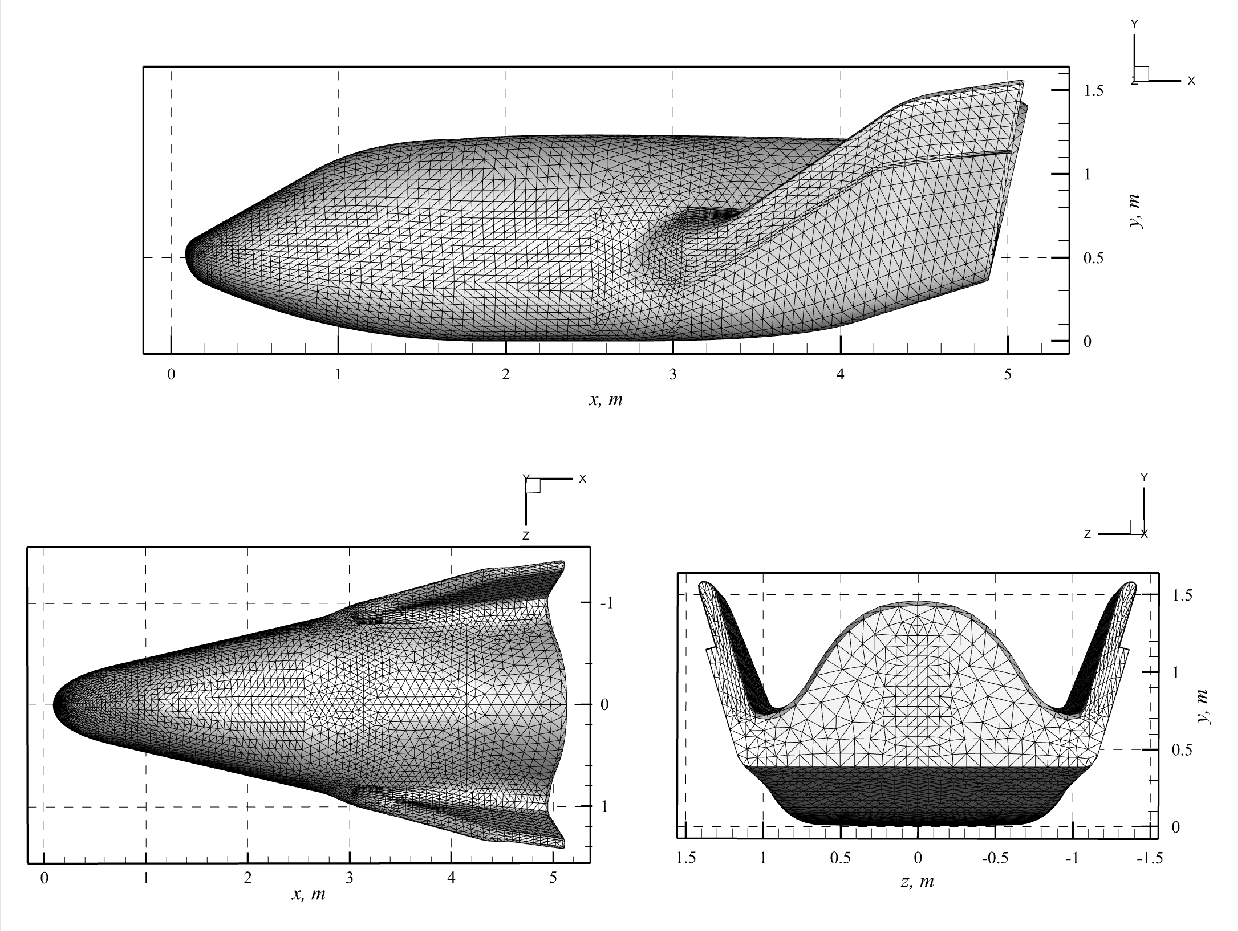}
    \caption{\label{3D-X38-geom} Sketch of the X38-like vehicle (17,414 triangles in the surface mesh).}
\end{figure}

To further evaluate the practical applicability of the Pullin model in aerospace flows, the hypersonic flow around the X38-like vehicle is investigated. A $1{:}16.7$ scaled configuration at four Knudsen numbers with $\mathrm{Ma}=8$, as well as the full-scale configuration at altitudes of 70-100 km with $\mathrm{Ma}=10$ and $20$, were previously studied by Jiang et al.~\cite{jiang2019implicit, jiang2022nonlinear} using the UGKS and DSMC methods. In the present study, the full-scale X38-like vehicle at altitudes of 90 km and 100 km is simulated using the DSMC method coupled with the Pullin model. A schematic of the X38-like vehicle is shown in Fig.~\ref{3D-X38-geom}, where the reference length $L_{ref}$ is $4.67 m$ and the reference area $A_{ref}$ is $5.86 m^2$. The corresponding freestream parameters are summarized in Table \ref{X38-Parameters}. The working gas is air, with a reference diameter of 4.19\r{A} and a viscosity index of 0.77 for the VHS model at 273 K. The freestream Mach number is set to $\mathrm{Ma}=10$, with an angle of attack of $\alpha=20^{\circ}$, and both translational and rotational temperatures are initialized to $T_{tr}=T_{rot}=T_{\infty}$. The wall temperature is maintained at $T_{wall}=300 ,\mathrm{K}$ under a fully diffusive boundary condition. The rotational collision number in the BL model is set to $Z_{\mathrm{BL}}=5$, while the corresponding parameter $Z$ in the Pullin model is evaluated from Eq.~(\ref{eq_Z_and_ZBL}). At an altitude of 100 km, the computational domain is discretized into 6,750,000 cells, with 20 simulated particles initialized per cell; whereas at 90 km, the domain is discretized into 125,000,000 cells, with 10 simulated particles per cell. All results are obtained by averaging over 40,000 time steps after reaching steady state.

\begin{table}
    \centering
    \caption{\label{X38-Parameters} Freestream conditions for hypersonic flow around an X38-like vehicle at altitudes of 90 and 100 km.}
    \begin{tabular}{p{70pt}<{\centering} p{30pt}<{\centering} p{30pt}<{\centering} p{80pt}<{\centering} p{48pt}<{\centering} p{48pt}<{\centering} p{30pt}<{\centering} p{65pt}<{\centering} p{30pt}<{\centering}}
        \hline
        \hline
        Altitude(km) & Gas & $\mathrm{Ma}$ & $n_{\infty}$($m^{-3}$) & $T_{\infty}$($\mathrm{K}$) & $T_{wall}$($\mathrm{K}$) & $\alpha$ & $R$($\mathrm{J/(kg \cdot K)}$) & $Z_{\mathrm{BL}}$\\
        \hline
        90 & Air & 10 & $7.116\times10^{19}$ & 186.867 & 300 & $20^{\circ}$ & 286.71 & 5.0 \\
        100 & Air & 10 & $1.189\times10^{19}$ & 195.081 & 300 & $20^{\circ}$ & 286.71 & 5.0 \\
        \hline
        \hline
    \end{tabular}
\end{table}

\begin{figure}
	\centering
	\subfigure[]{\label{X38-90-Ttr}\includegraphics[width=0.45\textwidth]{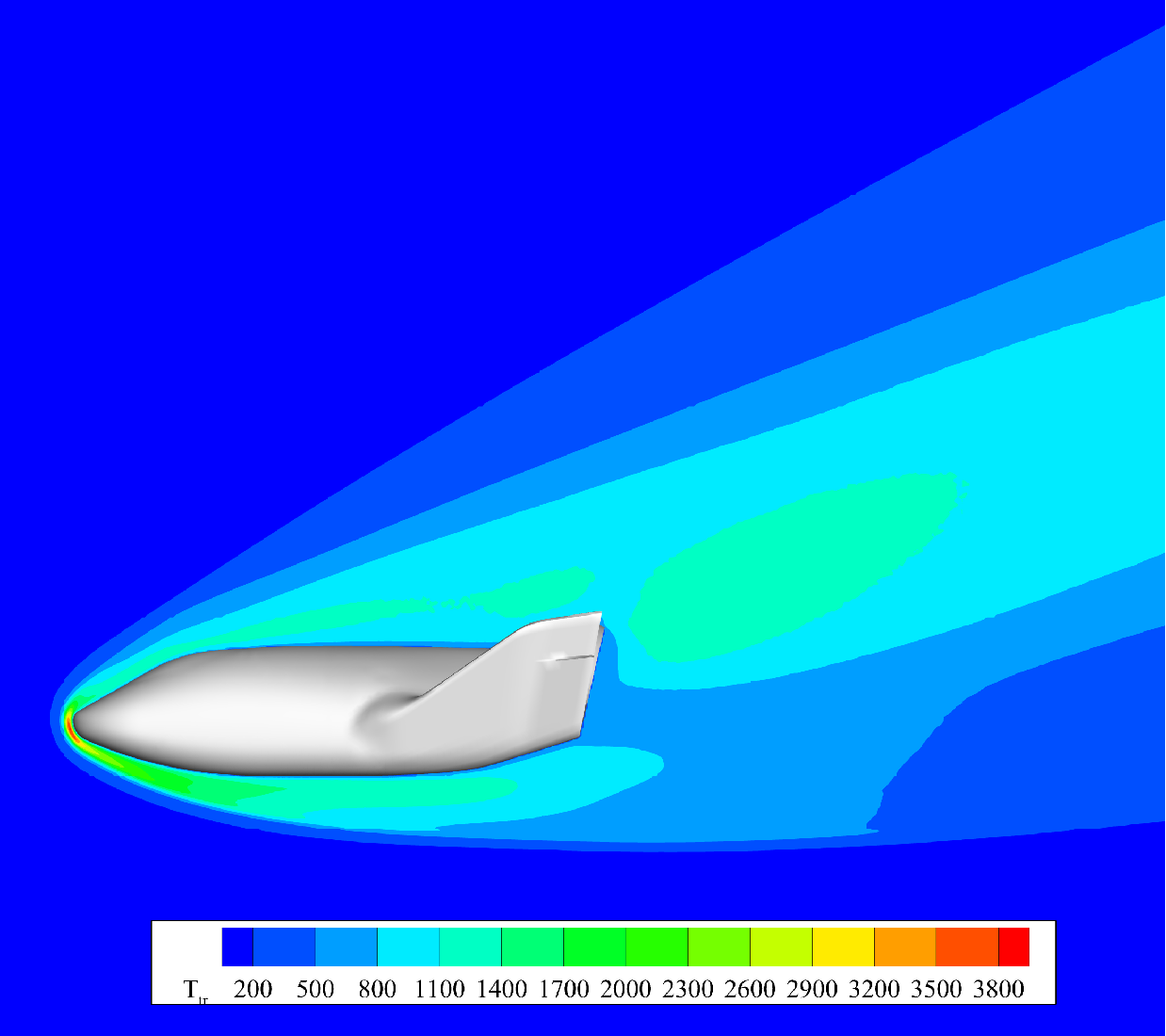}}
	\subfigure[]{\label{X38-90-Trot}\includegraphics[width=0.45\textwidth]{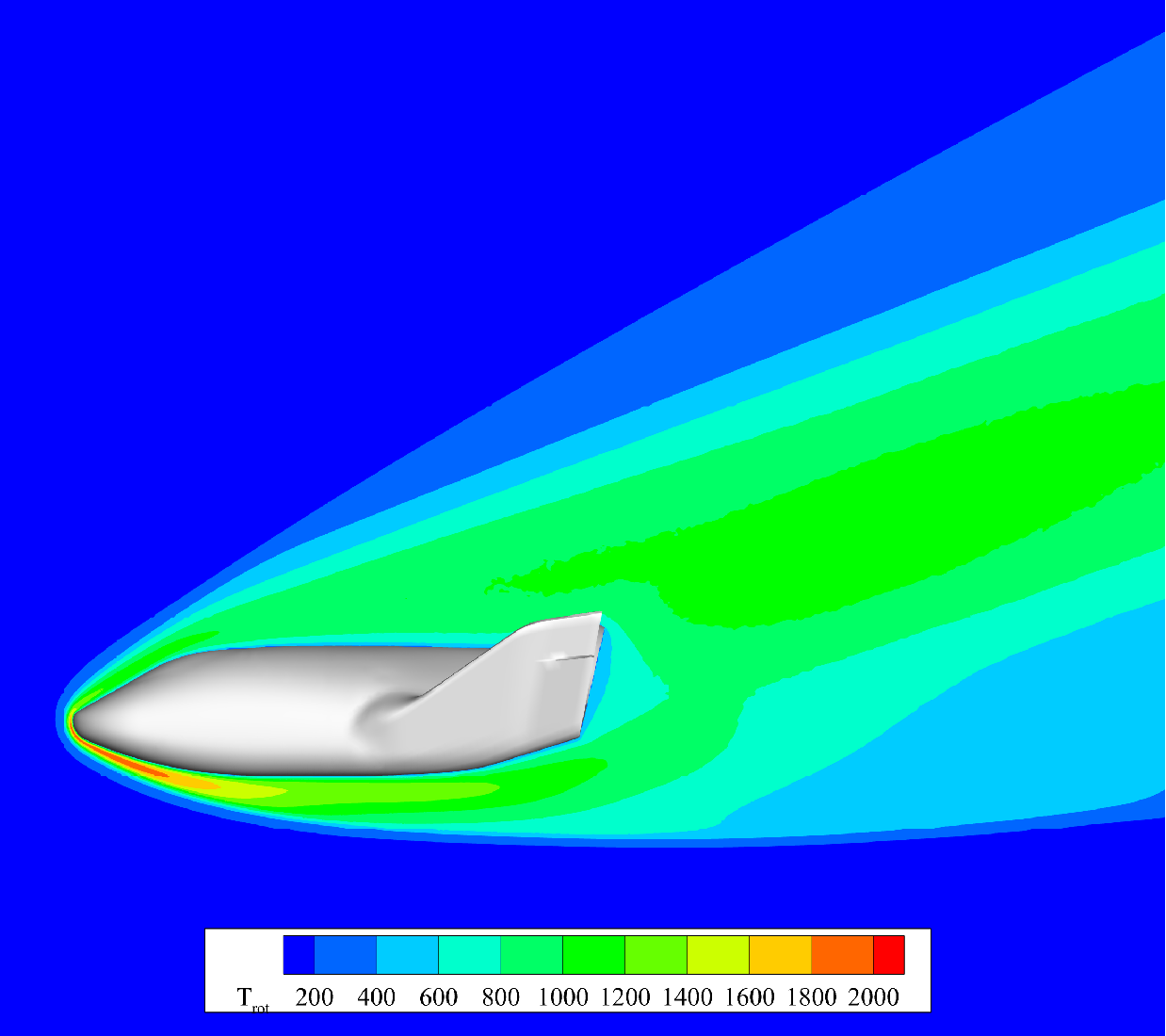}}
	\caption{\label{X38-90-T} Temperature contour results for hypersonic flow around an X38-like vehicle at 90 km altitude: (a) translational temperature, (b) rotational temperature.}
\end{figure}

\begin{figure}
	\centering
	\subfigure[]{\label{X38-100-Ttr}\includegraphics[width=0.45\textwidth]{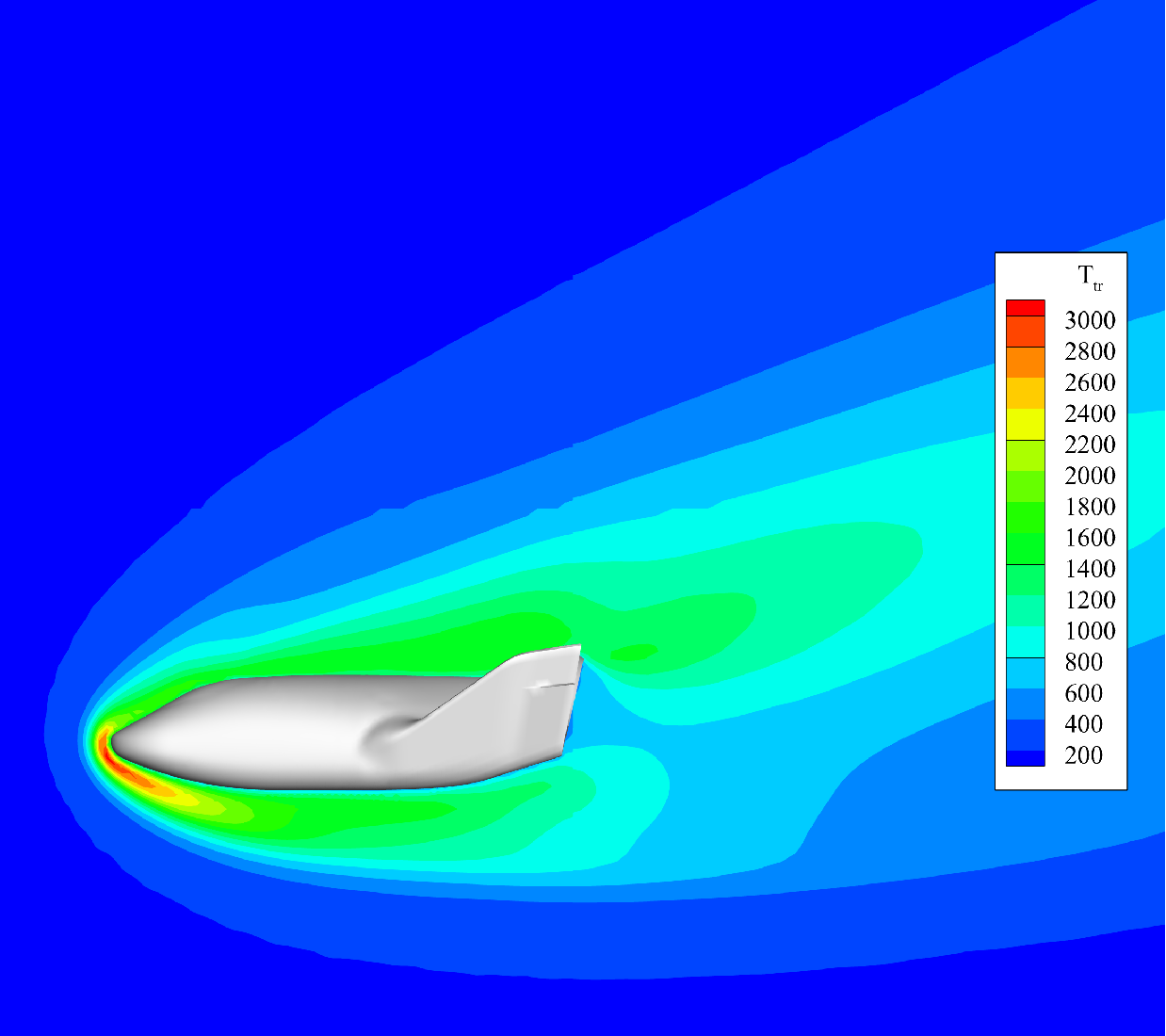}}
	\subfigure[]{\label{X38-100-Trot}\includegraphics[width=0.45\textwidth]{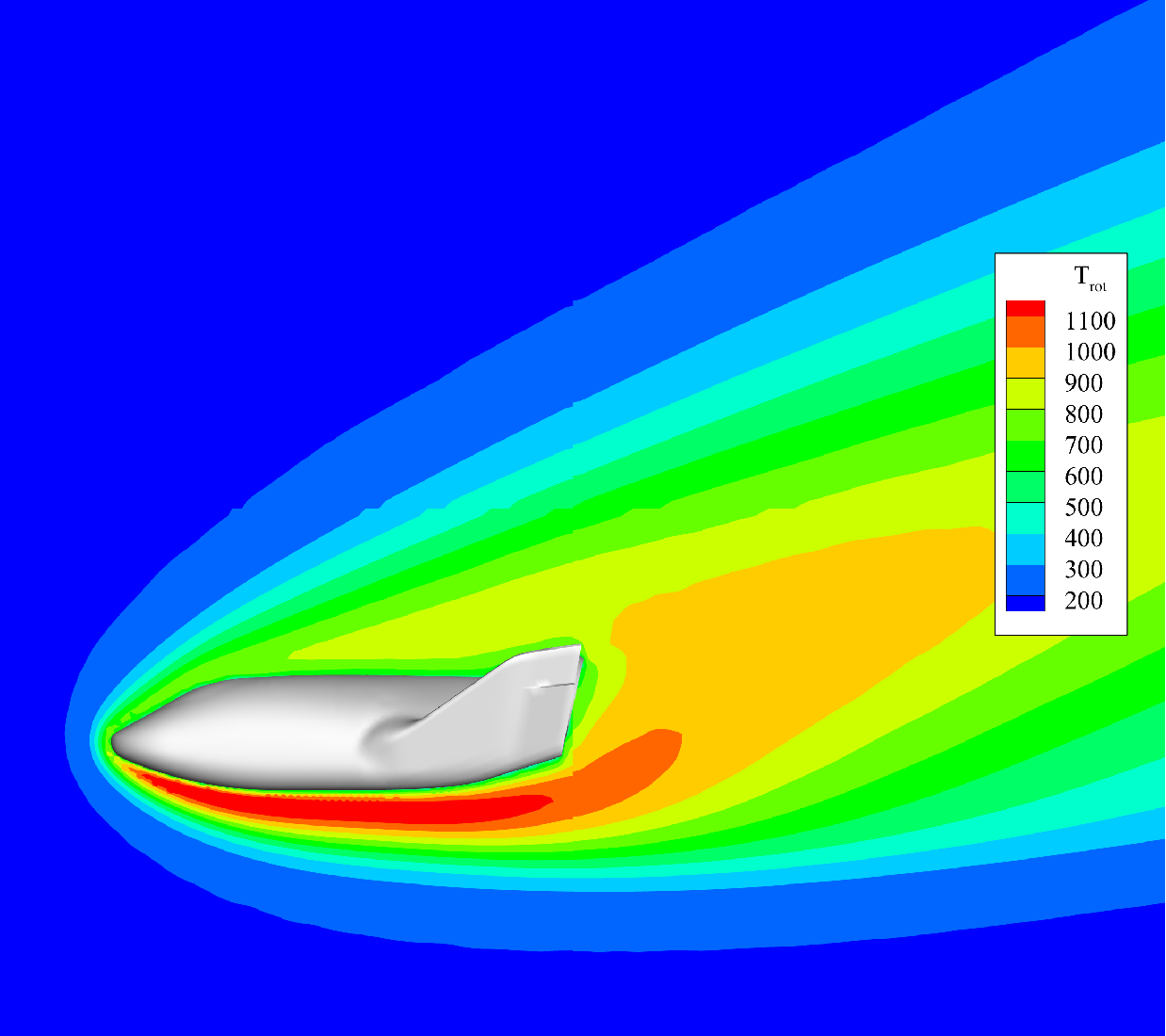}}
	\caption{\label{X38-100-T} Temperature contour results for hypersonic flow around an X38-like vehicle at 100 km altitude: (a) translational temperature, (b) rotational temperature.}
\end{figure}

\begin{figure}
	\centering
	\subfigure[]{\label{X38-90-pres}\includegraphics[width=0.45\textwidth]{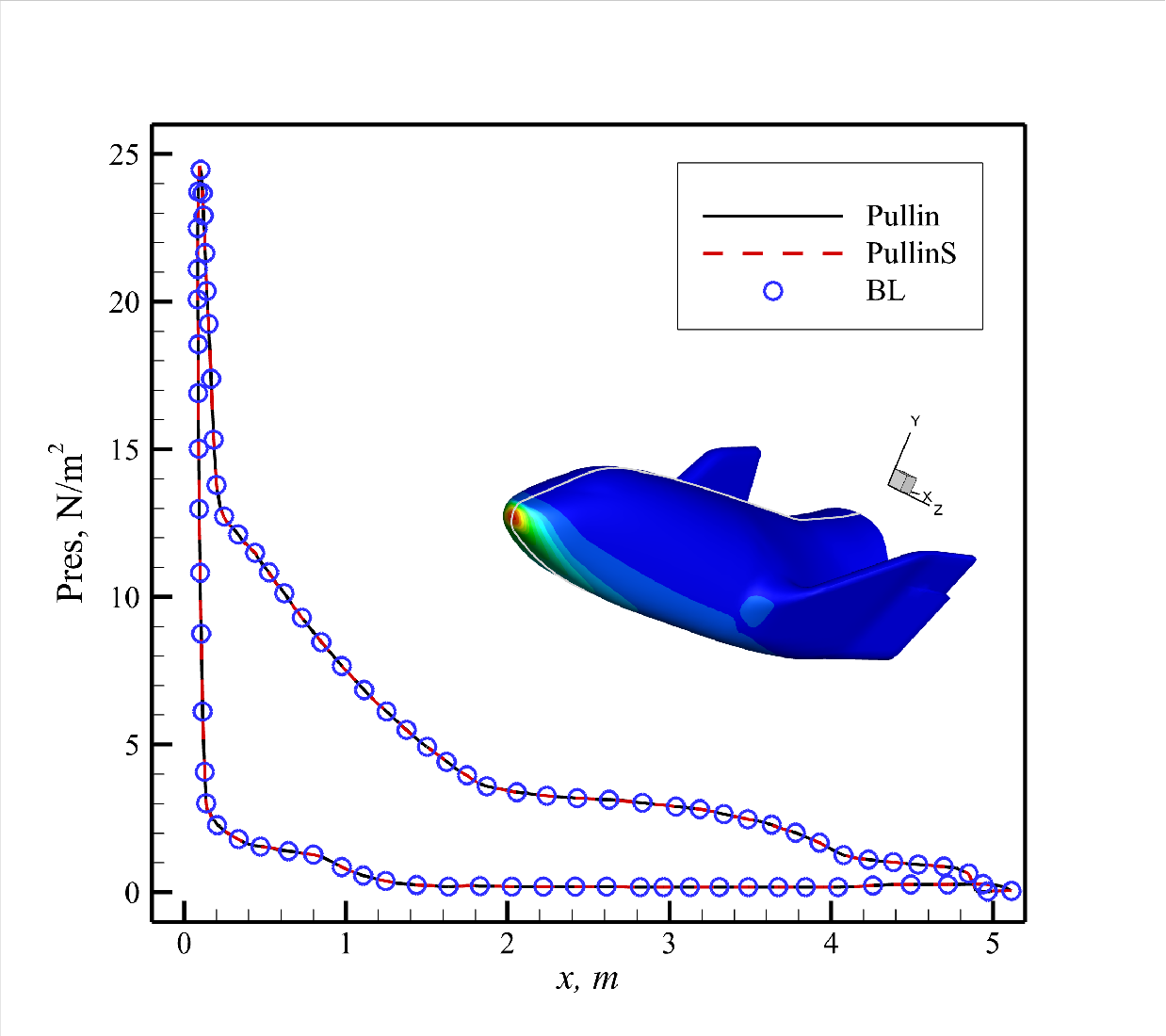}}
	\subfigure[]{\label{X38-90-tau}\includegraphics[width=0.45\textwidth]{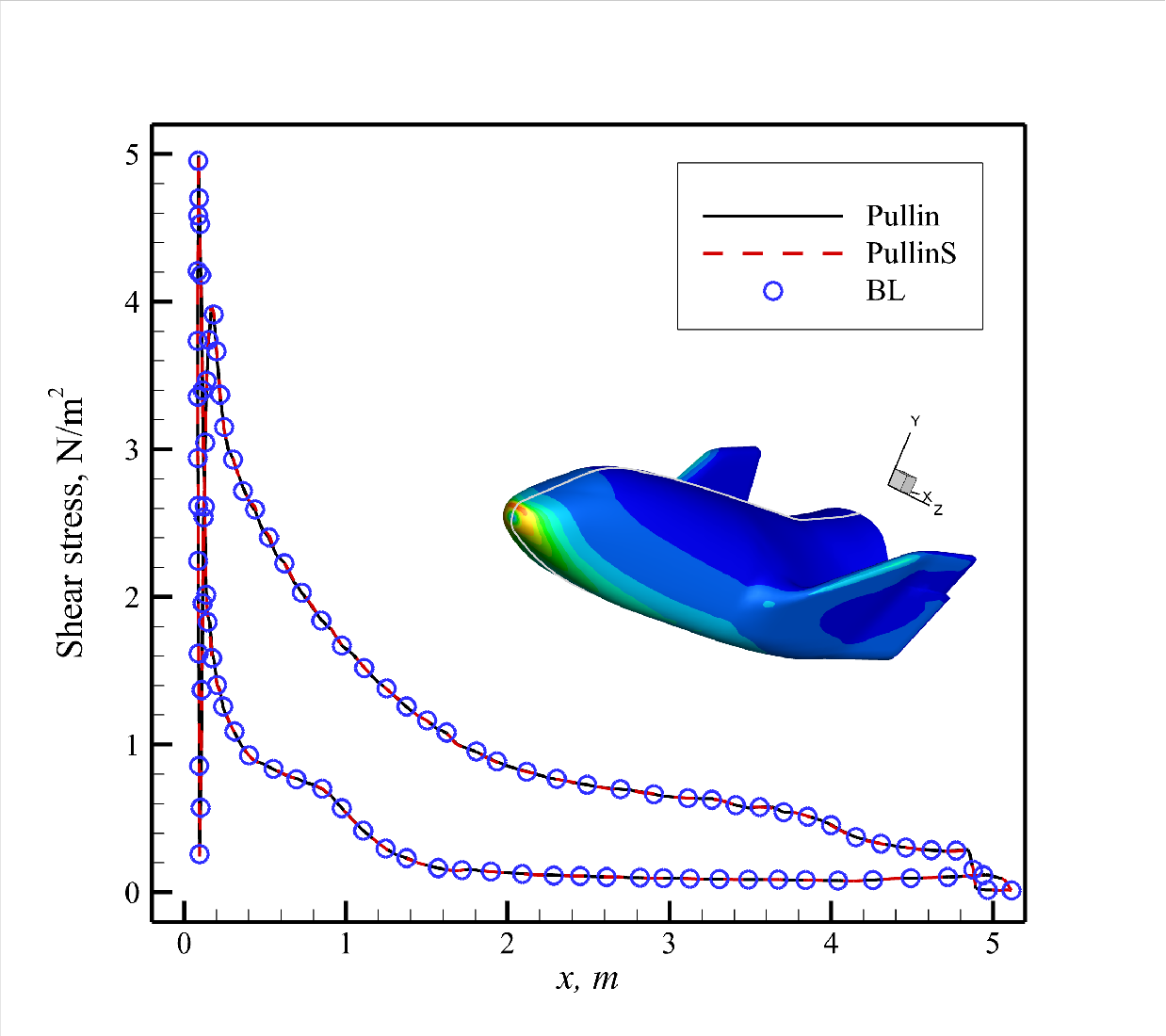}}
	\subfigure[]{\label{X38-90-q}\includegraphics[width=0.45\textwidth]{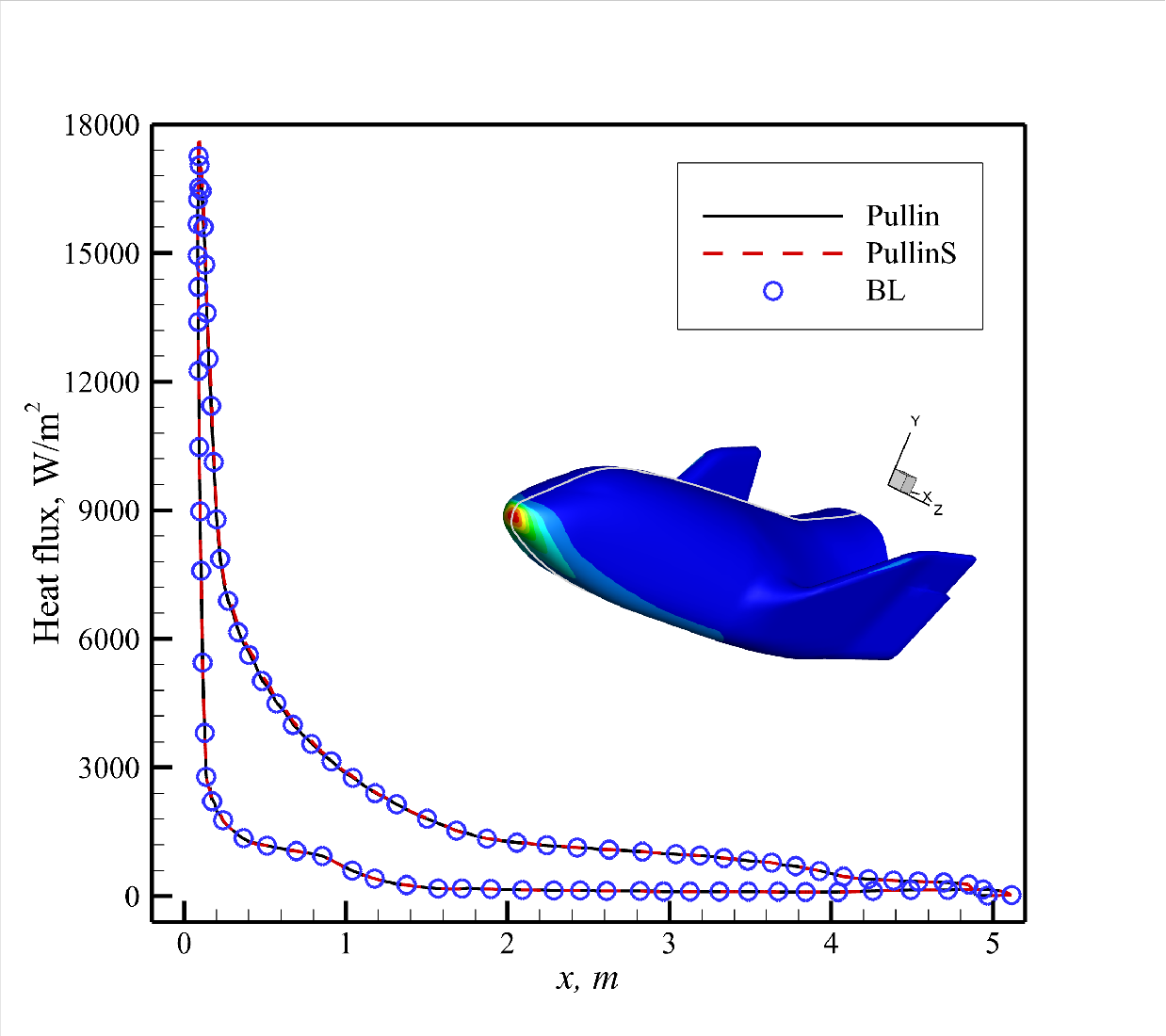}}
	\caption{\label{X38-90-surf} Surface results for hypersonic flow around an X38-like vehicle at 90 km altitude: (a) pressure, (b) shear stress, (c) heat flux. Solid line: Pullin model; dashed line: simplified Pullin model; circles: BL model.}
\end{figure}

\begin{figure}
	\centering
	\subfigure[]{\label{X38-100-pres}\includegraphics[width=0.45\textwidth]{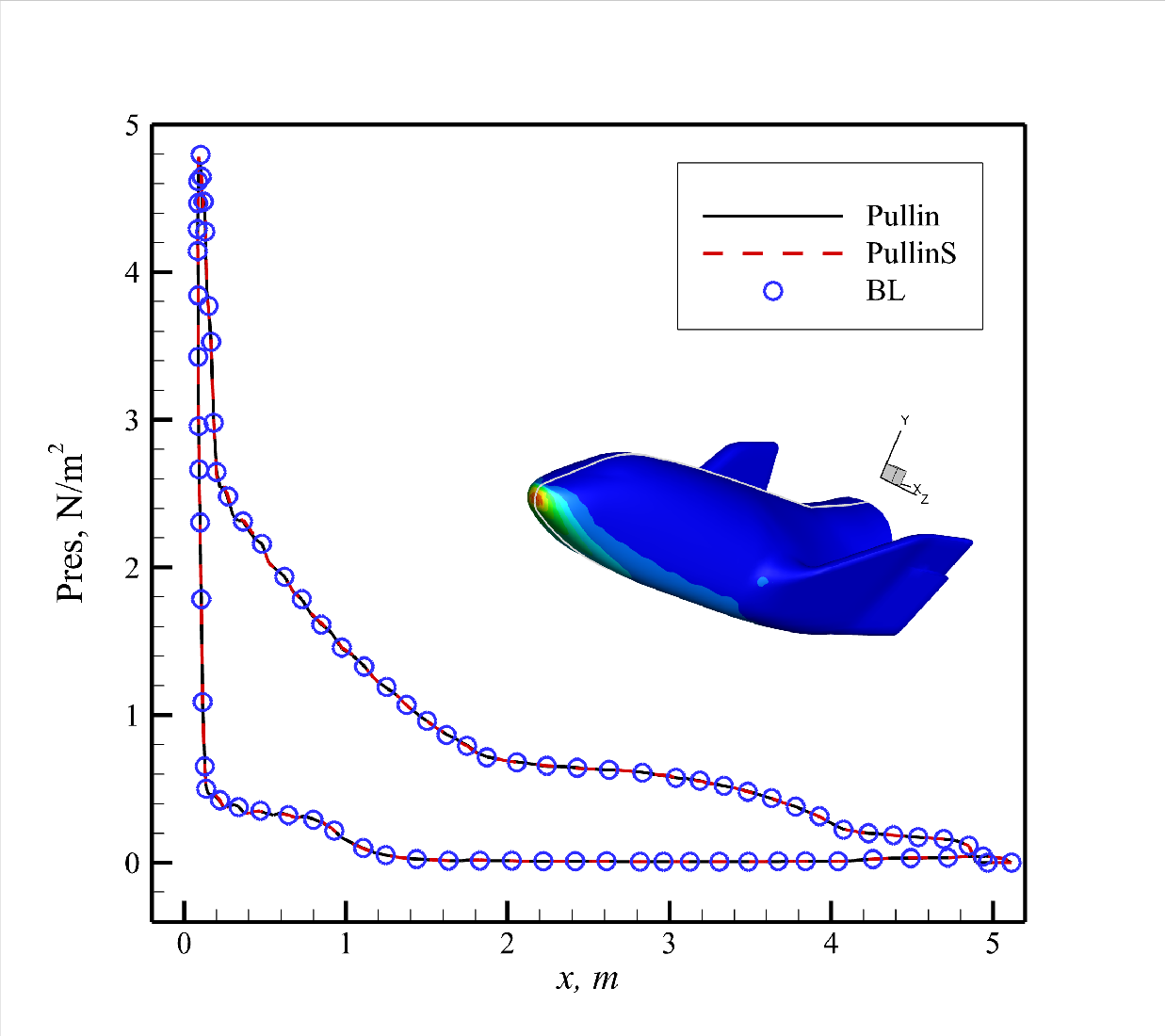}}
	\subfigure[]{\label{X38-100-tau}\includegraphics[width=0.45\textwidth]{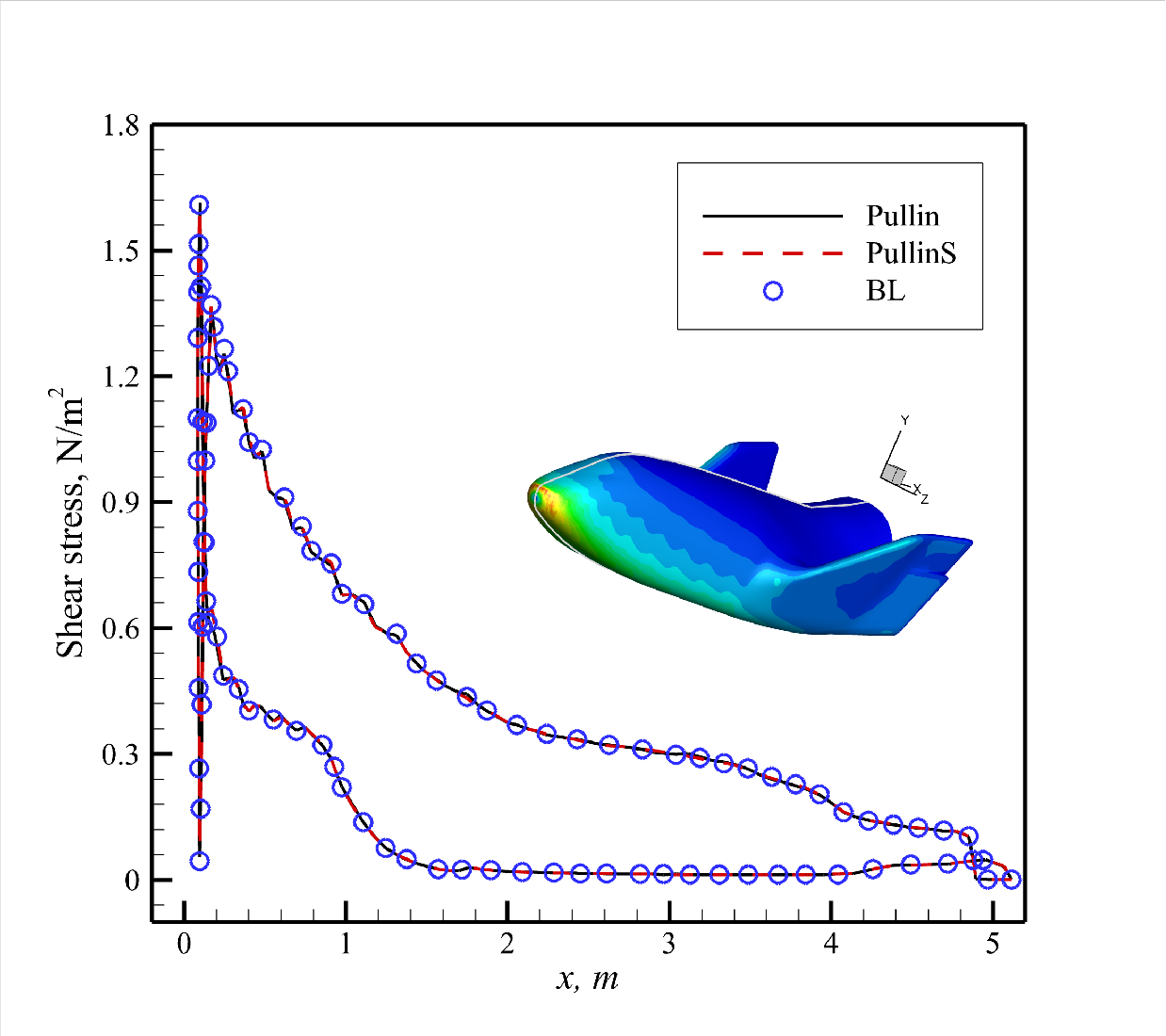}}
	\subfigure[]{\label{X38-100-q}\includegraphics[width=0.45\textwidth]{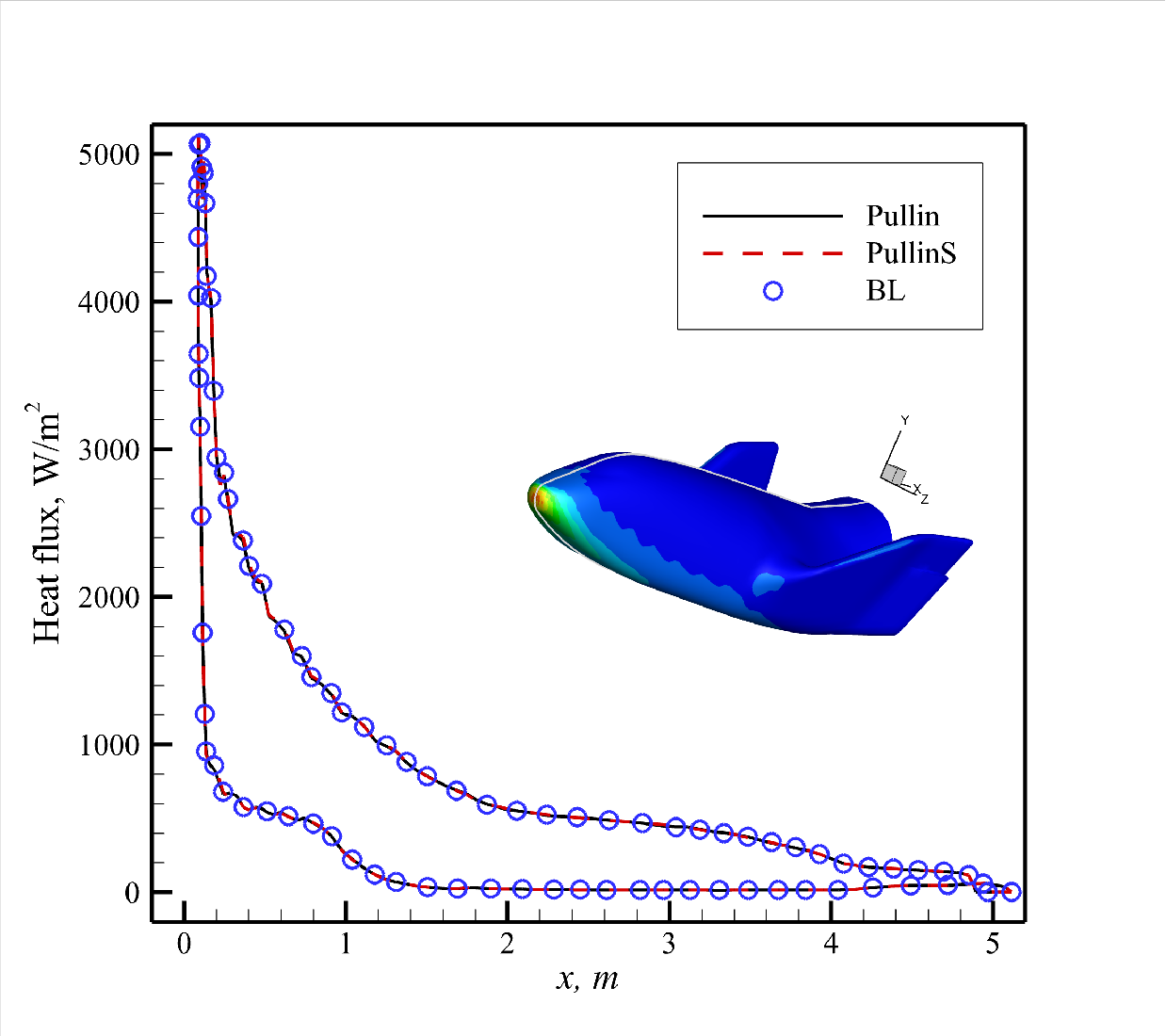}}
	\caption{\label{X38-100-surf} Surface results for hypersonic flow around an X38-like vehicle at 100 km altitude: (a) pressure, (b) shear stress, (c) heat flux. Solid line: Pullin model; dashed line: simplified Pullin model; circles: BL model.}
\end{figure}

Figures \ref{X38-90-T} and \ref{X38-100-T} present the rotational and translational temperature contours predicted by the Pullin model at altitudes of 90 km and 100 km, respectively. At 90 km, the translational and rotational temperatures exhibit notable differences, indicating a significant degree of thermal nonequilibrium. As the flight altitude rises to 100 km, the maximum rotational temperature shifts away from the stagnation point, and the high-temperature region extends downstream, reflecting the strong nonequilibrium characteristics of the flow. Figures \ref{X38-90-surf} and \ref{X38-100-surf} compare the surface distributions of pressure, shear stress, and heat flux predicted by the three internal energy relaxation models at altitudes of 90 km and 100 km, respectively. The results obtained from the Pullin model and its simplified variant show good agreement with those predicted by the BL model. The overall aerodynamic coefficients of the X38-like vehicle are listed in Table \ref{X-38 coeff}. The dimensionless coefficients are defined as
\begin{equation}
    \begin{aligned}
        & C_{L}=\frac{L}{0.5{{\rho }_{\infty }}U_{^{\infty }}^{2}{{A}_{ref}}}, \\
        & C_{d}=\frac{D}{0.5{{\rho }_{\infty }}U_{^{\infty }}^{2}{{A}_{ref}}}, \\
    \end{aligned}
\end{equation}
where $L$ and $D$ denote the lift and drag forces, respectively. The Pullin model and its simplified variant show excellent agreement with BL results, with maximum relative errors of only $0.06\%$ for the lift coefficient and $0.05\%$ for the drag coefficient.

The computational times for all test cases are summarized in Table \ref{Time of X38}. All computations are performed on the Computing Center in Xi'an using high-performance computing nodes, each equipped with dual Hygon 7285H 32C CPUs (2.5 GHz) and 256 GB memory. At an altitude of 90 km, the Pullin model is approximately $20.81\%$ slower than the BL model, while its simplified variant is $13.4\%$ slower. As the flight altitude increases to 100 km, the differences are reduced to $8.36\%$ and $4.1\%$, respectively. For simulations above 100 km, the choice of model has a negligible impact on computational efficiency.

\begin{table}
    \centering
    \caption{\label{X-38 coeff} Comparison of aerodynamic coefficients of the X38-like vehicle at Mach 10.}
    \begin{threeparttable}
        \begin{tabular}{p{65pt}<{\centering} p{70pt}<{\centering} p{50pt}<{\centering} p{50pt}<{\centering} p{70pt}<{\centering} p{50pt}<{\centering} p{70pt}<{\centering}}
            \hline
            \hline
            Altitude(km) & Coefficients & BL & Pullin & Relative error & PullinS & Relative error   \\
            \hline
            \multirow{2}*{90} & $C_{L}$   & 0.3142     & 0.3142    & - & 0.3140   & $-0.06 \%$ \\
            ~ & $C_{d}$   & 0.2708  & 0.2708    & -
            & 0.2709    & $-0.04 \%$ \\
            \hline
            \multirow{2}*{100} & $C_{L}$   & 0.39     & 0.3902    & $0.05 \%$  & 0.3901   & $0.03 \%$ \\
            ~ & $C_{d}$   & 0.5140       & 0.5140    & -
            & 0.5142    & $0.04 \%$ \\
            \hline
            \hline
        \end{tabular}
    \end{threeparttable}
\end{table}

\begin{table}
    \centering
    \caption{\label{Time of X38} Comparison of the computational time required for different models for the hypersonic flow around X38-like vehicle.}
    \begin{threeparttable}
        \begin{tabular}{p{65pt}<{\centering} p{60pt}<{\centering}  p{85pt}<{\centering} p{55pt}<{\centering} p{60pt}<{\centering} p{55pt}<{\centering} p{55pt}<{\centering}}
            \hline
            \hline
            Altitude(km) & Models & No. of particles & $N_{\mathrm{step}}$ & CPU cores & Time(h) & CPU hours \\
            \hline
            \multirow{3}*{90} & Pullin & 1 270 009 678 & 80 000 & 300 & 59.69 & 17907\\
            ~ & PullinS & 1 270 009 737 & 80 000 & 300 & 56.03 & 16809\\
            ~ & BL & 1 269 992 103 & 80 000 & 300 & 49.41 & 14823\\
            \hline
            \multirow{3}*{100} & Pullin & 136 497 915 & 80 000 & 150 & 6.61 & 991.5\\
            ~ & PullinS & 136 495 963 & 80 000 & 150 & 6.35 & 952.5\\
            ~ & BL & 136 489 077 & 80 000 & 150 & 6.1 & 915\\
            \hline
            \hline
        \end{tabular}
    \end{threeparttable}
\end{table}

\section{Conclusions}\label{sec5_Conclusion}
In this work, a new parameterization of the Pullin model for VHS molecules has been proposed, establishing a direct link between the partition parameter and the rotational collision number. Compared with the widely used BL model, the Pullin framework--derived from gas kinetic theory--offers improved physical fidelity by enabling rotational relaxation of internal energy in all simulated particles. With the proposed parameterization, both the full and simplified Pullin models have been successfully extended to diatomic VHS gases. Since the energy exchange requires sampling only three or five Beta-distributed variates, these models can be implemented easily within the DSMC framework.

The accuracy and efficiency of the Pullin model and its simplified variant have been validated through a series of numerical test cases--including zero-dimensional rotational relaxation of nitrogen, one-dimensional planar Couette flow and normal shock wave, two-dimensional hypersonic flow past a cylinder, and three-dimensional hypersonic flow around an X38-like vehicle--and systematically compared with the BL model. In all cases, both the full and simplified Pullin models show excellent agreement with reference solutions from theory, experiments, and the BL model. In the near-continuum regimes (Knudsen number Kn = 0.01, or altitudes below 90 km), the Pullin model is approximately $20\sim 40\%$ slower than the BL model due to the additional computational cost of sampling Beta-distributed variates. However, in the highly rarefied flow regimes of DSMC simulations (Knudsen number greater than 1, or altitudes above 100 km), the simplified Pullin model exhibits performance comparable to the BL model.

Overall, the Pullin model and its simplified variant offer a robust and practical alternative to the BL model within the DSMC framework, providing improved physical fidelity in representing rotational energy relaxation with only a modest computational cost. Their demonstrated accuracy across a wide range of test cases underscores their suitability for hypersonic aerothermodynamic applications. Future work will extend the simplified Pullin model by incorporating vibrational relaxation, thereby enhancing its applicability to high-enthalpy nonequilibrium flows.

%%%%%%%%%%%%%%%%%%%%%%%%%%%
\section*{Acknowledgements}
The authors thank Mr. Rui Zhang and Mr. Jianfeng Chen for their valuable discussions on kinetic models of internal energy relaxation. This work was financially supported by the National Natural Science Foundation of China (Grants 12172301), and the Program of Introducing Talents of Discipline to Universities (111 Project of China, Grant B17037).

\section*{DATA AVAILABILITY}
The data that support the findings of this study are available from the corresponding author upon reasonable request.
	
\clearpage

\bibliography{Reference}

\end{document}